\documentclass[10pt,journal,compsoc]{IEEEtran}
\usepackage{amsmath,amsfonts}
\usepackage{algorithm}
\usepackage{array}

\usepackage[caption=false,font=footnotesize,labelfont=rm,textfont=rm]{subfig}

\usepackage{textcomp}
\usepackage{stfloats}
\usepackage{url}
\usepackage{verbatim}
\usepackage{graphicx}
\usepackage{cite}
\usepackage{colortbl}
\usepackage{pifont}
\usepackage{textcomp}
\usepackage{stfloats}
\usepackage{threeparttable}
\usepackage{url}
\usepackage{verbatim}
\usepackage{graphicx}
\usepackage{cite}
\usepackage{diagbox}
\usepackage{bm}
\usepackage{bbm}
\usepackage{soul}
\usepackage{amsfonts}

\usepackage[colorlinks=true,linkcolor=blue,citecolor=blue,urlcolor=blue,backref=page]{hyperref}

\usepackage{multirow}
\usepackage{algpseudocode}
\usepackage{makecell}
\usepackage{booktabs}
\usepackage{mdframed}

\usepackage{CJKutf8}
\newcommand{\cn}[1]{\begin{CJK}{UTF8}{gbsn}#1\end{CJK}} 

\usepackage{booktabs}
\usepackage{bbding}
\usepackage[top=0.4in,bottom=0.4in,left=0.4in,right=0.4in]{geometry}

\usepackage{array}
\usepackage{ulem}
\usepackage{ragged2e}
\usepackage{amsthm}
\usepackage{utfsym}
\usepackage{wasysym}

\usepackage{color}
\definecolor{s1}{RGB}{13,169,89}
\definecolor{s2}{RGB}{189,2,0}

\definecolor{t1}{RGB}{255,25,25}
\definecolor{t2}{RGB}{25,25,255}

\definecolor{mycolor}{RGB}{169,189,211}

\newcommand{\etal}{\textit{et al}. }

\newcommand{\ie}{\textit{i}.\textit{e}., }
\newcommand{\eg}{\textit{e}.\textit{g}., }

\begin{document}
	
	\title{\textcolor{black}{S$^4$ST: A Strong, Self-transferable, faSt, and Simple Scale Transformation for Data-free Transferable Targeted Attack}}
	
	\author{Yongxiang Liu, Bowen Peng, Li Liu, Xiang Li
		\IEEEcompsocitemizethanks{
			\IEEEcompsocthanksitem The authors are with the College of Electronic Science and Technology, National University of Defense Technology (NUDT), Changsha 410073, China. Email: lyx\_bible@sina.com, pbow16@nudt.edu.cn, liuli\_nudt@nudt.edu.cn, lixiang01@vip.sina.com.
		}
		\thanks{Corresponding Author: Yongxiang Liu, Li Liu and Xiang Li.}
		\thanks{This work was supported partially by the National Key Research and Development Program of China under Grant 2021YFB3100800, the National Natural Science Foundation of China (NSFC) under Grant 62376283 and 62531026, the Innovation Research Foundation of National University of Defense Technology (JS2023-03), and the Fundamental and Interdisciplinary Disciplines Breakthrough Plan of the Ministry of Education of China (JYB2025XDXM110).}
	}
	
	\markboth{Accepted by IEEE TPAMI}{Yongxiang Liu}

	\IEEEtitleabstractindextext{
		\begin{abstract}
			\justifying
			Transferable Targeted Attacks (TTAs) face significant challenges due to severe overfitting to surrogate models. \textcolor{black}{Recent breakthroughs heavily rely on large-scale training data of victim models, while data-free solutions, \textit{i.e.}, image transformation-involved gradient optimization, often depend on black-box feedback for method design and tuning. These dependencies violate black-box transfer settings and compromise threat evaluation fairness.} In this paper, we propose two blind estimation measures, self-alignment and self-transferability, to analyze per-transformation effectiveness and cross-transformation correlations under strict black-box constraints. Our findings challenge conventional assumptions: (1) Attacking simple scaling transformations uniquely enhances targeted transferability, outperforming other basic transformations and rivaling leading complex methods; (2) Geometric and color transformations exhibit high internal redundancy despite weak inter-category correlations. These insights drive the design and tuning of S$^4$ST (Strong, Self-transferable, faSt, Simple Scale Transformation), which integrates dimensionally consistent scaling, complementary low-redundancy transformations, and block-wise operations. \textcolor{black}{Extensive evaluations across diverse architectures, training distributions, and tasks show that S$^{4}$ST achieves state-of-the-art effectiveness-efficiency balance without data dependency. We reveal that scaling's effectiveness stems from visual data's multi-scale nature and ubiquitous scale augmentation during training, rendering such augmentation a double-edged sword. Further validations on medical imaging and face verification confirm the framework's strong generalization.} 
			Codes are available at \url{https://github.com/scenarri/S4ST}.
		\end{abstract}
		}

	\maketitle
	\IEEEdisplaynontitleabstractindextext
	\IEEEpeerreviewmaketitle

	\section{Introduction}
	
	\IEEEPARstart{D}{eep} Neural Networks (DNNs) have been shown to be vulnerable to adversarial examples (AEs)~\cite{szegedy2013intriguing}, which refer to images with small, intentionally crafted perturbations that are usually imperceptible to the human visual system but can fool DNNs. The presence of AEs raises significant concerns when deploying DNNs in security-critical domains, including autonomous driving \cite{wang2022fca,cao2019adversarial}, face recognition \cite{ODI}, and remote sensing \cite{peng2025physical}, regarding both digital and real-world \cite{song2018physical} scenarios. Therefore, finding, understanding, and defending AEs have received significant attention \cite{madry2018towards,tramer2018ensemble,10214340,10478545}. Crucially, AEs exhibit transferability \cite{harnessing2015goodfellow}, successfully fooling unseen victims based on a surrogate model. This property enables black-box transfer attacks, driving efforts to enhance transferability \cite{gusurvey}.
	
	While untargeted transferability is well-studied, yielding a substantial body of research \cite{momentum2018dong,evading2019dong,inputdiversity2019xie,SIA,BSR,decowa}, Targeted Transferable Attacks (TTAs) are significantly more challenging and consequently less explored. This is primarily due to the difficulties of driving samples into specific target decision spaces across unknown architectures. To form robust target semantics \cite{domainfeatureuap, naseer2021generating}, representative methods optimize universal perturbations \cite{weng2023exploring}, generators \cite{naseer2021generating, zhao2023minimizing}, or the surrogate model itself \cite{zhu2021rethinking, zhu2022toward, Wu_2024_CVPR}. However, relying on extensive data to model distributions categorizes them as \textit{data-reliant TTAs}, presenting two major drawbacks:
	\begin{enumerate}
		\item They often exploit victim training/testing data, causing data leakage that violates strict black-box settings and impedes fair threat evaluation.
		\item They require massive training resources. For instance, generative attacks \cite{zhao2023minimizing} train class-specific generators, hindering broader applications like cross-dataset transfer or adversarial training.
	\end{enumerate}
	
	In this paper, we focus on \textit{data-free TTAs}, also known as the gradient-based attacks \cite{momentum2018dong,evading2019dong}, which efficiently optimize any \textit{(sample, label)} pair but suffer from severe overfitting and poor transferability \cite{liu2016delving}. While often attributed to cross-entropy gradient vanishing, prompting alternative loss functions \cite{PoTrip,logit,marginangleT,AOSBLL}, input transformations actually play a more critical role \cite{inputdiversity2019xie,wei2023rethinking}. As Zhao \etal \cite{logit} showed, AEs rarely transfer without them. Crucially, despite being termed ``data-free", existing transformation methods still rely on heuristic choice or implicit black-box feedback for design and tuning. Guidelines for strict black-box transformation design remain absent, leaving their potential largely untapped.
	
	\begin{figure*}[tbp]
		\includegraphics[width=1\linewidth]{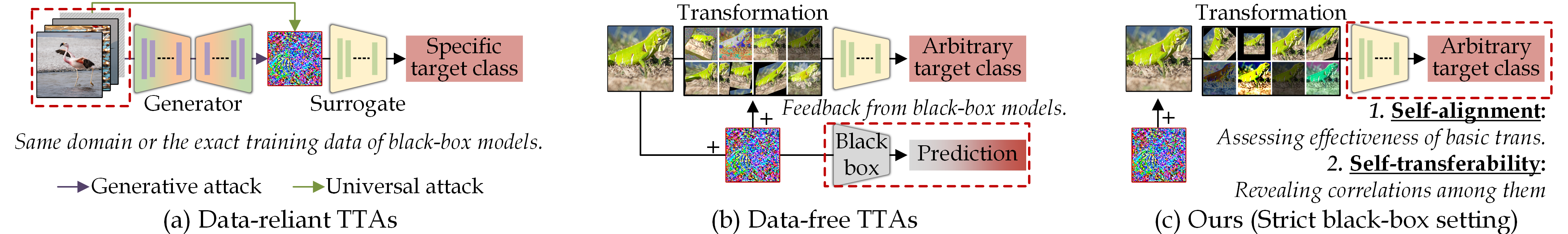}
		\vspace{-8mm}
		\caption{\textcolor{black}{What drives the method design and training distinguishes this study from existing work: (a) Data-reliant TTAs, \eg generative \cite{naseer2021generating, zhao2023minimizing} and universal \cite{weng2023exploring} attacks require large-scale data (often the very training data of the black-box models) and extensive training for each target label; (b) Data-free TTAs, \eg transformation-based gradient attacks, typically depend on query access or feedback from the black-box model for method design and tuning; (c) Our solution introduces two estimation metrics, self-alignment and self-transferability, to analyze the effectiveness and correlations of basic transformations. This enables the design of composite transformations and facilitates effective attacks without any additional data or access to the black-box model.}}
		\label{overviewfig}
		\vspace{-4mm}
	\end{figure*}
	
	\begin{figure}[tbp]
		\includegraphics[width=1\linewidth]{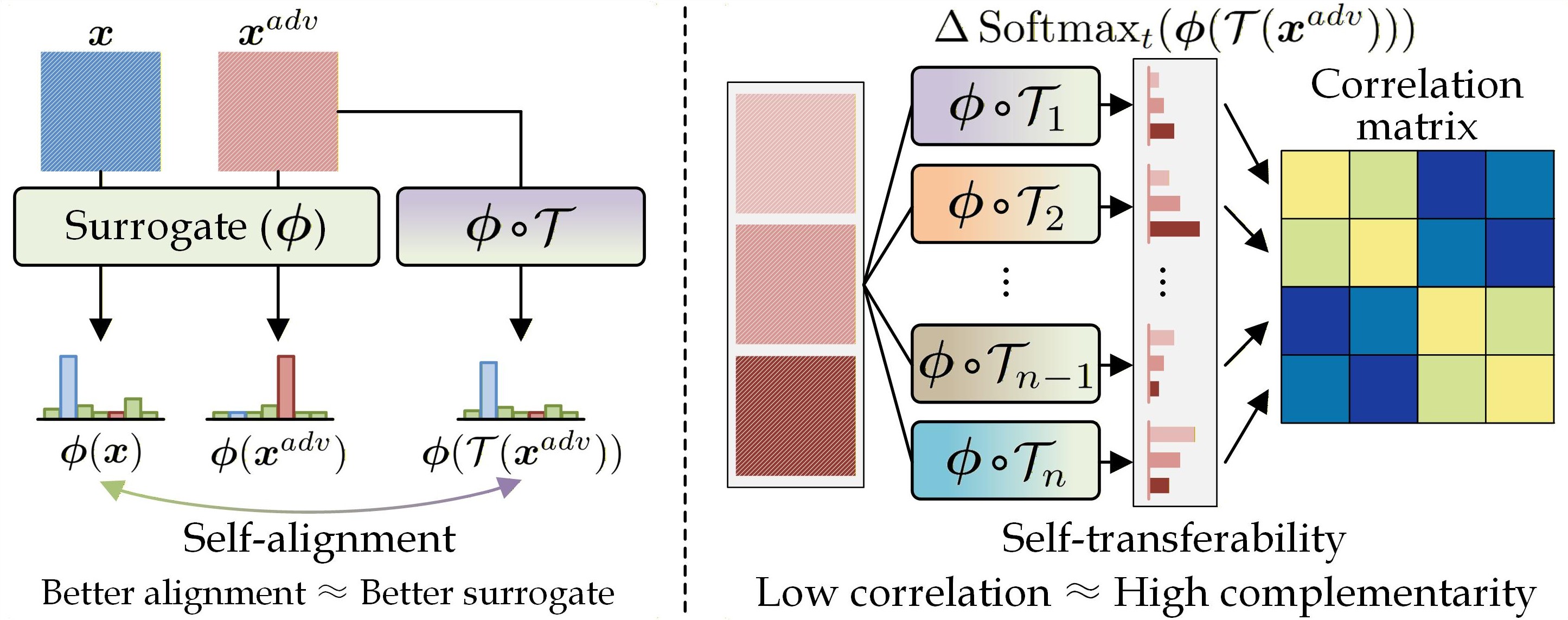}
		\vspace{-8mm}
		\caption{\textcolor{black}{Conceptual illustration of the proposed self-alignment (left) and self-transferability (right) blind estimation measures.}}
		\label{illu_fig}
		\vspace{-5mm}
	\end{figure}
	
	This work explores TTAs under strict black-box constraints using only a domain-matched surrogate model \cite{evading2019dong,momentum2018dong,naseer2021generating,zhao2023minimizing}. Unlike existing solutions (Fig. \ref{overviewfig}), we aim to design data-free transformations to prevent data leakage and ensure fair threat assessment. This requires integrating basic transformations effectively by analyzing per-transformation effectiveness and cross-transformation correlations. \textcolor{black}{As conceptually illustrated in Fig. \ref{illu_fig}, since overfitting bottlenecks transferability, we first leverage the surrogate's self-alignment to estimate a transformation's ability to restore the representation distribution, thereby preserving transferability toward unknown black-box models. Intuitively, this can be viewed in an adversarial learning fashion: we seek the most challenging virtual victim, the surrogate prepended with a transformation, $\bm{\phi}\circ\mathcal{T}$, for the AE optimization. Furthermore, we evaluate cross-transformation correlations via self-transferability. Attacking the surrogate with complex transformations and evaluating against basic transformations (virtual defenses) allows us to objectively identify synergies and eliminate redundancy.} These surrogate-internal measures reveal key findings:
	
	\begin{enumerate}
		\item Simple scaling outstandingly promotes targeted transferability. It achieves stronger surrogate self-alignment and operates effectively across broader intensity ranges than other basic transformations.
		\item Geometric and color transformations show high internal similarity but weak inter-category correlation. Strong redundancy exists even between seemingly unrelated transformations (\eg scaling and rotation), necessitating cautious selection for complementary enhancements.
	\end{enumerate}
	
\begin{figure}[tbp]
		\includegraphics[width=1\linewidth]{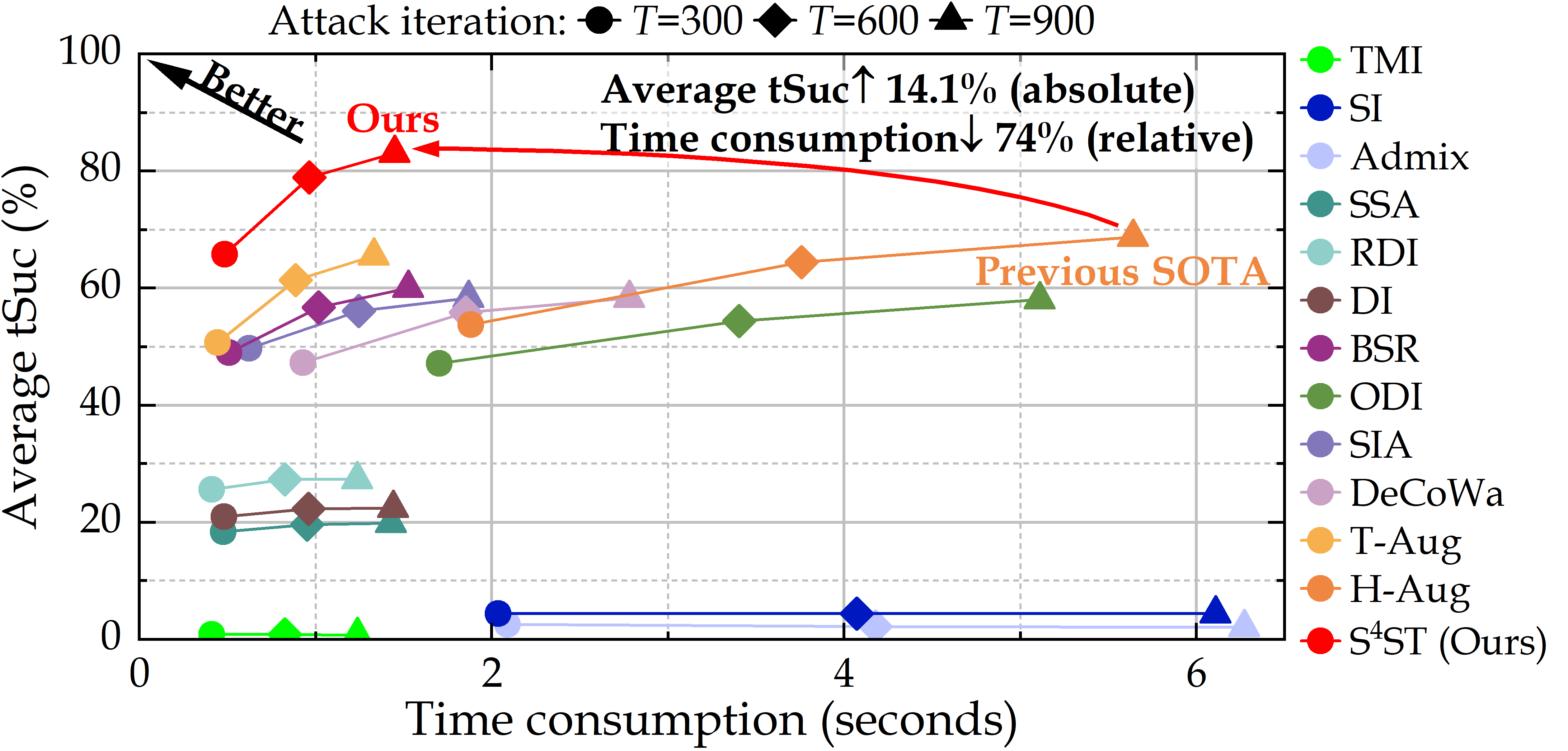}
		\vspace{-8mm}
		\caption{Comparison against existing transformation methods at incremental attack iterations (ImageNet-Compatible dataset). Compared with the previous SoTA, H-Aug \cite{wei2023rethinking}, S$^4$ST yields an absolute improvement of 14.2\% on average tSuc to 83.0\%, and a relative reduction of 74.3\% on time consumption to 1.45s. Compared with the previous setting ($T=300$), more attack iterations benefit a lot for potent transformations like our S$^4$ST. tSuc denotes targeted attack success rate and experimental details are in Section \ref{singlesurroatt}.}
		\label{t-s-plot}
		\vspace{-5mm}
	\end{figure}

	These findings drive the development of S$^4$ST (Strong, Self-transferable, faSt, and Simple Scale Transformation), a novel scaling-centered transformation for data-free TTA. S$^4$ST enhances simple scaling via three efficient strategies: \textit{1)} dimensionally consistent modifications injecting diversity while reducing computation; \textit{2)} integration with low-correlation complementary transformations; and \textit{3)} block-wise scaling. Notably, S$^4$ST determines complementary transformations and tunes parameters entirely via self-transferability analysis in a strict black-box manner, requiring no extra data while achieving a superior effectiveness-efficiency balance (Fig. \ref{t-s-plot}).
	
	\textcolor{black}{We further discover that scaling's distinct effectiveness roots in visual data's multi-scale nature and modern deep vision practices. Standard training universally employs scale augmentations (\eg RandomResizedCrop), embedding scale invariance as a shared inductive bias across most vision models. S$^4$ST strategically exploits this to craft highly transferable AEs without violating black-box constraints.}
	
	In summary, we demonstrate that under strict black-box conditions, prohibiting any access to victim models and their training data, it is possible to design and tune transformation methods that achieve state-of-the-art (SoTA) targeted transferability. Our main contributions are:
	\begin{enumerate}
		\item We propose self-alignment and self-transferability as blind estimation measures. They serve as effective proxies to analyze basic transformations' effectiveness and synergies without accessing victims or extra data, reducing the reliance on empirical choices common in prior art.
		\item We pioneer the discovery of simple scaling's superior efficacy in enhancing targeted transferability. This stems from visual data's inherent nature and the universal adoption of scale augmentation during training, revealing a dual-edged sword: practices enhancing generalization simultaneously introduce transfer attack vulnerabilities.
		\item We propose S$^4$ST, an advanced scaling-centered transformation integrating modified scaling, complementary transformations, and block-wise operations under strict black-box constraints.
		\item  Extensive evaluations across natural images, medical imaging, and face verification validate our framework's transferability. S$^4$ST outperforms existing transformation methods and data-reliant SoTA TTAs (using 50k-1.2M samples), showing robust transferability to commercial APIs and vision-language models (VLMs).
	\end{enumerate}
	
	Aligning with advancements in adversarial machine learning, this research aims to deepen our understanding of adversarial vulnerabilities, thereby contributing to the development of trustworthy artificial intelligence systems \cite{ortiz2021optimism,10478135}. The concept of targeted transferability is not only instrumental in enhancing privacy and intellectual property protection \cite{fowl2021adversarial,chen2023selfensemble} and enabling robust inference \cite{madry2018towards} but also critical for identifying potential risks in high-stakes applications such as autonomous driving and VLMs \cite{liu2025survey}. Although malicious exploitation is possible, we hope this work catalyzes the development of robust defenses and positive societal impacts.
	
	The remainder is organized as follows: Section \ref{pre} introduces the problem, related works, and existing consensus validations. Section \ref{sec4} details our motivation, analysis, and methodology. Sections \ref{exp} and \ref{discussion} present experiments and discussions. Section \ref{conclusion} concludes the paper.
	
	\section{Background and Preliminary}\label{pre}
	
	\subsection{Problem Setting and Baseline Approach}\label{baseline}
	Consider an image classifier $F$ that predicts a label among $K$ categories, $\bm{y}=\{y_i|y_i\in\mathbb{Z}, 0\leq i<K\}$, for an input with RGB channels, $\bm{x}\in \mathbb{R}^{3 \times H \times W}$, where $H$ and $W$ denote the image height and width, respectively. Without loss of generality, we denote $F=C \circ \bm{\phi}$, where $\bm{\phi}(\bm{x})\in \mathbb{R}^{K}$ represents the logits and $C(\cdot)$ indexes the label with the highest probability. $f(\bm{x})=\operatorname{Softmax}(\bm{\phi}(\bm{x}))=\bm{p}_{\bm{\phi}}\in[0,1]^{K}$ represents the final predction distribution.  TTA aims to manipulate black-box classifiers, $G=C\circ \bm{\psi}, g(\bm{x})=\bm{p}_{\bm{\psi}}$, to produce an arbitrary targeted prediction, $y_t$, for any input: 
	\begin{equation}
		\label{obj}
		\centering
		G(\bm{x}^{adv}) = y_t, \quad \text{s.t.} \quad  \|\bm{x}^{adv}-\bm{x}\|_{p} \leq \epsilon,
	\end{equation}
	where $\bm{x}^{adv}$ is the AE crafted to be within a $p$-norm ball centered at $\bm{x}$ with radius $\epsilon$ (the perturbation budget). In the context of transfer attacks, $\bm{x}^{adv}$ is generated using the surrogate model $f$:
	\begin{equation}
		\label{obj-2}
		\centering \bm{x}^{adv}=\underset{\bm{x}^{\prime}:\left\|\bm{x}^{\prime}-\bm{x}\right\|_p \leq \epsilon}{\arg \min } \mathcal{L}\left(f(\bm{x}^{\prime}), y_t\right),
	\end{equation}
	where $\mathcal{L}$ is the loss function designed to induce the targeted misclassification. In line with the established protocol \cite{logit}, we employ the gradient-based method, TMI \cite{evading2019dong,momentum2018dong}, as the baseline for date-free attacks:
	\begin{equation}
		\centering
		\label{TMI_eq}
		\begin{aligned}
			&\bm{x}^{adv}_0 = \bm{x}, \quad \bm{g}_0 = \bm{0},\\
			\bm{g}_{i+1}=\mu \cdot \bm{g}_i&+\frac{\bm{W}*\nabla_{\bm{x}} \mathcal{L}(f(\mathcal{T}(\bm{x}^{adv}_i), y_t))}{\left\|\bm{W}*\nabla_{\bm{x}} \mathcal{L}(f(\mathcal{T}(\bm{x}^{adv}_i), y_t))\right\|_1},\\
			\bm{x}^{adv}_{i+1} &= \Pi_{\epsilon}(\bm{x}^{adv}_{i}-\alpha \cdot \operatorname{sign} (\bm{g}_{i+1})).
		\end{aligned}
	\end{equation}
	The step size $\alpha$ and perturbation budget $\epsilon$ are configured to 2/255 and 16/255, respectively. The accumulation decay factor $\mu$ is set to 1 for momentum $\bm{g}$ \cite{momentum2018dong}, and a 5$\times$5 Gaussian kernel \cite{evading2019dong} $\bm{W}$ is used for TI. We adopt the margin-based cross-entropy loss \cite{marginangleT} and the attack iteration is extended to $T=900$ (compared to 300 in previous work) in keeping with the increased diversity afforded by the most advanced transformation methods $\mathcal{T}$.

	\subsection{Transferable Targeted Attacks}
	
	\textcolor{black}{We categorize existing TTAs into data-free and data-reliant paradigms, analyzing their limitations to contextualize the distinct contributions of S$^{4}$ST.}

	\begin{figure}[tbp]
		\includegraphics[width=1\linewidth]{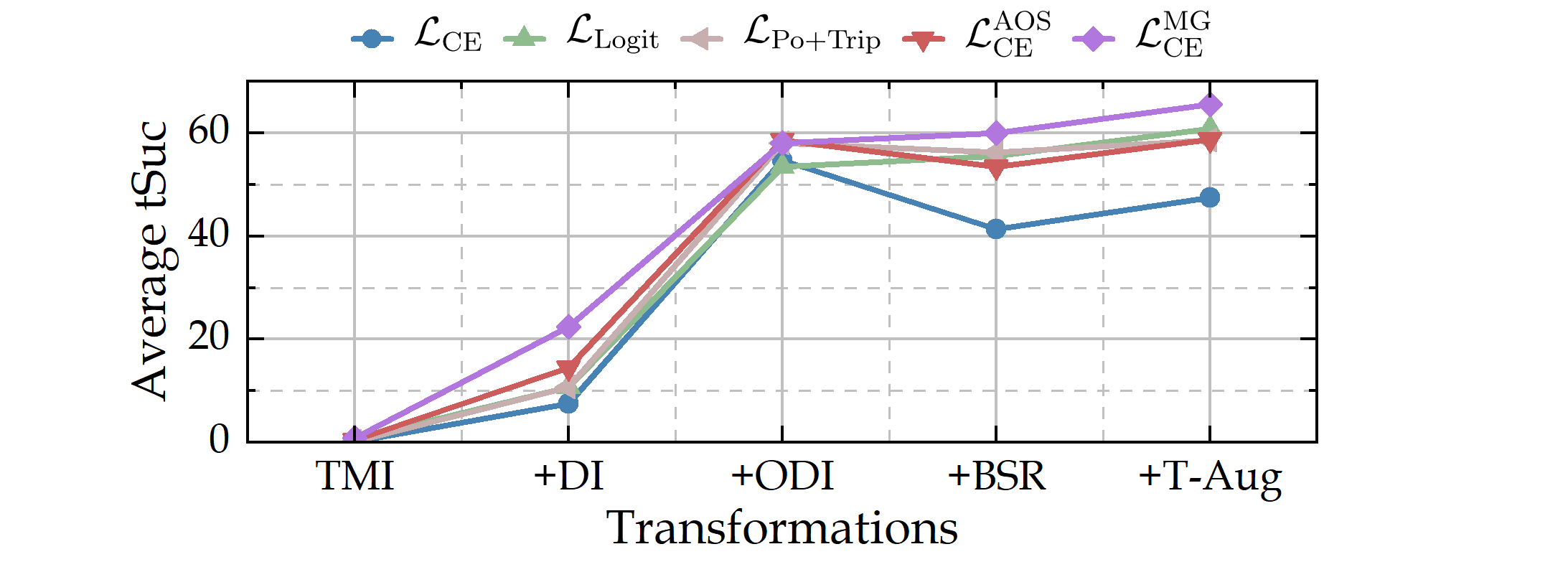}
		\vspace{-8mm}
		\caption{\textcolor{black}{Average tSuc evaluated on the ImageNet-Compatible. Input transformations are identified as key role in enabling effective data-free TTAs.}}
		\label{loss_trans_plot}
		\vspace{-5mm}
	\end{figure}

	\subsubsection{Data-free TTAs}\label{simtta}	
	Data-free TTAs typically rely on gradient-based iterative attacks \cite{kurakin2016adversarial,momentum2018dong,evading2019dong,madry2018towards} using a surrogate model. Overcoming gradient vanishing (see supplementary) is a key challenge, addressed by novel loss functions that replace (\eg PoTrip \cite{PoTrip}, Logit \cite{logit}) or soften the standard cross-entropy \cite{marginangleT,AOSBLL}.
	Crucially, image transformations are essential for data-free TTAs. As Fig. \ref{loss_trans_plot} shows, specialized losses fail without them. Although untargeted transformations like random scaling \& padding in DI \cite{inputdiversity2019xie} and RDI \cite{RDI} show promise \cite{CFM,logit}, understanding which transformations specifically drive targeted transferability remains scarce.
	
	\begin{figure}[tbp]
		\centering
		\includegraphics[width=1\linewidth]{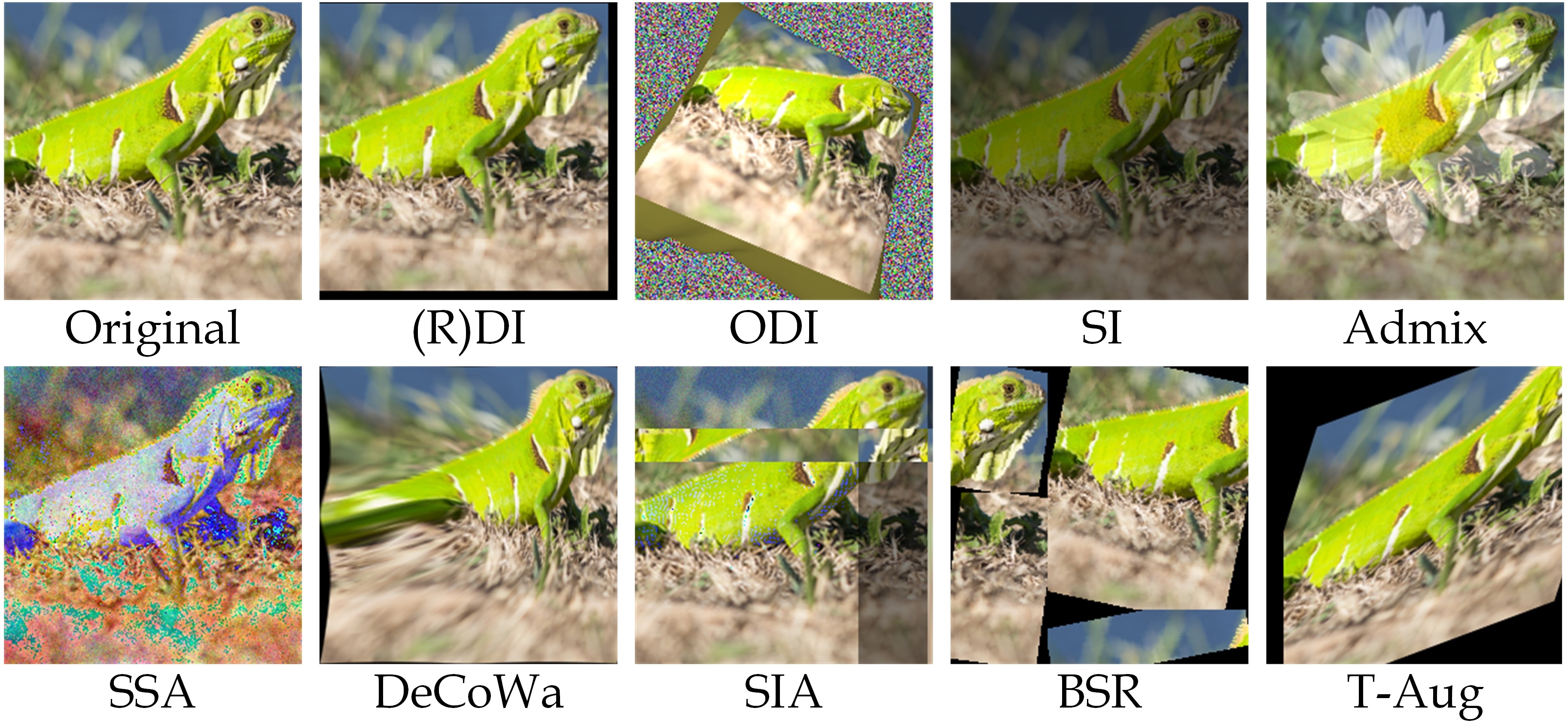}
		\vspace{-8mm}
		\caption{Illustration of the original image and its transformed versions.}
		\label{existing}
		\vspace{-5mm}
	\end{figure}
	
	Wei \etal \cite{wei2023rethinking} showed that DI \cite{inputdiversity2019xie} diversifies surrogate discrimination regions across iterations, expanding them in black-box models, and proposed T/H-Aug using random combinations of attention-deviation transformations. Byun \etal introduced Object-based DI (ODI) \cite{ODI}, hypothesizing that AEs transferable across 3D shapes and views better compromise black-box models.
	
	However, these techniques rely on empirical observations or victim access, lacking a principled black-box understanding. To address this, our framework introduces self-alignment and self-transferability metrics to systematically evaluate transformations, challenge existing consensuses (Section \ref{failures_current_cons}), and analytically guide S$^{4}$ST's design. We also include prominent untargeted attack transformations (Fig. \ref{existing}, details in supplementary) to validate whether prevalent design consensuses apply to TTAs.

	\subsubsection{Data-reliant TTAs}
	
	Data-reliant TTAs leverage large-scale data to enhance transferability via three main paradigms: \textit{1) Targeted universal attacks} \cite{domainfeatureuap,weng2023exploring} optimize a single perturbation across large datasets to universally deceive models on unseen samples; \textit{2) Surrogate refinement} \cite{springer2021little,Wu_2024_CVPR,zhu2021rethinking,zhu2022toward} fine-tunes surrogate models for improved smoothness or data distribution matching; and \textit{3) Generative attacks} \cite{naseer2021generating,zhao2023minimizing,yang2022boosting} train generators to model data distributions, facilitating AE transfer.
	
	Despite their efficacy, these methods incur prohibitive data and computational costs. We demonstrate that such dependencies are unnecessary. For instance, while the generative M3D \cite{zhao2023minimizing} requires $\sim$10 hours and 50K samples to train a per-label generator (see supplementary), our data-free S$^{4}$ST achieves superior transferability with minimal overhead and ultimate flexibility for any \textit{(sample, label)} pair.

	\subsection{Limitations of Existing Design Consensuses}\label{failures_current_cons}
	We scrutinize the prevailing consensuses in developing transformations to assess their robustness and utility. We evaluate the black-box transferability of various transformations across default and expanded hyperparameter configurations. This transferability is measured by the target class ($t$) probability variance between adversarial (generated by $\mathcal{T}$-TMI) and clean samples:
	\begin{equation}
		\centering
		\textit{T}_{\mathcal{T}} = \mathbb{E}_{\bm{x}\in\mathcal{X},g \in\mathcal{G}}\left(g(\bm{x}_{\mathcal{T}}^{adv},t) - g(\bm{x},t)\right).
		\label{transferabilty}
	\end{equation}
	
	We then examine correlations between $\textit{T}_{\mathcal{T}}$ and established consensus metrics: \textit{Diversity} \cite{admix,inputdiversity2019xie,nestrov2019lin,evading2019dong,SIA} (average cross-entropy loss on transformed inputs), \textit{Attention Deviation} \cite{evading2019dong,BSR,wei2023rethinking,10251606,11018095} (changes in model attention maps), and \textit{Gradient Magnitude} \cite{logit,SelfU,AOSBLL,marginangleT} (average $\ell_{2}$-norm of attack gradients). Details are in the supplementary material.
	
	\begin{figure*}[tbp]
		\includegraphics[width=1\linewidth]{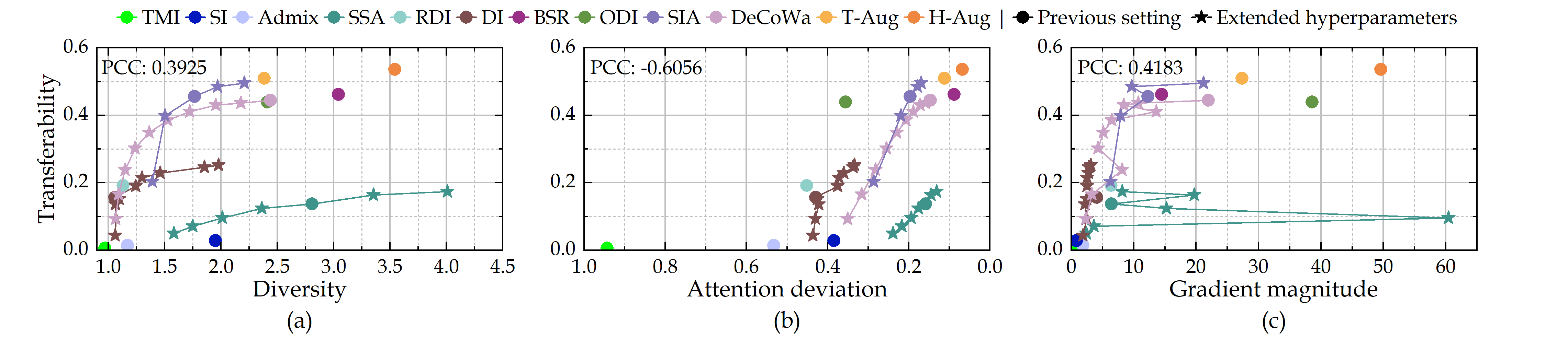}
		\vspace{-8mm}
		\caption{Scatter diagrams depicting relationships between {\textit{black-box transferability}} and (a) {\textit{diversity}}, (b) {\textit{attention deviation}}, and (c) {\textit{gradient magnitude}}. As indicated by the Pearson Correlation Coefficients (PCCs) in the top left corner of each figure, all three metrics exhibit weak correlation to transferability in the cross-method case, impeding their application in identifying what transformations are effective and designing a complex transformation from scratch.}
		\label{existing_censensus_analysis}
		\vspace{-3mm}
	\end{figure*}
	
	As shown in Fig. \ref{existing_censensus_analysis}, intra-method analyses across hyperparameter configurations often reveal positive correlations, suggesting these metrics guide parameter scaling within specific transformations. However, severe inter-method confusion complicates designing novel transformations from scratch. This lack of reliable cross-method guidance underscores the need for a fundamental analysis, motivating our proposed self-alignment and self-transferability measures.

	\section{The Proposed Method}\label{sec4}
	
	This section delineates our proposed method. We introduce two strategies to gauge the per-transformation effectiveness and cross-transformation correlations among 12 basic transformations, culminating in the powerful yet efficient S$^{4}$ST for data-free TTAs.
	
	\subsection{Motivation}
	
	We aim to develop robust blind estimators for designing data-free TTA transformations, answering two core questions: \textit{1) which transformations are most effective?} and \textit{2) how to determine their synergistic effects?} Our goal is to outperform data-reliant methods in efficacy while maintaining flexibility for any \textit{(sample, label)} pair without extra data or training. By adhering to strict black-box constraints, these estimators ensure applicability to broader data beyond natural images. We begin by analyzing individual basic transformations before exploring their combinations (details and intensities $\mathcal{S}$ are in the supplementary).
	
	\subsection{Assessing Effectiveness via Self-Alignment}
	
	Evaluating basic transformations' effectiveness in enhancing targeted transferability is challenging due to the inaccessibility of black-box models and the complex dynamics of randomized parameters across hundreds of attack iterations. To address this, we propose leveraging the AE overfitting phenomenon itself as a proxy shortcut.
	
	\begin{figure}[bp]
		\vspace{-5mm}
		\includegraphics[width=1\linewidth]{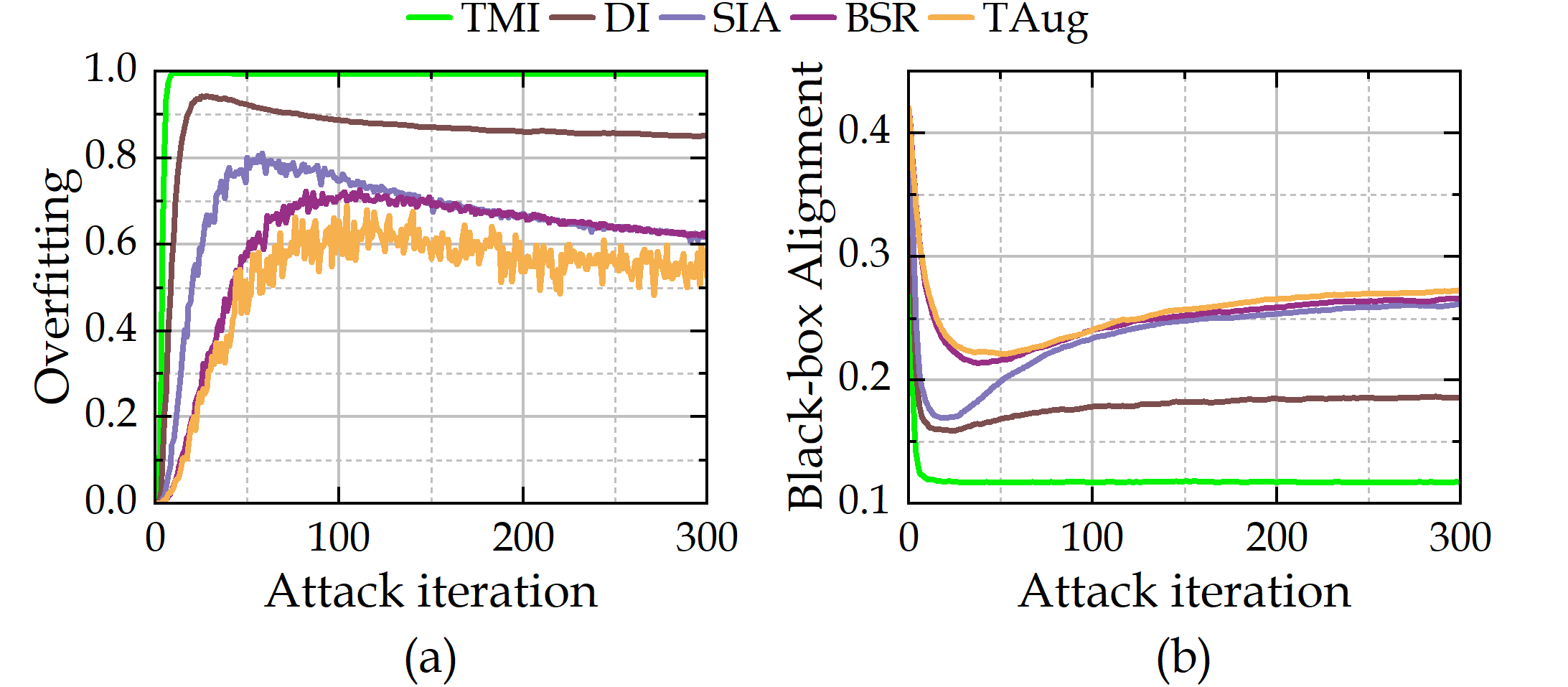}
		\vspace{-8mm}
		\caption{\textcolor{black}{Overfitting and alignment metrics between the surrogate model and black-box models on targeted AEs, which reflects the black-box transferability resulting from different transformations (see Fig. \ref{t-s-plot}). The vanilla gradient update, \ie TMI, quickly misaligns the surrogate from black-box models and leads to poor transferability.}}
		\label{cross_alignment}
	\end{figure}

	\subsubsection{Feature-space Model Alignment as Proxy}\label{model_alignment_intro}
	
	\textcolor{black}{During vanilla attacks (\eg TMI), the surrogate's target-label confidence rapidly saturates ($f(\bm{x}_{\mathcal{T}}^{adv},t)\rightarrow1$), causing gradient vanishing and severe overfitting (Fig. \ref{cross_alignment}a), formalized as $\mathbb{E}_{\bm{x}\in\mathcal{X},g \in\mathcal{G}}\left(f(\mathcal{T}(\bm{x}_{\mathcal{T}}^{adv}),t)-g(\bm{x}_{\mathcal{T}}^{adv},t)\right)$. Conversely, applying transformations mitigates this.}
	
	Why are transformations effective? Just as the best defensive transformations restore predictions toward clean samples \cite{xie2018mitigating}, the best surrogate transformations during attacks recalibrate the surrogate's feature extraction to resemble that of clean samples. Beyond alleviating confidence saturation, this crucially \textit{realigns feature representations between surrogate and black-box models}. This is grounded in \textit{representation convergence}: well-trained models in the same domain share similar feature spaces due to common data manifolds \cite{ben2006analysis,huhposition,zhang2016understanding}. AEs that preserve this inherent alignment transfer better, whereas gradient descent for a specific target disrupts it. Transformations thus help recalibrate the surrogate's features to exploit the original alignment with black-box models.
	
	
	While gradient alignment \cite{zhang2024does}, $\mathbb{E}_{\bm{x}\in\mathcal{X}} \text{CosSim}\left(\nabla_{\bm{x}}\mathcal{L}_f, \nabla_{\bm{x}}\mathcal{L}_g\right)$, directly measures transferability (Eq. \eqref{TMI_eq}), we adopt feature-space alignment, $\mathbb{E}_{\bm{x}\in\mathcal{X}} \text{CosSim}\left(\bm{\phi}(\bm{x}), \bm{\psi}(\bm{x})\right)$, as a relaxed, label-agnostic proxy suitable for black-box settings with disparate label sets. To capture local geometry explicitly, we utilize a $k$-nearest neighbors overlap measure \cite{huhposition}
	\begin{equation}
		\centering
		\begin{aligned}
			\textcolor{black}{\operatorname{Alignment}\left(\Phi, \Psi\right) =} & \\ 	\textcolor{black}{\mathbb{E}_{g\in\mathcal{G},(\bm{\phi},\bm{\psi})\in(\Phi,\Psi)}}&\textcolor{black}{\frac{1}{k}\left|d_\text{knn}(\bm{\phi},\Phi\setminus\bm{\phi})\cap d_\text{knn}(\bm{\psi},\Psi\setminus\bm{\psi})\right|,}
		\end{aligned}
		\label{mutualalignment}
	\end{equation}
	\textcolor{black}{where $\Phi = \{\bm{\phi}(\bm{x}_i)\}_{i=1}^b$ and $\Psi = \{\bm{\psi}(\bm{x}_i)\}_{i=1}^b$ denote features from $f$ and $g$ for dataset $\mathcal{X}$. Function $d_\text{knn}$ returns the indices of the $k$-nearest neighbors, with $k=100$ for $b=1,000$ images.}
	
	\begin{figure}[tbp]
		\centering	
		\includegraphics[width=1\linewidth]{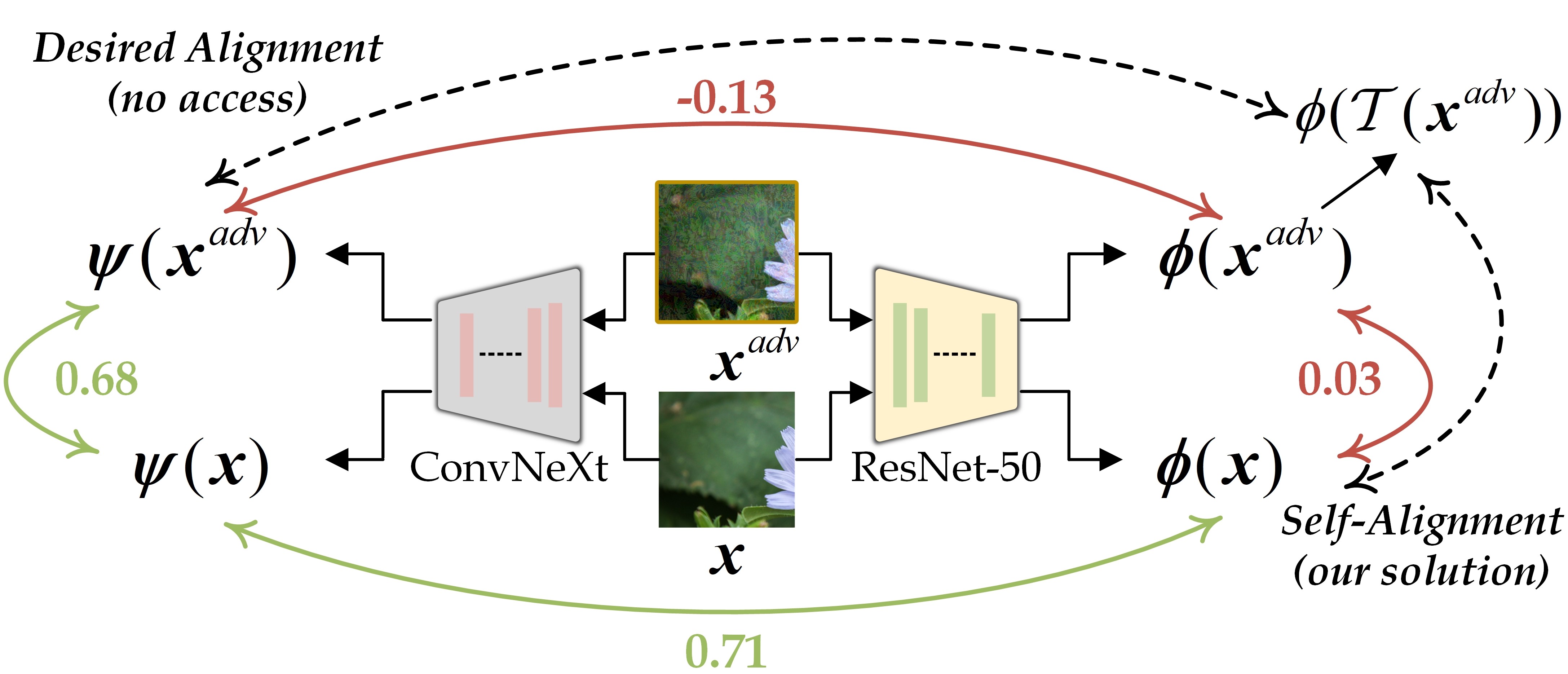}
		\vspace{-7mm}
		\caption{The colored numbers indicate the cosine similarity between features generated by the surrogate model (ResNet-50) and the black-box model (ConvNeXt), where the AE $\bm{x}^{adv}$ is created based on the TMI algorithm. On the clean sample, these two models exhibit a high degree of alignment. However, an AE with weak transferability has limited impact on the black-box model, which suggests that we can introduce an image transformation $\mathcal{T}$ to make $\bm{\phi}(\mathcal{T}(\bm{x}^{adv}))$ more closely approximate $\bm{\phi}(\bm{x})$ ({\textit{self-alignment}}), thereby enhancing the approximation towards $\bm{\psi}(\bm{x}^{adv})$ ({\textit{alignment}}).}
		\label{individualrep}
		\vspace{-4mm}
	\end{figure}
	
	We validate this proxy by probing inter-model alignment $\operatorname{Alignment}(\Psi^{adv}_{\mathcal{T}},\Phi^{adv}_{\mathcal{T}})$ across attack iterations. Here, $(\cdot)^{adv}_{\mathcal{T}}$ denotes features of adversarial counterparts crafted via $\mathcal{T}$-TMI on surrogate $f$. As Fig. \ref{cross_alignment}(b) shows, vanilla surrogate gradients cause rapid misalignment and poor transferability. Conversely, effective transformations maintain alignment. Crucially, this alignment metric reliably predicts the actual attack transferability (Fig. \ref{t-s-plot}), validating its utility as an assessment proxy.
	
	\subsubsection{From Black-box Alignment to Self-Alignment}
	
	To circumvent dynamic probing complexities under black-box constraints, we propose \textit{self-alignment}. This measures how transformations restore \textit{single-step feature consistency on overfitted AEs} within the surrogate, revealing their efficacy in mitigating overfitting.
	
	Illustrated in Fig. \ref{individualrep}, sample-level feature similarity intuitively justifies this. \textcolor{black}{Given inherent alignment $\bm{\phi}(\bm{x}) \approx \bm{\psi}(\bm{x})$, a weakly-transferable AE $\bm{x}^{adv}$ (\eg via TMI) barely impacts the black-box model $g$. Thus, $\bm{\psi}(\bm{x}^{adv}) \approx \bm{\psi}(\bm{x}) \approx \bm{\phi}(\bm{x})$, all diverging markedly from the overfitted surrogate feature $\bm{\phi}(\bm{x}^{adv})$ as attack iterations progress.} To enhance transferability by bringing $\bm{\psi}(\bm{x}^{adv})$ closer to $\bm{\phi}(\mathcal{T}(\bm{x}^{adv}))$, we extend this strategy under strict black-box principles to pursue \textit{self-alignment}: $\bm{\phi}(\mathcal{T}(\bm{x}^{adv}))\approx\bm{\phi}(\bm{x})$.
	
	We validate this condition at the dataset level across multiple black-box models (Table \ref{modelalignment}, where subscript-free sets denote vanilla TMI features). The high and comparable scores of $\operatorname{Alignment}(\Phi,\Psi)$ and $\operatorname{Alignment}(\Psi,\Psi^{adv})$ confirm our analysis. Consequently, surrogate self-alignment on severely overfitted AEs, $\operatorname{Alignment}(\Phi^{adv}_{\mathcal{T}},\Phi)$, emerges as a viable proxy for the desired black-box alignment, $\operatorname{Alignment}(\Phi^{adv}_{\mathcal{T}},\Psi^{adv})$.
	
	\begin{table}[tbp]
		\caption{Alignment between surrogate model and multiple black-box models.}
		\vspace{-4mm}
		\label{modelalignment}
		\centering
		\begin{tabular}{lc}
			\Xhline{1pt}
			Case	& Alignment \\
			\hline
			$\operatorname{Alignment}(\Phi,\Psi)$ & 0.42 \\
			$\operatorname{Alignment}(\Psi,\Psi^{adv})$ & 0.32 \\
			$\operatorname{Alignment}(\Psi^{adv},\Phi^{adv})$ & 0.12 \\
			$\operatorname{Alignment}(\Phi^{adv},\Phi)$ & 0.10 \\
			\Xhline{1pt}
		\end{tabular}
		\vspace{-2mm}
	\end{table}
	
	\begin{figure}[tbp]
		\centering	
		\includegraphics[width=1\linewidth]{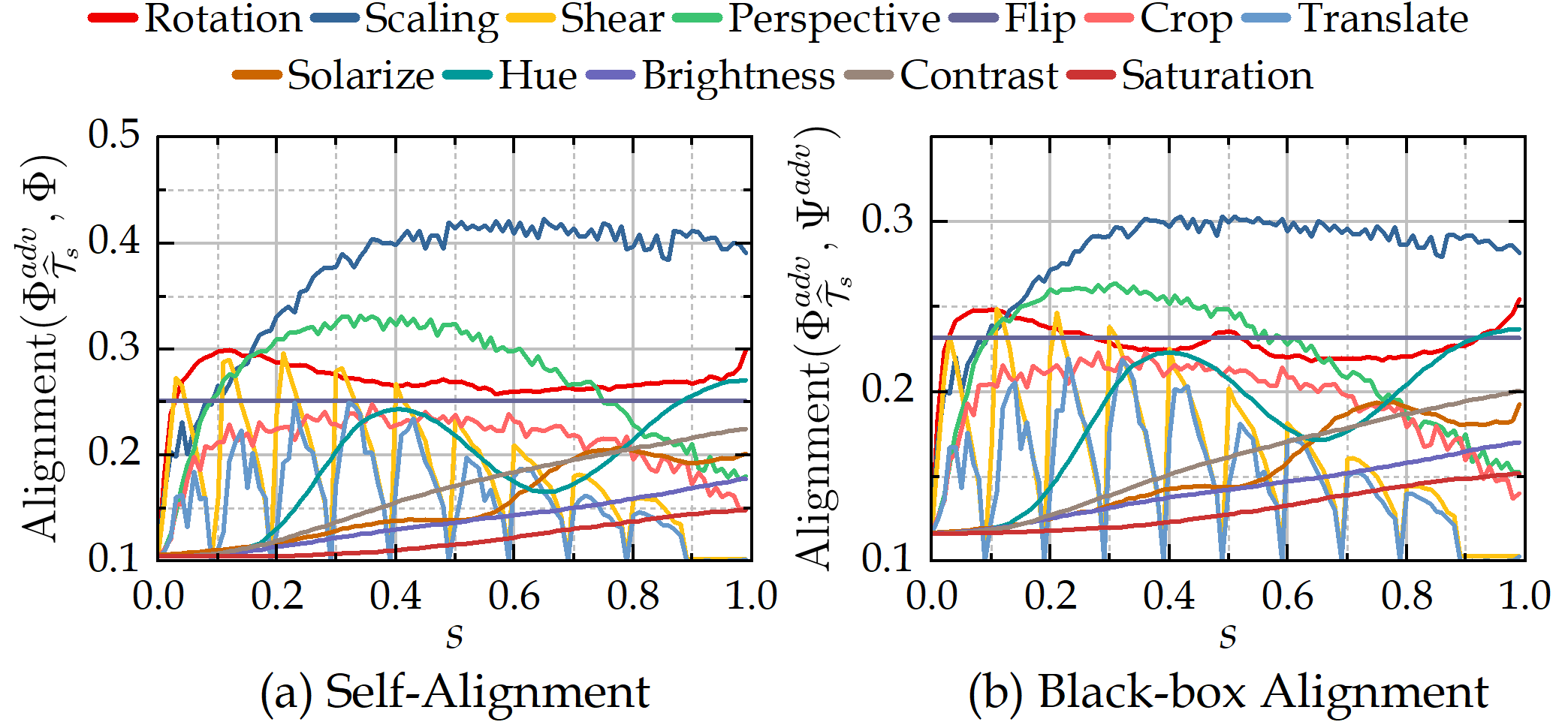}
		\vspace{-7mm}
		\caption{The metrics of {\textit{self-alignment}} and {\textit{black-box alignment}} exhibit a high degree of similarity. Notably, the scaling transformation demonstrates distinctive effectiveness, simultaneously inducing high alignment strength and a large effective range of transformation intensities.}
		\label{alignment}
		\vspace{-5mm}
	\end{figure}
	
	\textbf{Assessing effectiveness via self-alignment}. Fig. \ref{alignment}(a) plots $\operatorname{Alignment}(\Phi^{adv}_{\widehat{\mathcal{T}}_s},\Phi)$ for basic transformations $\widehat{\mathcal{T}}$ across intensities $s$. While geometric transformations (rotation, scaling, perspective) promote self-alignment, \textit{scaling uniquely achieves significantly higher self-alignment over a broader effective intensit y range, marking it as a key transformation.} Though self-alignment is a single-step estimation on overfitted AEs, full-attack experiments in Section \ref{basetranscomp} consistently corroborate scaling's superior transferability.
	
	
	\textbf{Effectiveness of the self-alignment measure}. Figs. \ref{alignment}(a) and \ref{alignment}(b) confirm that self-alignment accurately mirrors black-box alignment towards unknown models, capturing both intensity trends and relative transformation magnitudes. \textcolor{black}{This reliability is rooted in \textit{representation convergence}, well-trained domain models develop aligned feature spaces regardless of architecture or training distributions. Extensive cross-model evaluations (see supplementary) substantiate this core assumption, providing strong empirical support for self-alignment as a robust proxy.}

	\subsection{Revealing Correlation via Self-Transferability}
	
	To investigate cross-transformation correlations under strict black-box constraints, merely measuring feature distribution similarities of overfitted AEs (akin to self-alignment) is insufficient, as it fails to capture the full dynamics of AE optimization.

	Thus, we propose \textit{self-transferability} for a more robust black-box estimation. We generate targeted AEs using diverse complex transformations (see Section \ref{failures_current_cons} and supplementary) and compute their PCCs when attacking the surrogate defended by basic transformations. Treating these basic transformations as defensive layers \cite{li2020ghost,xie2018mitigating}, their empirical expectations across intensities are then a proxy of black-box transferability:
	\begin{equation}
		\textit{ST}_{\widehat{\mathcal{T}}} = \mathbb{E}_{\bm{x}\in\mathcal{X}, s\in\mathcal{S}}\left(f( \widehat{\mathcal{T}}_s(\bm{x}_{\mathcal{T}}^{adv}),t) - f( \widehat{\mathcal{T}}_s(\bm{x}),t)\right).
		\label{selftrans}
	\end{equation}
	Highly correlated self-transferability between two basic transformations indicates redundancy, providing a foundation for synergistic combinations.

	\begin{figure}[bp]
		\centering	
		\vspace{-3mm}
		\includegraphics[width=1\linewidth]{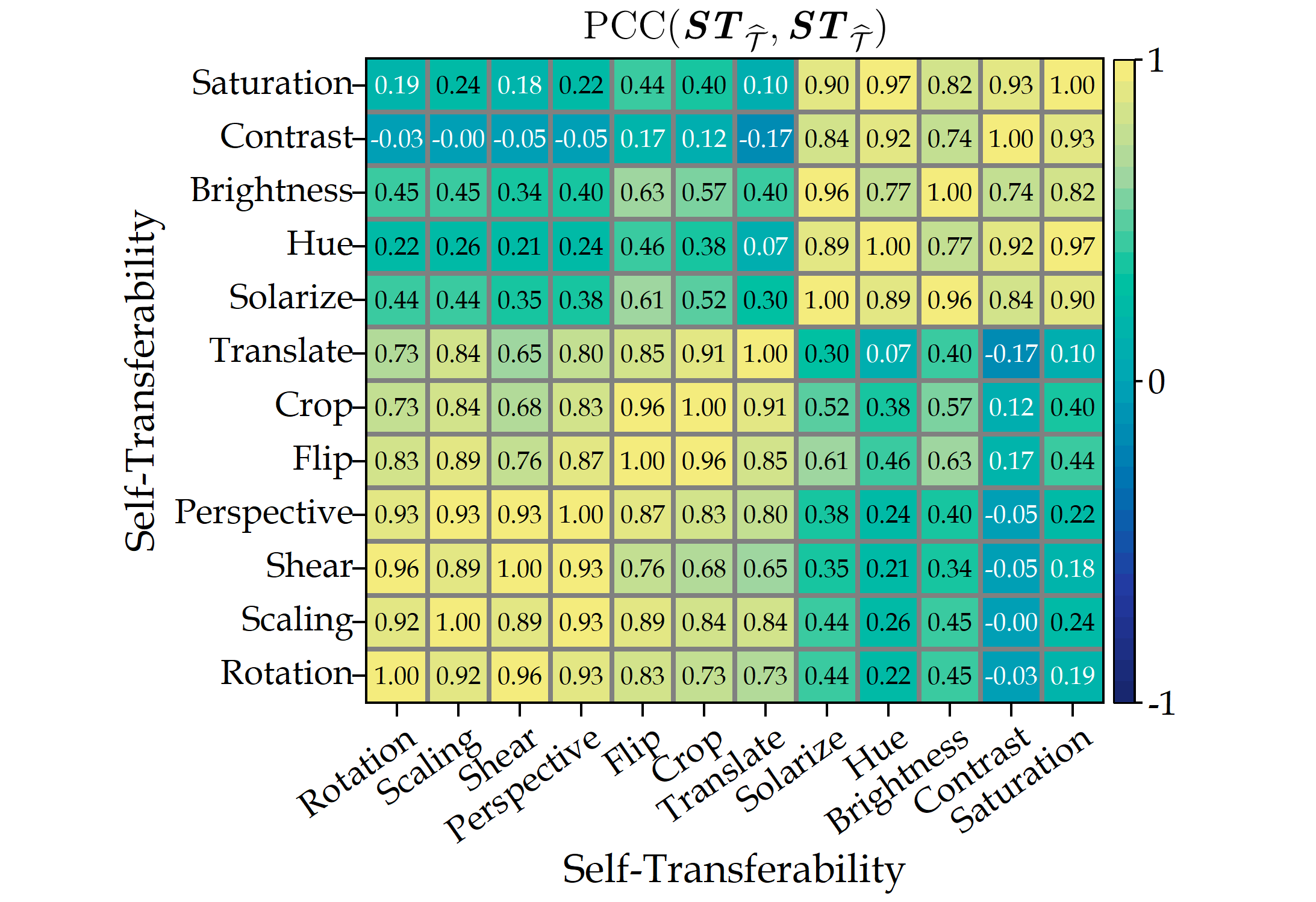}
		\vspace{-8mm}
		\caption{PCCs between the \textit{self-transferability} against 12 basic transformations, based on AEs obtained from 12 complex transformations under a total of 40 hyperparameter configurations.
		}
		\label{marix}
	\end{figure}
	
	\textbf{Revealing correlation via self-transferability}. Fig. \ref{marix} details the PCC tests (calculation details in supplementary). \textit{It reveals strong intra-category correlations within geometric and color transformations, but weak inter-category correlations.} Consequently, it is more strategic to combine highly effective, distinct transformations, such as pairing the potent scaling transformation with color adjustments, rather than indiscriminately stacking redundant geometric operations. This measure rigorously guides our augmentation design (Section \ref{augdesign}).

	\begin{figure}[tbp]
		\centering	
		\includegraphics[width=1\linewidth]{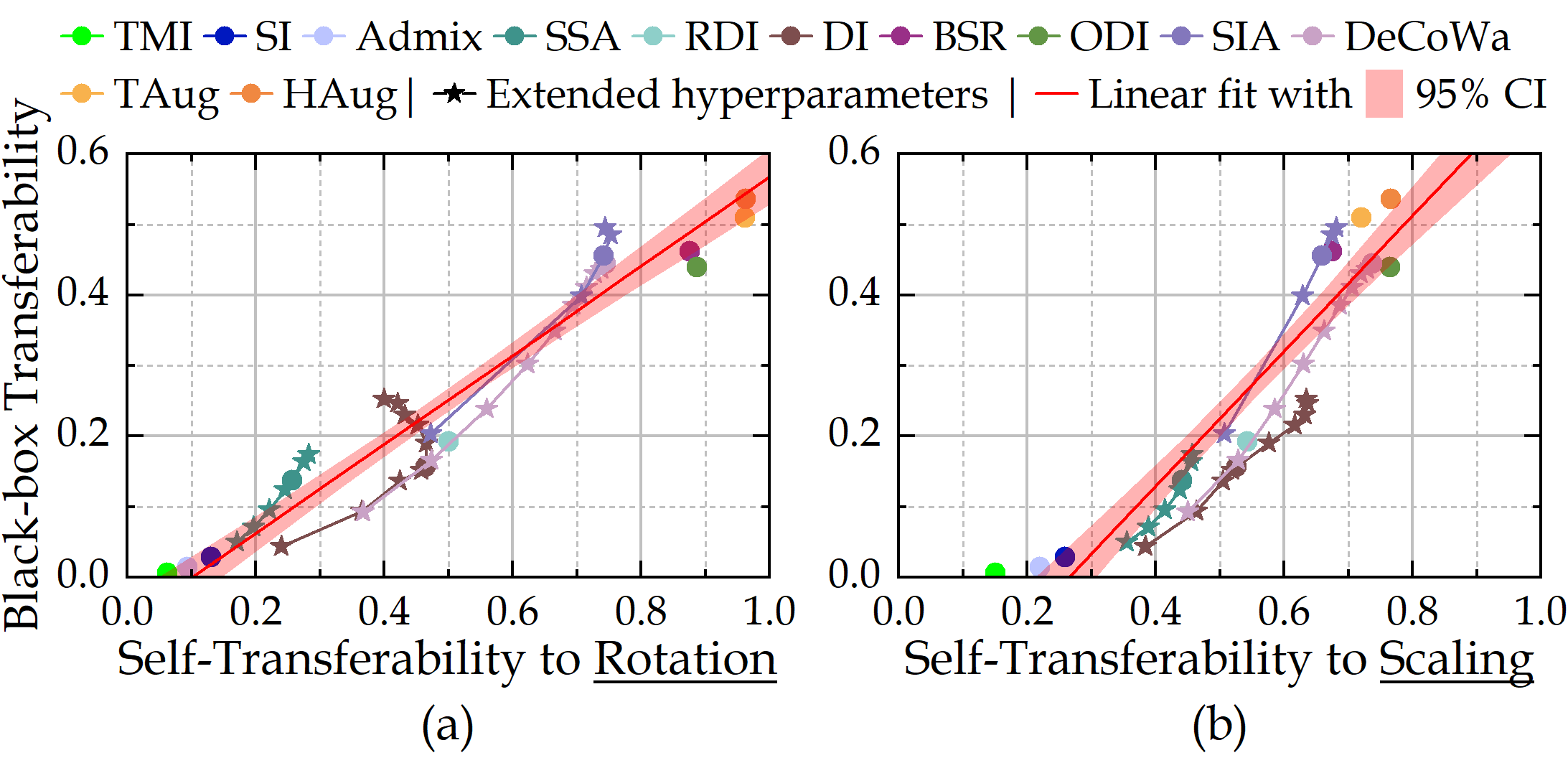}
		\vspace{-8mm}
		\caption{Scatter diagrams depicting relationships between {\textit{black-box transferability}} and {\textit{self-transferability}} to (a) rotation and (b) scaling transformations.}
		\label{lineartest}
		\vspace{-4mm}
	\end{figure}
	
\begin{figure}[tbp]
		\centering	
		\includegraphics[width=1\linewidth]{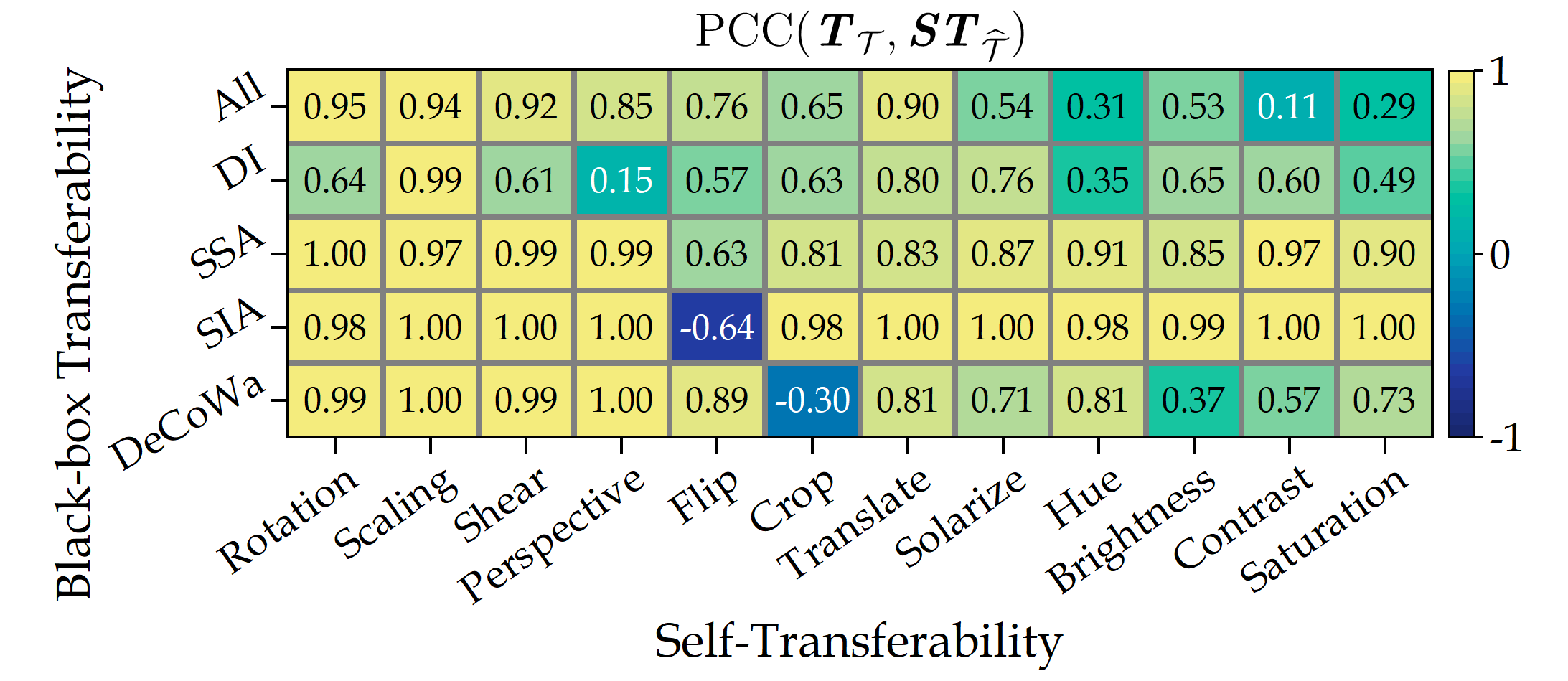}
		\vspace{-8mm}
		\caption{PCCs calculated between the {\textit{black-box transferability}} against 14 black-box models and the {\textit{self-transferability}} against various basic transformations, based on AEs obtained from 12 existing attacks  under a total of 40 hyperparameter configurations.}
		\label{heatmap}
		\vspace{-4mm}
	\end{figure}

	\textbf{Effectiveness of the self-transferability measure.} We validate self-transferability (Eq. \eqref{selftrans}) as a high-quality estimator for black-box transferability (Eq. \eqref{transferabilty}). Scatter plots for rotation and scaling (Fig. \ref{lineartest}) reveal notably tighter cross-method correlations than prior metrics (Fig. \ref{existing_censensus_analysis}), effectively reflecting black-box trends. Analyzing the PCCs across all basic transformations (Fig. \ref{heatmap}) yields key findings:
	\begin{enumerate}
		\item Self-transferability against most geometric transformations demonstrates consistently high-quality linear correlations with black-box transferability. Notably, seemingly disparate attacks, like frequency-domain distortions \cite{SSA}, elastic deformations \cite{decowa}, and random block-wise transformations \cite{SIA}, uniformly impact multi-scale feature exploitation, maintaining strong correlations across geometric transformations like rotation, shear, and translation.
		\item Self-transferability against color transformations exhibits weak cross-method correlations but robust intra-method correlations (\eg SSA and SIA, both involving image color manipulation).
	\end{enumerate}
	
	These results confirm that self-transferability against basic geometric transformations (\eg rotation, scaling) acts as an excellent blind estimator, enabling strict data-free method design and parameter tuning. Note that to strictly preserve the black-box setting, this specific analytical finding is excluded from our main attack experiments.
	
	\subsection{S$^\textbf{4}$ST}
	
\begin{figure}[tbp]
		\centering
		\includegraphics[width=1\linewidth]{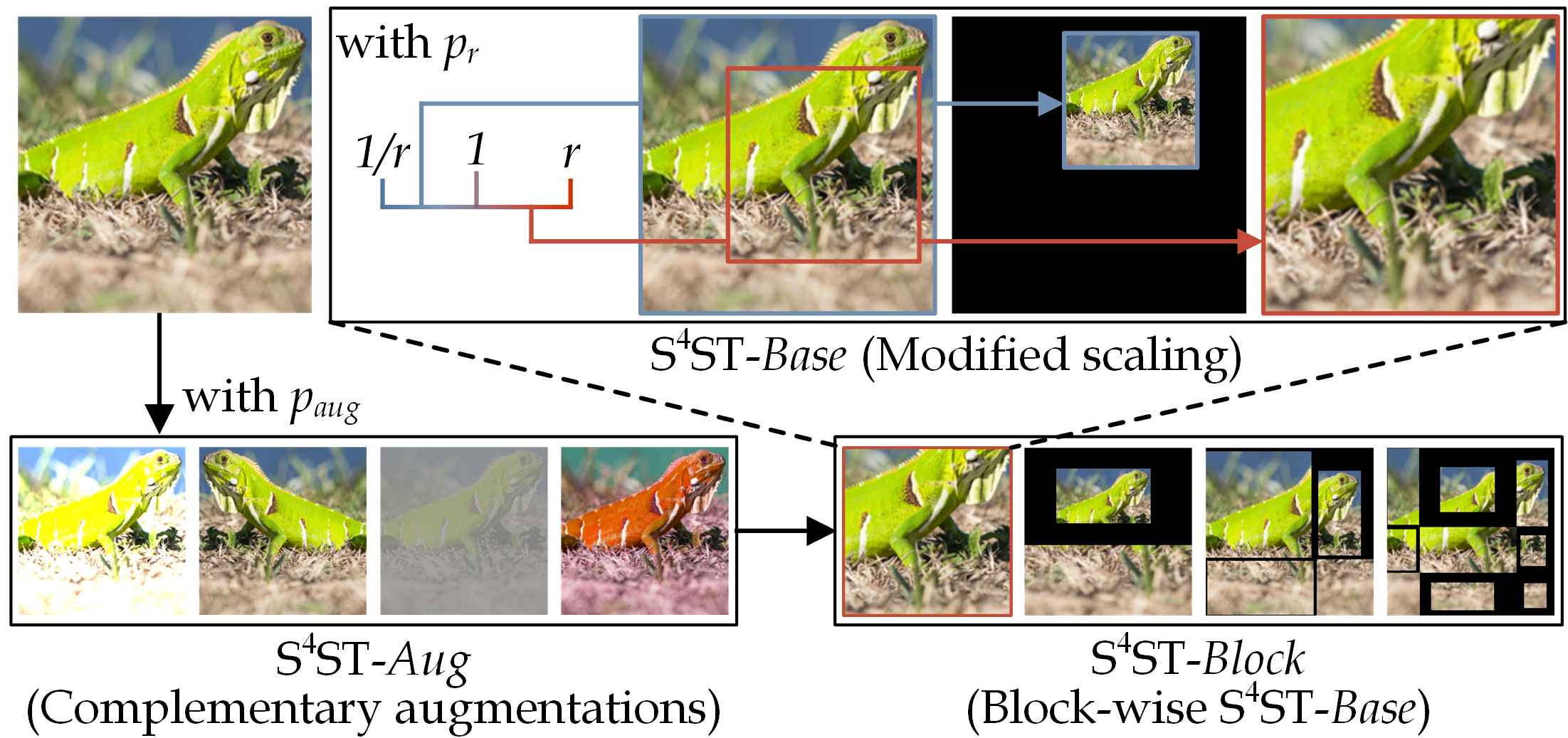}
		\vspace{-8mm}
		\caption{The proposed S$^4$ST encompasses the \textit{Base} operation, which is based on scaling, complemented by two enhancements—\textit{Aug} and \textit{Block}.}
		\label{s4st}
		\vspace{-4mm}
	\end{figure}

	Drawing on our analytical findings, we introduce Strong, Self-transferable, Fast, and Simple Scale Transformation (S$^4$ST), a novel scaling-centered transformation designed to enhance data-free TTAs. Fig. \ref{s4st} illustrates its schematic overview. S$^{4}$ST is rooted in an improved scaling transformation and synergized with complementary operations and block-wise manipulation strategies, comprising three distinct components: S$^{4}$ST-\textit{Base}, S$^{4}$ST-\textit{Aug}, and S$^{4}$ST-\textit{Block}.
	
	\subsubsection{S$^{4}$ST-\textit{Base}}
	
\begin{figure}[bp]
		\centering	
		\vspace{-4mm}
		\includegraphics[width=1\linewidth]{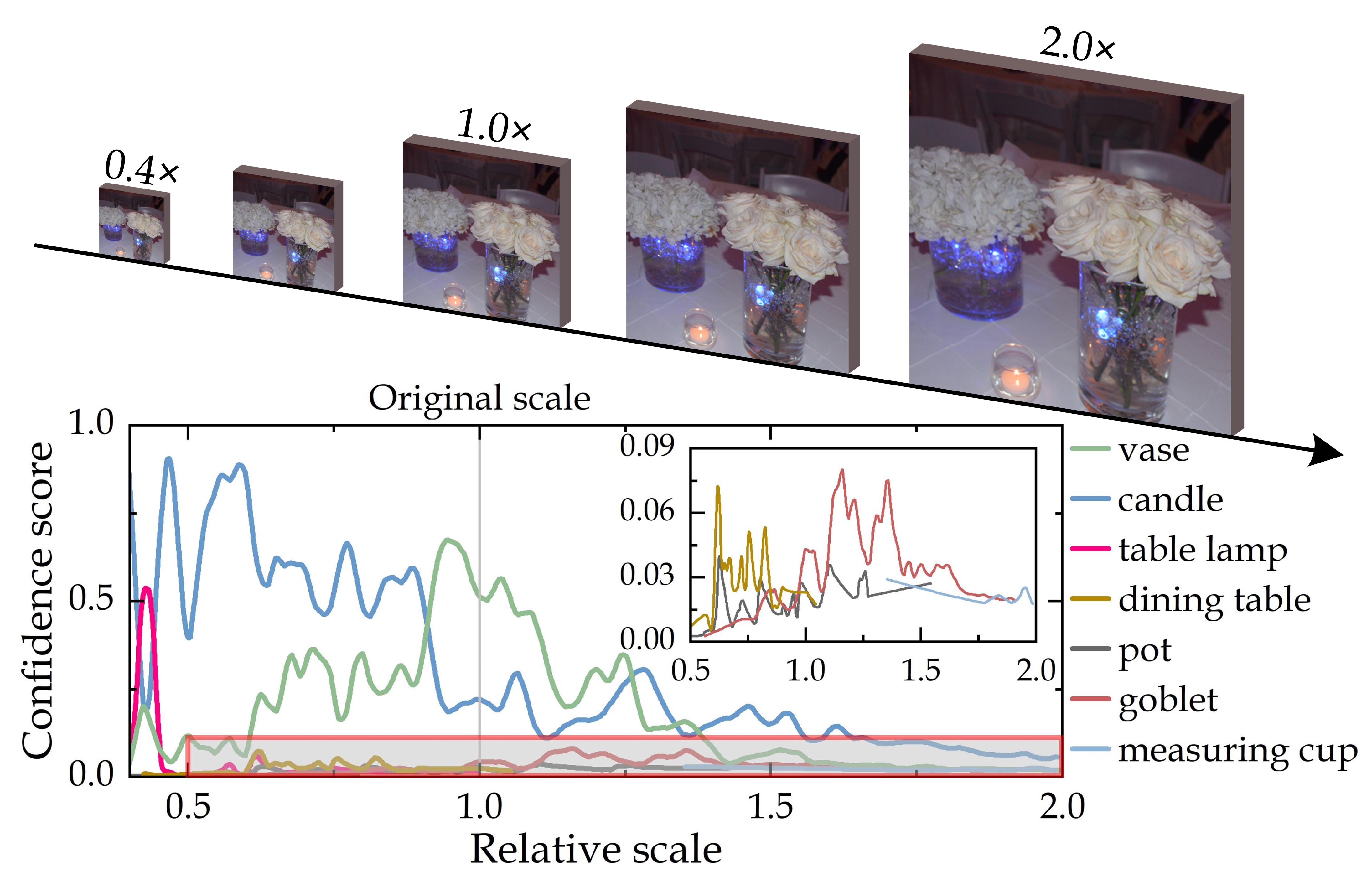}
		\vspace{-8mm}
		\caption{\textcolor{black}{Model focus varies significantly across scales. Ground truth: \texttt{vase}.}}
		\label{illu_scaling}
	\end{figure}

	\textcolor{black}{Our preliminary analysis reveals scaling's robust self-alignment across broad intensities. To unpack the mechanistic impact of scaling directionality, Fig. \ref{illu_scaling} provides a quantitative case study. We observe distinct transitions in predictive behavior across scales: large scales diminish the primary class's dominance, allowing multiple latent categories to emerge with competing confidence scores; small scales induce highly volatile confidence distributions, reflecting the scale sensitivity that local objects dominate the global semantics. Both regimes effectively exploit the image's inherent multi-object/multi-class information.}

	\textcolor{black}{Driven by these insights, S$^{4}$ST-\textit{Base} acts as a unified bidirectional scaling operation synergizing DI \cite{inputdiversity2019xie} and RDI \cite{RDI}. Crucially, instead of naive DI upscaling, we randomly crop a local patch and resize it back to original dimensions. This mitigates the severe computational overhead of large-scale inputs while amplifying local semantic exploitation, thereby boosting adversarial transferability.}
	
	In specific, we define the intensity of S$^{4}$ST-\textit{Base} by $r$. Consider a sample $\bm{x}$ with dimensions $H \times W$ and an intensity $r > 1$. With a probability $p_r$, S$^{4}$ST-\textit{Base} selects a relative scale factor $r^{\prime}$ from a uniform distribution $\mathcal{U}(\frac{1}{r}, r)$. If $r^{\prime} < 1$, the sample $\bm{x}$ is downscaled to dimensions $r^{\prime}H \times r^{\prime}W$, followed by random zero-padding to restore its original size. In contrast, if $r^{\prime} > 1$, a square patch of dimensions $\frac{1}{r^{\prime}}H \times \frac{1}{r^{\prime}}W$ is randomly cropped from the image and then upscaled to the initial dimensions.
	
	\subsubsection{S$^{\textit{4}}$ST-\textit{Aug}}\label{augdesign}
	
	\begin{figure}[tbp]
		\centering	
		\includegraphics[width=1\linewidth]{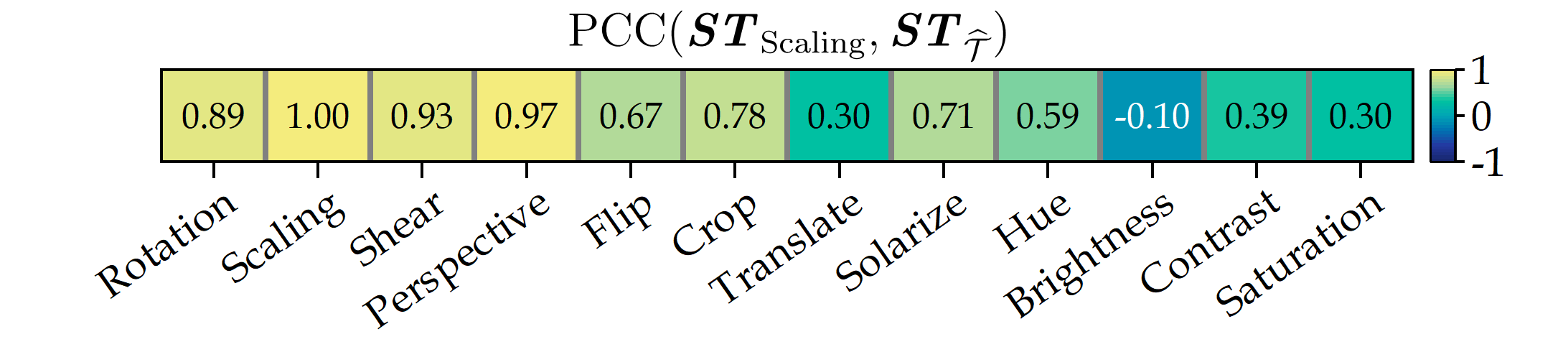}
		\vspace{-8mm}
		\caption{PCCs between the self-transferability to scaling and other basic transformations, derived from 11 intensity levels using S$^{4}$ST-\textit{Base}. This evaluation did not involve any black-box models.}
		\label{comptransfig}
		\vspace{-5mm}
	\end{figure}
	
	S$^4$ST-\textit{Aug} integrates complementary basic transformations to boost efficacy. With probability $p_{\textit{aug}}$, it applies a randomly chosen transformation from a curated pool to the entire image. To compile this pool, we evaluate S$^4$ST-\textit{Base} AEs across intensities $r \in [1.02, 1.04, 1.06, 1.08, 1.1, 1.2, 1.4, 1.6, 1.8, 2.0, 2.2]$, assessing their self-transferability correlations against various basic transformations (Fig. \ref{comptransfig}). We select five transformations exhibiting the lowest PCCs with $\textit{ST}_{\text{Scaling}}$, deliberately omitting translation and cropping due to their high overlap with S$^{4}$ST-\textit{Base}. Following \textit{torchvision}\footnote{\url{https://github.com/pytorch/vision/tree/main/torchvision/transforms}} parameterization, the S$^{4}$ST-\textit{Aug} pool includes:
	\begin{itemize}
		\item \textbf{Flip} randomly flips the image vertically or horizontally.
		\item \textbf{Brightness}/\textbf{Contrast}/\textbf{Saturation} randomly alters the \texttt{brightness}/\texttt{contrast}/\texttt{saturation} factor from 1 to $s^{\prime}\sim\mathcal{U}(0,2)$.
		\item \textbf{Hue} randomly adjusts the \texttt{hue} factor from 0 to $s^{\prime}\sim\mathcal{U}(-0.5, 0.5)$.
	\end{itemize}
	
	\subsubsection{S$^{\textit{4}}$ST-\textit{Block}}
	
	Block-wise transformations \cite{SIA,BSR,11079882} enhance transferability by increasing transformation diversity (Fig. \ref{existing_censensus_analysis}(a)). Validating its favorable correlation with self-transferability (SIA-legend in Fig. \ref{lineartest}), we advance this concept via S$^4$ST-\textit{Block}.
	
	It randomly divides the image into a non-overlapping $m_h \times m_w$ grid, ensuring a minimum block size for effectiveness. The dimension-preserving S$^4$ST-\textit{Base} is then independently applied to each block, diversifying the attack while maintaining coherent final adversarial examples.
	
	Note that block-wise application serves as an auxiliary technique rather than a conventional transformation. Its efficacy likely stems from the model's inherent tolerance to local transformations, driven by data distribution characteristics. Relevant analysis can be found in Section \ref{analysis_why_scaling}.
	
	\subsubsection{Black-box Parameter Selection Strategy}
	
	Strictly adhering to black-box constraints, S$^{4}$ST tunes hyperparameters without victim model access. First, our optimization objective maximizes average self-transferability across the 12 basic transformations at varying intensities, substituting black-box transferability. Second, we employ Bayesian Optimization (BO) to jointly search all four parameters $[p_{\textit{r}}, r, p_{\textit{aug}}, m]$, avoiding sub-optimalities from isolated component tuning. \textcolor{black}{Evaluated on a 50-image subset of ImageNet-Compatible, the 100-trial BO process (including 10 initial random explorations) takes $\sim$2 hours and 15 minutes on a single NVIDIA GeForce RTX 4090 GPU ($\sim$80 seconds/trial)}. The resulting configuration is $p_{\textit{r}}=0.9, r=1.7, p_{\textit{aug}}=1.0, m=6$.
	
\begin{figure}[tbp]
		\centering	
		\includegraphics[width=1\linewidth]{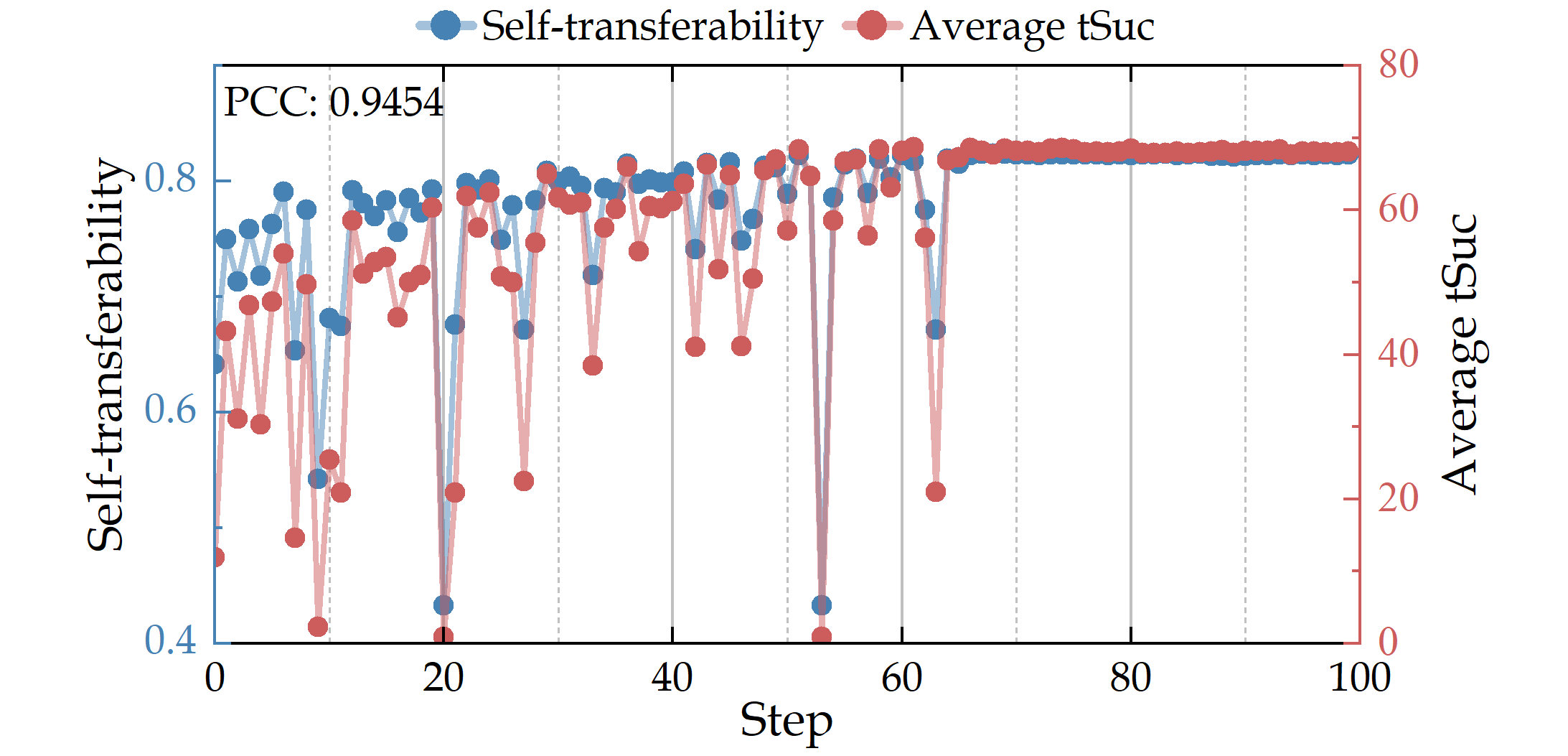}
		\vspace{-8mm}
		\caption{\textcolor{black}{Self-transferability as a blind estimator for S$^{4}$ST parameter search in BO. Average tSuc is calculated over 14 black-box models with $T=300$.}}
		\label{bayesoptim}
		\vspace{-5mm}
	\end{figure}

	Fig. \ref{bayesoptim} depicts self-transferability tracking the true black-box tSuc (across 14 models) during the search. As a surrogate objective, self-transferability effectively captures critical fluctuations, trends, and relative magnitudes of true transferability, successfully guiding the BO toward high-quality parameters.
	
	\textcolor{black}{\textbf{Usage and Boundary Cases.} S$^4$ST's hyperparameter tuning is domain-specific yet dataset- and task-agnostic. The parameters optimized here can be directly applied to multiple natural image datasets and tasks, and even extend to chest X-ray classification. However, because S$^4$ST relies on scaling and transformations derived from natural image analysis, severe domain shifts, such as face recognition, which involves fundamentally different data distributions and learning objectives, pose challenges (detailed in Section \ref{faceveri}).}
	
	\subsubsection{Relation to Existing Methods}
	
	While sharing implementation-level similarities with existing methods, S$^{4}$ST differs fundamentally in its analytical foundation and design rationale.
	
	\textbf{Scaling-centric Design.} While DI~\cite{inputdiversity2019xie} and RDI~\cite{RDI} empirically utilize scaling, S$^{4}$ST offers a principled, bidirectional mechanism. \textcolor{black}{First, our self-alignment analysis analytically justifies scaling's dominance over other transformations. Second, unlike unidirectional approaches, S$^{4}$ST-\textit{Base} leverages both scale enlargement (triggering fine-grained class competition) and reduction (exploiting global semantic sensitivity). Finally, replacing naive upscaling with a random crop-and-upscale strategy eliminates quadratic overhead while enforcing localized feature exploitation, yielding a superior effectiveness-efficiency balance over existing variants} (Section \ref{scalevariations}).
	
	\textbf{Complementary Augmentation and Block Processing.} Though inspired by composite methods like SIA \cite{SIA} and BSR \cite{BSR}, our design eschews empirical heuristics for principled analysis (Section \ref{failures_current_cons}). Specifically, S$^{4}$ST-\textit{Aug}'s complementary transformations are objectively selected via black-box self-transferability correlations (Fig.~\ref{comptransfig}), favoring color adjustments over intuitive geometric choices. Additionally, S$^{4}$ST-\textit{Block} exclusively employs self-alignment-verified block-wise scaling, contrasting with SIA's arbitrary transformations or BSR's empirical rotation (see Section~\ref{comptrans} for detailed comparisons).
	
	\section{Experiments}\label{exp}
	
	In this section, we detail the experimental setup, and scrutinize the analysis and the ablation of components for our proposed method, showcasing its novel aspects in contrast to existing techniques. This is followed by extensive experiments designed to compare our method with the existing transformation techniques and SoTA approaches, covering both in-domain and cross-domain scenarios.
	
	\subsection{Experimental Settings}\label{expdetails}
	
	\textbf{Dataset}. Our experiments are conducted on the ImageNet-Compatible dataset\footnote{\url{https://github.com/cleverhans-lab/cleverhans/tree/master/cleverhans_v3.1.0/examples/nips17_adversarial_competition/dataset}}, a benchmark commonly utilized for evaluating TTA performance \cite{logit,SelfU,ODI}. Released for the NIPS 2017 adversarial competition, it encompasses 1,000 images with dimensions of 299$\times$299, each paired with a targeted label. 
	
	\textbf{Models}. In line with prior studies, we generate AEs using ResNet-50 \cite{resnet2016kh} and transfer them to a variety of dissimilar victim models. These include CNN architectures \{MobileNet-v2 (MNv2) \cite{Sandler_2018_CVPR}, EfficientNet-b0 (EN) \cite{efficientnet}, ConvNeXt-small (CNX) \cite{liu2022convnet}, Inception-v3 (INv3) \cite{szegedy2016rethinking}, Inception-v4 (INv4) \cite{szegedy2017inception}, Inception ResNet-v2 (IRv2), and Xception (Xcep) \cite{chollet2017xception}\} and Vision Transformers (ViTs) \{ViT \cite{dosovitskiy2020image}, SwinT \cite{Liu_2021_ICCV}, MaxViT \cite{tu2022maxvit}, Twins \cite{chu2021twins}, PiT \cite{heo2021rethinking}, TNT \cite{NEURIPS2021_854d9fca}, and DeiT \cite{pmlr-v139-touvron21a}\}. We source their pre-trained weights from the \textit{torchvision}\footnote{\url{https://github.com/pytorch/vision/tree/master/torchvision/models}} and \textit{timm} \cite{rw2019timm} libraries. Supplemental black-box transfer experiments are performed on defense models, real-world APIs, VLMs, dense prediction models, with further details provided in subsequent sections.

	\textbf{Metrics}. We measure the efficacy of different methods by their average tSuc, \ie the proportion of successful transfers across the dataset from the surrogate to the victim models. \textcolor{black}{We also measure the average $\ell_{2}$ norm of generated perturbations to avoid unfair comparison \cite{chen2023measuring}.} Additionally, we evaluate the time required by different transformations to craft AEs. All experiments are conducted on an NVIDIA GeForce RTX 4090 GPU using the PyTorch \cite{NEURIPS2019_bdbca288} framework. For consistency, the batch size is set to 10 across all experiments, ensuring that no method surpasses the memory threshold, thereby maintaining fairness in the comparison of time consumption metrics.
	
	\subsection{Analysis on S$^{\textbf{4}}$ST}\label{methodanalysis}
	
	\subsubsection{Component Ablation}
	
	Table \ref{componentsAblation} details the component ablation. S$^{4}$ST-\textit{Aug} and S$^{4}$ST-\textit{Block} exhibit strong synergy, collectively driving the average tSuc to a notable 83.0\%. Furthermore, S$^{4}$ST-\textit{Aug}'s five-transformation pool demonstrates robust complementarity; excluding any single transformation yields minimal performance drop.	
	
	\begin{table}[bp]
		\vspace{-6mm}
		\caption{Component ablation of S$^{4}$ST.}
		\vspace{-4mm}
		\label{componentsAblation}
		\centering
		\text{(a) Results by different S$^{4}$ST combinations} \\ \vspace{1mm}
		\begin{tabular}{ccc|c|cc}
			\Xhline{1pt}
			S$^{4}$ST-\textit{Base} & -\textit{Aug} &-\textit{Block} & $[p_{\textit{r}}, r, p_{\textit{aug}}, m]$ & tSuc & Time \\ \hline
			\usym{1F5F8} &  &  & $[0.9, 1.7, 0., 1]$ & 58.2 & 1.24 \\
			\usym{1F5F8} & \usym{1F5F8} &  & $[0.9, 1.7, 1., 1]$ &  76.8 & 1.40 \\
			\usym{1F5F8} &  & \usym{1F5F8} & $[0.9, 1.7, 0., 6]$ & 74.3 & 1.31 \\
			\usym{1F5F8} & \usym{1F5F8} & \usym{1F5F8} & $[0.9, 1.7, 1., 6]$ & 83.0 & 1.45 \\
			\Xhline{1pt}
		\end{tabular}
		\\ \vspace{2mm}
		\text{(b) Results by excluding complementary augmentations from S$^{4}$ST} \\ \vspace{1mm}
		\begin{tabular}{c|ccccc}
			\Xhline{1pt}
			S$^{4}$ST & -Flip & -Contrast & -Brightness & -Saturation & -Hue \\ \hline
			83.0 & 80.8 & 82.6 & 82.7 & 82.7 & 82.6 \\
			\Xhline{1pt}
		\end{tabular}
	\end{table}
	
	\begin{figure}[tbp]
		\includegraphics[width=1\linewidth]{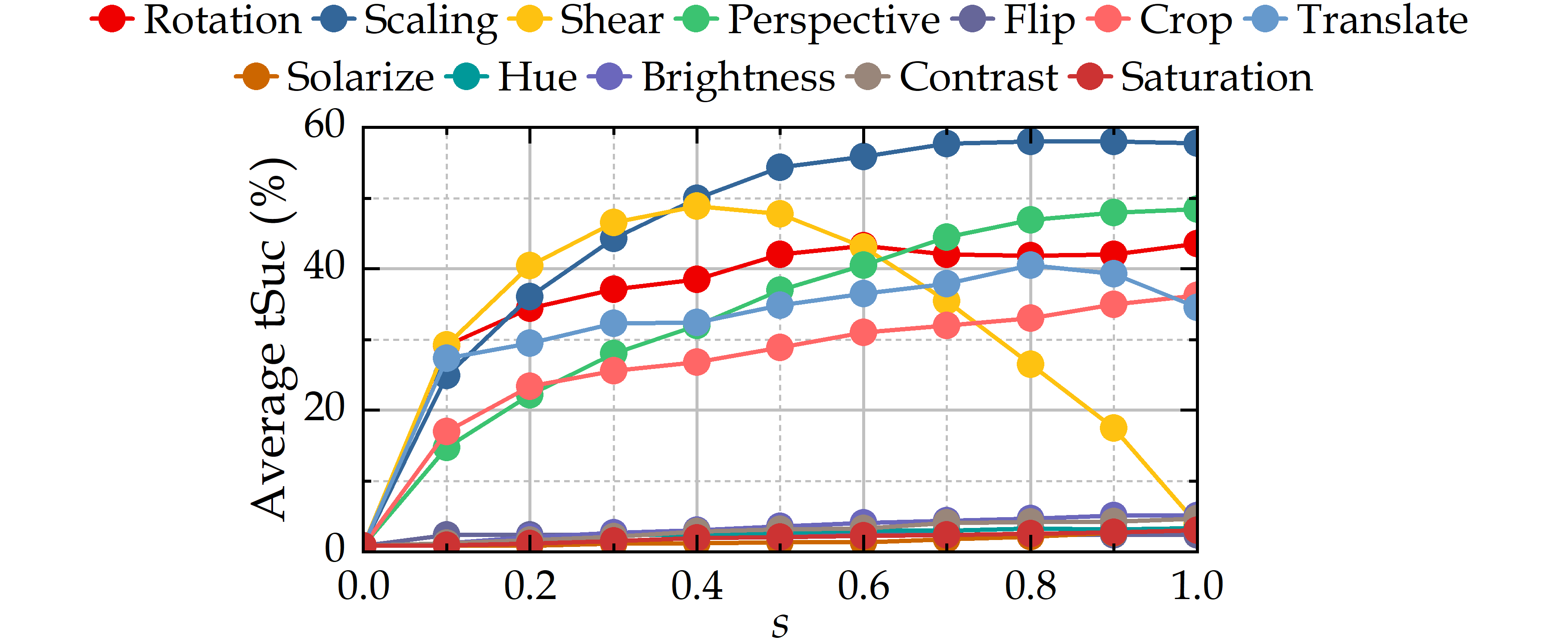}
		\vspace{-8mm}
		\caption{Results by applying various basic transformations with incremental intensities. The scaling transformation exhibits a distinctive superior enhancement in targeted transferability.}
		\label{atk_basic_trans}
		\vspace{-2mm}
	\end{figure}
	
	\begin{figure}[tbp]
		\vspace{-2mm}
		\includegraphics[width=1\linewidth]{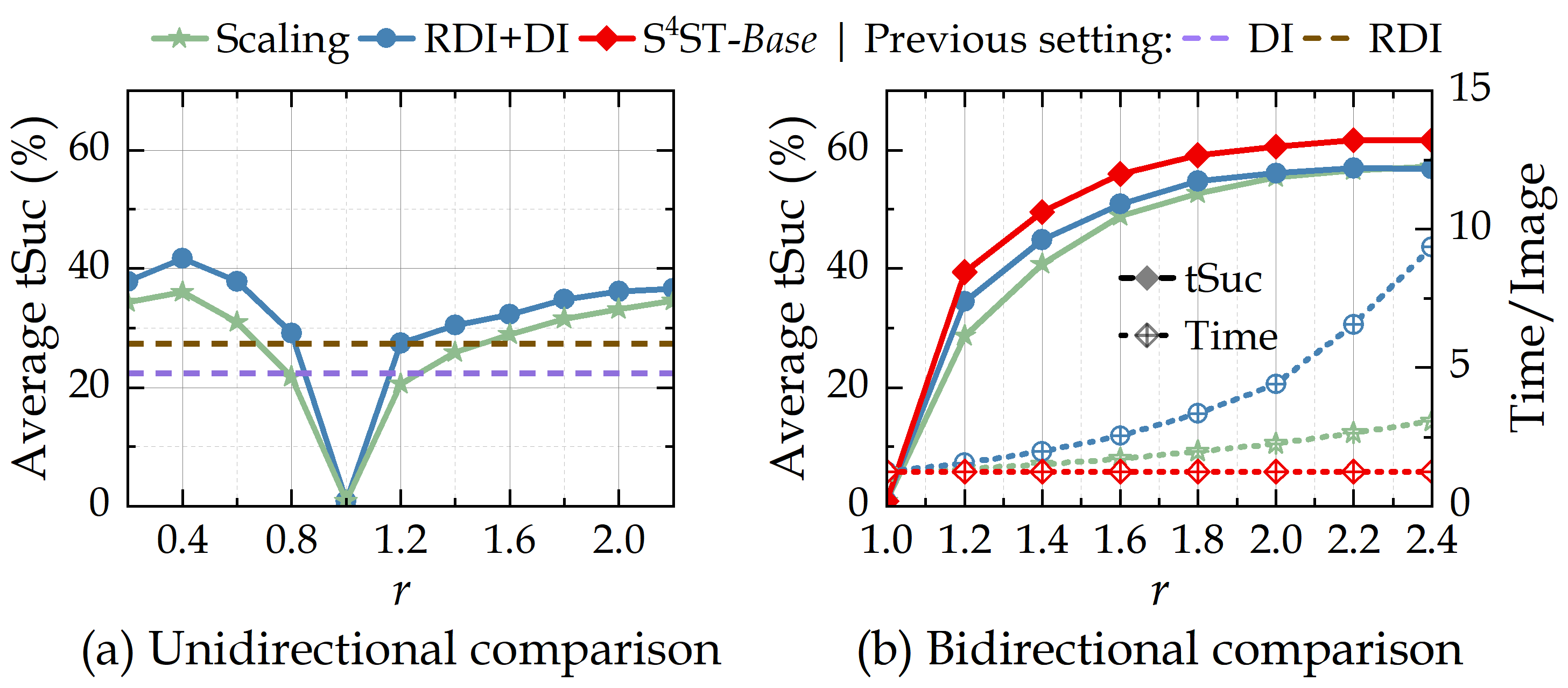}
		\vspace{-8mm}
		\caption{Our analysis uncovers significant untapped potential in existing scaling-related methods. We illustrate that prior achievements can be substantially enhanced by extending the scaling range and incorporating bidirectional scaling. Tailored refinements to the S$^{4}$ST-\textit{Base} further amplify this effect. The additional time required by our S$^{4}$ST is trivial, significantly outstripping the simple scaling and the integration of RDI \cite{RDI} and DI \cite{inputdiversity2019xie} methods in terms of computational speed.}
		\label{casestudy}
		\vspace{-2mm}
	\end{figure}
	
	\subsubsection{Scaling versus Other Basic Transformations}\label{basetranscomp}
	
	To validate simple scaling's superiority objectively, we evaluate AEs generated using studied basic transformations. Executing each with probability 0.9 and random intensity $s^{\prime} \sim \mathcal{U}(0, s)$, Fig. \ref{atk_basic_trans} shows scaling uniquely achieves an unparalleled 58.0\% average tSuc at $s=0.8$, outperforming shear (48.8\%) and perspective (48.4\%) by nearly 10\%. Notably, shear's efficacy wanes significantly at higher intensities. 
	
	\begin{figure}[bp]
		\vspace{-4mm}
		\includegraphics[width=1\linewidth]{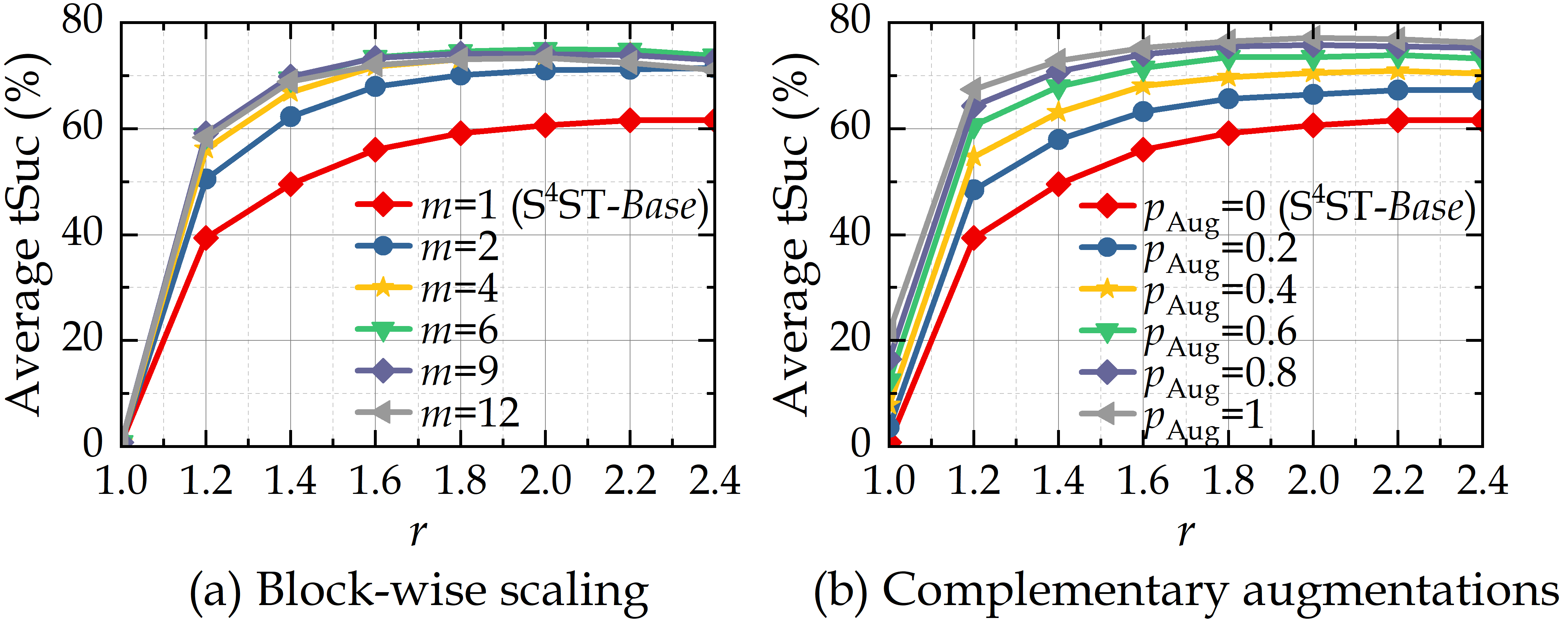}
		\vspace{-8mm}
		\caption{Further enhancements to S$^{4}$ST-\textit{Base}  achieved by (a) S$^{4}$ST-\textit{Block} and (b) S$^{4}$ST-\textit{Aug}, both of which demonstrate significant improvements.}
		\label{compoeffects}
	\end{figure}

	\subsubsection{Comparison among Scaling Variations}\label{scalevariations}

	Fig. \ref{casestudy} compares (R)DI methods under unidirectional and bidirectional settings. Under prior TTA configurations, DI and RDI attain merely 22.4\% and 27.3\% average tSuc, respectively. Minor parameter tuning elevates these to 36.6\% and 41.7\%. Synergizing both upscale and downscale directions yields 56.8\% at $r=2.4$. While comparable to simple scaling (57.2\%), our S$^{4}$ST-\textit{Base} fundamentally outperforms both, boosting tSuc to 61.6\%. This underscores our method's superiority in unlocking and augmenting basic scaling. Furthermore, Fig. \ref{casestudy}(b) confirms S$^{4}$ST-\textit{Base}'s computational efficiency: its incremental time investment over the baseline ($r=1.0$) is minimal, significantly outpacing rival integration methods.

	\subsubsection{Effectiveness of S$^{\textit{4}}$ST Components}
	
	Fig. \ref{compoeffects} illustrates the benefits of S$^{4}$ST-\textit{Aug} and S$^{4}$ST-\textit{Block} built upon S$^{4}$ST-\textit{Base} (fixing $p_{\textit{r}}=0.9$). For S$^{4}$ST-\textit{Block}, increasing the grid blocks $m$ markedly enhances efficacy, peaking at $m=6 \, (2\times 3)$; further increases impede optimization convergence within 900 iterations. For S$^{4}$ST-\textit{Aug}, higher application frequencies ($p_{\textit{Aug}}$) steadily amplify performance. Crucially, integrating the five complementary transformations with scaling ($r\geq1.2$) yields substantially better outcomes than their solitary application ($r=1.0$), reinforcing the soundness of synergizing a primary transformation with complementary enhancements.
	
	\begin{table*}[htbp]
		\caption{Comparison between input transformations based on ResNet-50 as the surrogate model. $^{{\dag}}$ strands for TTA-oriented methods. The best and the second best results are in \textcolor{t1}{\textbf{Red}} and \textcolor{t2}{\textbf{Blue}}, respectively.}
		\vspace{-4mm}
		\label{rn50_trans}
		\centering
		\resizebox{\linewidth}{!}{
			\begin{tabular}{l|ccc|cccccccc|cccccccc}
				\Xhline{1pt}
				Attack & tSuc & \textcolor{black}{$\ell_{2}$} & Time & MNv2 & EN & CNX  & INv3 & INv4 & IRv2 & Xcep & tSuc$_{\text{CNNs}}$ & ViT  & Swin & MaxViT & Twins & PiT & TNT & DeiT & tSuc$_{\text{ViTs}}$  \\ \hline
				TMI$_{\text{CVPR19}}$ \cite{momentum2018dong,evading2019dong} & 0.8 & 23.15 & 1.24 & 2.0 & 1.5 & 2.4 & 0.6 & 0.7 & 0.7 & 0.9 & 1.3 &  0.2 & 0.9 & 0.4 & 0.1 & 0.1 & 0.2 & 0.2 & 0.3 \\ 
				+Admix$_{\text{ICCV21}}$ \cite{admix} & 2.1 & 23.29 & 6.27 & 5.9 & 4.7 & 6.3 & 0.9 & 1.6 & 2.4 & 2.5 & 3.5 & 0.3 & 2.0 & 1.0 & 0.5 & 1.1 & 0.6 & 0.2 & 0.8 \\  
				+SI$_{\text{ICLR20}}$ \cite{nestrov2019lin} & 4.4 & 23.32 & 6.11 & 8.5 & 9.6 & 8.4 & 5.7 & 5.3 & 6.3 & 7.0 & 7.3 &  1.1 & 2.4 & 1.5 & 1.2 & 2.5 & 1.3 & 0.6 & 1.5 \\ 
				+SSA$_{\text{ECCV22}}$ \cite{SSA} & 19.8 & 23.80 & 1.43 & 35.0 & 32.0 & 35.1 & 22.5 & 30.4 & 35.1 & 33.0 & 31.9 &  4.8 & 11.1 & 7.1 & 7.4 & 10.8 & 7.2 & 6.0 & 7.8 \\ 
				+DI$_{\text{CVPR19}}$ \cite{inputdiversity2019xie} & 22.4 & 23.42 & 1.44 & 29.6 & 33.7 & 48.8 & 15.4 & 35.4 & 38.1 & 28.3 & 32.8 &  5.9 & 17.1 & 10.8 & 11.1 & 19.1 & 11.6 & 8.3 & 12.0 \\ 
				+RDI$_{\text{ECCV20}}$ \cite{RDI}& 27.3 & 23.45 & 1.24 & 37.1 & 42.9 & 47.4 & 25.8 & 39.8 & 45.6 & 36.9 & 39.4 & 11.0 & 21.7 & 13.2 & 13.6 & 22.3 & 15.0 & 10.2 & 15.3 \\ 
				+ODI$^{{\dag}}_\text{CVPR22}$ \cite{ODI} & 58.1 & 23.52 & 5.11 & 73.2 & 77.1 & 75.3 & {64.3} & 75.7 & 77.0 & 73.2 & 73.7 & 37.6 & 54.1 & 37.9 & 38.4 & 54.5 & 43.0 & 31.7 & 42.5 \\ 
				+SIA$_{\text{ICCV23}}$ \cite{SIA} & 58.3 & 23.57 & 1.55 & \textcolor{t2}{\textbf{82.5}} & 81.3 & {84.3} & 51.1 & 74.2 & 76.2 & 72.0 & 74.5 & 30.8 & {58.5} & {42.6} & 39.3 & 53.2 & 40.7 & 29.8 & 42.1 \\
				+DeCoWA$_{\text{AAAI24}}$ \cite{decowa} & 58.4 & 23.57 & 2.78 & 74.9 & 78.0 & 77.4 & 56.7 & 74.4 & 77.7 & 73.7 & 73.3 & {40.6} & 58.4 & 36.1 & 36.6 & 55.1 & 44.1 & 33.3 & {43.5} \\    
				+BSR$_{\text{CVPR24}}$ \cite{BSR} & 60.0 & 23.58 & 1.84 & 81.8 & 80.3 & 81.4 & 62.6 & {79.3} & {78.1} & {76.6} & {77.2} & 32.2 & 56.8 & 38.4 & {39.6} & {55.7} & {44.2} & {33.5} & 42.9 \\ 
				+T-Aug$^{{\dag}}_\text{ArXiv23}$ \cite{wei2023rethinking} & 65.5 & 23.57 & 1.33 & 82.1 & \textcolor{t2}{\textbf{82.7}} & 81.9 & \textcolor{t2}{\textbf{72.4}} & 84.1 & \textcolor{t2}{\textbf{84.4}} & 82.8 & 81.5 & 41.7 & 61.9 & 47.5 & 45.2 & \textcolor{t2}{\textbf{64.2}} & 50.2 & 36.4 & 49.6 \\ 
				+H-Aug$^{{\dag}}_\text{ArXiv23}$ \cite{wei2023rethinking} & \textcolor{t2}{\textbf{68.8}} & 23.29 & 5.64 & 81.8 & 82.4 & \textcolor{t2}{\textbf{85.9}} & 69.7 & \textcolor{t2}{\textbf{86.0}} & 82.8 & \textcolor{t2}{\textbf{84.7}} & \textcolor{t2}{\textbf{81.9}} & \textcolor{t2}{\textbf{43.1}} & \textcolor{t2}{\textbf{73.4}} & \textcolor{t2}{\textbf{56.7}} & \textcolor{t2}{\textbf{59.0}} & 63.0 & \textcolor{t2}{\textbf{54.1}} & \textcolor{t2}{\textbf{40.4}} & \textcolor{t2}{\textbf{55.7}} \\ 
				+S$^4$ST$^{{\dag}}_\text{Ours}$ & \textcolor{t1}{\textbf{83.0}}
				& 23.56 & 1.45 & \textcolor{t1}{\textbf{93.5}} & \textcolor{t1}{\textbf{93.9}} & \textcolor{t1}{\textbf{93.9}} & \textcolor{t1}{\textbf{85.9}} & \textcolor{t1}{\textbf{93.0}} & \textcolor{t1}{\textbf{92.9}} & \textcolor{t1}{\textbf{93.1}} & \textcolor{t1}{\textbf{92.3}} & \textcolor{t1}{\textbf{63.0}} & \textcolor{t1}{\textbf{86.3}} & \textcolor{t1}{\textbf{74.8}} & \textcolor{t1}{\textbf{73.8}} & \textcolor{t1}{\textbf{80.1}} & \textcolor{t1}{\textbf{75.1}} & \textcolor{t1}{\textbf{63.2}} & \textcolor{t1}{\textbf{73.8}} \\ 
				\Xhline{1pt}
		\end{tabular}}
		\vspace{-3mm}
	\end{table*}

	\begin{table}[tbp]
		\caption{Results by alternative surrogate models. $^{{\dag}}$ strands for TTA-oriented methods}
		\vspace{-4mm}
		\label{alter_trans}
		\centering
		\resizebox{1\linewidth}{!}{
			\begin{tabular}{l|ccc|ccc|ccc}
				\Xhline{1pt}
				Surrogate$\rightarrow$ & \multicolumn{3}{c|}{DenseNet-121 \cite{densenet2017huang}} & \multicolumn{3}{c|}{VGG16-bn \cite{vgg2015sk}} & \multicolumn{3}{c}{ViT \cite{dosovitskiy2020image}} \\ \hline
				Attack$\downarrow$ & tSuc$_{\text{CNNs}}$ & tSuc$_{\text{ViTs}}$ & Time &  tSuc$_{\text{CNNs}}$ & tSuc$_{\text{ViTs}}$ & Time &  tSuc$_{\text{CNNs}}$ & tSuc$_{\text{ViTs}}$ & Time \\ \hline
				TMI \cite{momentum2018dong,evading2019dong} & 1.7 &  0.4 & 1.86 & 1.5 & 0.2 & 1.66 & 0.0 & 0.0 & 3.99 \\ 
				+ODI$^{{\dag}}$ \cite{ODI} & 61.2 & 31.3 & 5.57 & 51.0 & 25.7 & 5.26 & \textcolor{t2}{\textbf{20.0}} & \textcolor{t2}{\textbf{43.9}} & 7.70 \\ 
				+SIA \cite{SIA} & 65.2 & 34.5 & 2.39 & 44.2 & 23.2 & 2.30 & 7.9 & 38.8 & 4.63 \\ 
				+DeCoWA \cite{decowa} & 56.8 & 30.5 & 3.89 & 59.7 & 33.3 & 3.66 & 6.4 & 24.1 & 8.30 \\ 
				+BSR \cite{BSR} & 63.8 & 32.0 & 2.03 & 43.8 & 20.3 & 1.95 & 15.5 & 41.7 & 4.27 \\ 
				+T-Aug$^{{\dag}}$ \cite{wei2023rethinking} & 65.0 & 33.7 & 2.01 & \textcolor{t2}{\textbf{61.6}} & 34.7 & 1.75 & 15.0 & 41.4 & 4.06 \\ 
				+H-Aug$^{{\dag}}$ \cite{wei2023rethinking} & \textcolor{t2}{\textbf{72.8}} & \textcolor{t2}{\textbf{45.8}} & 4.51 & 60.7 & \textcolor{t2}{\textbf{36.0}} & 7.00 & - & - & - \\ 	
				+S$^4$ST$^{{\dag}}$ (Ours) & \textcolor{t1}{\textbf{86.7}} & \textcolor{t1}{\textbf{62.6}} & 2.15 & \textcolor{t1}{\textbf{76.8}} & \textcolor{t1}{\textbf{54.4}} & 1.83 & \textcolor{t1}{\textbf{50.3}} & \textcolor{t1}{\textbf{81.6}} & 4.05 \\ 
				\Xhline{1pt}
			\end{tabular}
		}
		\vspace{-5mm}
	\end{table}
	
	\subsection{Comparison Against Existing Transformations}\label{comptrans}
	
	\textbf{Competitors \& Setup}. We compare S$^{4}$ST with DI \cite{inputdiversity2019xie}, RDI \cite{RDI}, SI \cite{nestrov2019lin}, Admix \cite{admix}, SSA \cite{SSA}, ODI \cite{ODI}, SIA \cite{SIA}, BSR \cite{BSR}, T-Aug \cite{wei2023rethinking}, and H-Aug \cite{wei2023rethinking}. For fair comparison, we limit transformed copies per iteration to 1 for SSA, SIA, BSR, and DeCoWa, and set $m_{1}=1, m_{2}=5$ for Admix. Unless otherwise specified, all gradient-based attacks adopt the same baseline approach described in Section \ref{baseline}, which incorporates TI \cite{evading2019dong}, MI \cite{momentum2018dong}, and margin-based CE loss \cite{marginangleT}, with $T=900$.
	
	\subsubsection{Single Surrogate Attacks}\label{singlesurroatt}
	
	Table \ref{rn50_trans} details the single-surrogate (ResNet-50) evaluation. S$^{4}$ST exhibits dominant performance, surpassing H-Aug and ODI by 14.2\% and 24.9\% in average tSuc, respectively. It is also highly efficient, requiring only 25.7\% and 28.4\% of the computational time of H-Aug and ODI. Notably, aligning with Fig. \ref{casestudy}, even simple scaling or S$^{4}$ST-\textit{Base} achieves transferability rivaling all baselines except T/H-Aug, while block-wise scaling and complementary augmentations synergistically secure the final SoTA performance. Furthermore, perturbations from all methods saturate under the $\ell_{\infty}$ constraint, exhibiting negligible $\ell_{2}$ norm differences. This consistent superiority extends to alternative surrogate models (Table \ref{alter_trans}, without retuning) and varying attack iterations (Fig. \ref{t-s-plot}).

\begin{figure}[tbp]
		\vspace{-2mm}
		\includegraphics[width=1\linewidth]{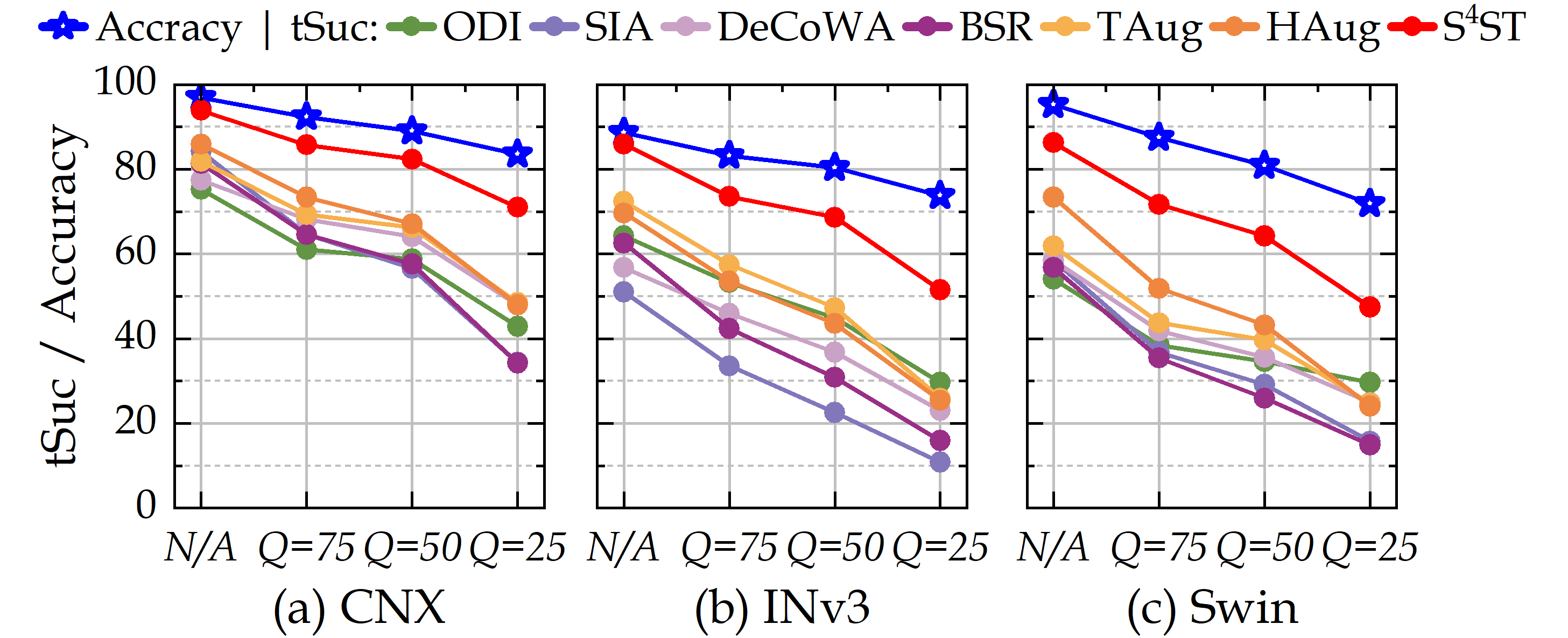}
		\vspace{-8mm}
		\caption{\textcolor{black}{JPEG \cite{guo2018jpeg} defense evaluation. \textit{N/A} indicates w/o defense and \textit{Q} is the JPEG compression quality. AEs are generated by ResNet-50 and transferred against (a) CNX, (b) INv3, and (c) Swin.}}
		\label{jpegresults}
		\vspace{-5mm}
	\end{figure}

	\textcolor{black}{We also evaluate robustness against JPEG compression \cite{guo2018jpeg}, a representative pre-processing defense. As shown in Fig. \ref{jpegresults}, RN50-generated AEs are subjected to varying JPEG qualities before victim model evaluation. Notably, S$^4$ST maintains a high tSuc even when aggressive compression severely degrades the victims' natural accuracy, demonstrating superior robustness. This resilience challenges the conventional generalization-robustness paradox, complicating defenders' efforts to mitigate attacks without compromising legitimate model utility.}

	\subsubsection{Ensemble-based Attacks and Against Robust Victims}
	
	Table \ref{secured_eval} presents the targeted (tSuc) and untargeted (uSuc) success rates for ensemble-based attacks against robust victims. We evaluate two strategies: self-ensemble \cite{SSA,SIA,decowa,BSR} (averaging gradients over multiple transformed copies per iteration) and cross-model ensemble \cite{momentum2018dong} (averaging logits across different models). Following standard protocols \cite{naseer2021generating,zhao2023minimizing}, our robust victim models include Augmix \cite{hendrycks2019augmix}, stylized training \cite{geirhos2018imagenet}, adversarial training (AT) \cite{madry2018towards,salman2020adversarially}, and Ensemble AT \cite{tramer2018ensemble}.
	
	S$^4$ST significantly outperforms existing methods across all scenarios. Crucially, it scales exceptionally well with increased computational effort and additional surrogate models. In the self-ensemble setting, S$^4$ST achieves performance rivaling the cross-model ensemble, a critical practical advantage, given the difficulty of obtaining diverse surrogate models in strict black-box scenarios. Furthermore, while adversarially trained models with larger perturbation budgets show stronger defenses, they remain highly susceptible to untargeted erroneous predictions induced by these targeted AEs.

	\vspace{-2mm}
	\subsection{Comparison with SoTA}\label{rsnours}
	
	\textbf{Competitors \& Setup}. We benchmark S$^{4}$ST-TMI against SoTA solutions: SASD-WS-DI-TMI \cite{Wu_2024_CVPR}, DRA-DI-TMI \cite{zhu2022toward}, IAA \cite{zhu2021rethinking}, Clean Feature Mixup (CFM-RDI-TMI) \cite{CFM}, Self-Universality (SU-DI-TMI) \cite{SelfU}, Little Robust Surrogate (LRS-DI-TMI) \cite{springer2021little}, Dominant Feature Attack (DFA) \cite{domainfeatureuap}, TTP \cite{naseer2021generating}, M3D \cite{zhao2023minimizing}, and C-GSP \cite{yang2022boosting}. Because training universal perturbations \cite{domainfeatureuap} or single-target generators \cite{naseer2021generating,zhao2023minimizing} for all ImageNet-Compatible labels is computationally prohibitive (see supplementary), we adopt the 10-Targets (all-source) setting \cite{naseer2021generating}. We average the tSuc on 10 target labels \{\texttt{Grey-Owl}, \texttt{Goose}, \texttt{Bulldog}, \texttt{Hippopotamus}, \texttt{Cannon}, \texttt{Fire-Truck}, \texttt{Model-T}, \texttt{Parachute}, \texttt{Snowmobile}, \texttt{Street sign}\} against the entire dataset. All methods target a ResNet-50 surrogate, evaluating across CNNs, ViTs, and robust models (Table \ref{agnostic_comp}). 
	
	\begin{table*}[tbp]
		\caption{Results of attacking robust models (tSuc/uSuc) and ensemble-based attacks. $^{{\dag}}$ stands for TTA-oriented methods. uSuc is the ratio (in \%) of the false predictions caused by the targeted AE by the original correct predictions.}
		\vspace{-4mm}
		\label{secured_eval}
		\centering
		\resizebox{.95\linewidth}{!}{
			\begin{tabular}{l|ccccccccccc|c|c}
				\Xhline{1pt}
				\multirow{2}{*}{Attack} & \multirow{2}{*}{{Augmix \cite{hendrycks2019augmix}}} & \multicolumn{2}{c}{Stylized \cite{geirhos2018imagenet}} & \multicolumn{4}{c}{Adversarial training (AT) \cite{madry2018towards}}  & \multicolumn{3}{c}{Ensemble AT \cite{tramer2018ensemble}}  & \multirow{2}{*}{Avg.} & \multirow{2}{*}{tSuc$_{\text{CNNs}}$} &  \multirow{2}{*}{tSuc$_{\text{ViTs}}$}\\  \cline{3-4}  \cline{5-8}  \cline{9-11}  
				
				&    & SIN & SIN-IN &$\ell_{2}=.1$ & $\ell_{2}=.5$ & $\ell_{\infty}=.5$ & $\ell_{\infty}=1.$ & INv3$_{ens3}$ & INv3$_{ens4}$ & IRv2$_{ens}$ &  &  &  \\ \hline
				
				\multicolumn{14}{c}{Surrogate: ResNet-50} \\ \hline
				TMI \cite{momentum2018dong,evading2019dong} & 3.4/55.8 & 0.5/51.6 & 33.7/83.4 & 0.2/42.4 & 0.0/28.6 & 0.0/24.3 & 0.0/19.1 & 0.0/17.8 & 0.0/17.9 & 0.0/12.4 & 3.8/35.3 & 1.3 & 0.3 \\ 
				+ODI$^{{\dag}}$ \cite{ODI} & 78.5/94.6 & 45.2/87.6 & 96.5/99.4 & 65.6/88.8 & 3.5/44.5 & 6.4/46.2 & 0.6/28.2 & 14.4/51.4 & 5.8/40.1 & 3.2/27.6 & 32.0/60.8 & 73.7 & 42.5 \\
				+SIA \cite{SIA} & \textcolor{t2}{\textbf{89.1}}/96.7 & 39.8/84.6 & \textcolor{t2}{\textbf{99.6}}/\textcolor{t1}{\textbf{100.0}} & 64.1/87.5 & 1.5/41.5 & 3.7/40.2 & 0.1/27.2 & 5.7/37.7 & 1.7/32.5 & 0.6/19.9 & 30.6/56.8 & 74.5 & 42.1 \\
				+DeCoWA \cite{decowa} & 82.8/95.7 & 45.2/87.9 & 97.7/99.8 & \textcolor{t2}{\textbf{72.2}}/\textcolor{t2}{\textbf{91.7}} & \textcolor{t2}{\textbf{4.0}}/\textcolor{t2}{\textbf{48.5}} & \textcolor{t2}{\textbf{8.2}}/\textcolor{t2}{\textbf{49.0}} & \textcolor{t2}{\textbf{0.7}}/\textcolor{t2}{\textbf{30.2}} & 16.8/\textcolor{t2}{\textbf{58.2}} & \textcolor{t2}{\textbf{8.8}}/\textcolor{t2}{\textbf{49.0}} & 4.0/\textcolor{t2}{\textbf{34.1}} & 34.0/\textcolor{t2}{\textbf{64.4}} & 73.3 & 43.5 \\
				+BSR \cite{BSR} & 86.5/96.1 & \textcolor{t2}{\textbf{48.0}}/87.0 & 98.9/\textcolor{t2}{\textbf{99.9}} & 69.6/89.9 & 2.8/44.9 & 5.5/46.9 & 0.3/25.7 & 11.0/48.7 & 3.8/38.5 & 2.1/27.6 & 32.9/60.5 & 77.2 & 42.9 \\
				+T-Aug$^{{\dag}}$ \cite{wei2023rethinking} & 87.4/\textcolor{t2}{\textbf{97.1}} & 44.2/\textcolor{t2}{\textbf{88.6}} & 97.6/99.8 & 71.8/90.0 & 3.7/46.8 & 7.0/47.2 & 0.3/28.8 & \textcolor{t2}{\textbf{19.5}}/57.1 & 8.0/46.1 & \textcolor{t2}{\textbf{4.6}}/33.4 & \textcolor{t2}{\textbf{34.4}}/63.5 & 81.5 & 49.6 \\
				+H-Aug$^{{\dag}}$ \cite{wei2023rethinking} & 80.7/94.5 & 34.8/81.0 & 93.4/97.8 & 62.0/86.7 & 2.8/43.9 & 5.6/44.4 & 0.3/27.9 & 13.0/49.6 & 5.4/39.0 & 2.5/26.6 & 30.1/59.1 & \textcolor{t2}{\textbf{81.9}} & \textcolor{t2}{\textbf{55.7}} \\
				+S$^4$ST$^{{\dag}}$ (Ours)  &  \textcolor{t1}{\textbf{94.9}}/\textcolor{t1}{\textbf{98.8}} & \textcolor{t1}{\textbf{64.1}}/\textcolor{t1}{\textbf{92.3}} & \textcolor{t1}{\textbf{99.8}}/\textcolor{t2}{\textbf{99.9}} & \textcolor{t1}{\textbf{89.2}}/\textcolor{t1}{\textbf{96.7}} & \textcolor{t1}{\textbf{13.7}}/\textcolor{t1}{\textbf{54.5}} & \textcolor{t1}{\textbf{21.2}}/\textcolor{t1}{\textbf{57.5}} & \textcolor{t1}{\textbf{2.2}}/\textcolor{t1}{\textbf{31.0}} & \textcolor{t1}{\textbf{42.3}}/\textcolor{t1}{\textbf{71.0}} & \textcolor{t1}{\textbf{20.7}}/\textcolor{t1}{\textbf{55.4}} & \textcolor{t1}{\textbf{12.4}}/\textcolor{t1}{\textbf{40.1}} & \textcolor{t1}{\textbf{46.1}}/\textcolor{t1}{\textbf{69.7}} & \textcolor{t1}{\textbf{92.3}} & \textcolor{t1}{\textbf{73.8}} \\
				\hline
				
				\multicolumn{14}{c}{Surrogate: ResNet-50 $\times$ 10 (Self-Ensemble)} \\ \hline
				TMI$_{\textit{Self-Ens}}$ & 3.4/55.8 & 0.5/51.6 & 33.7/83.4 & 0.2/42.4 & 0.0/28.6 & 0.0/24.3 & 0.0/19.1 & 0.0/17.8 & 0.0/17.9 & 0.0/12.4 & 3.8/35.3 & 1.3 & 0.3 \\ 
				+ODI$^{{\dag}}$ \cite{ODI} & 92.2/98.2 & \textcolor{t2}{\textbf{66.9}}/93.3 & 98.9/\textcolor{t1}{\textbf{99.9}} & \textcolor{t2}{\textbf{83.2}}/94.4 & 7.5/\textcolor{t2}{\textbf{51.1}} & 12.5/\textcolor{t2}{\textbf{52.2}} & \textcolor{t2}{\textbf{1.1}}/30.6 & 27.8/\textcolor{t2}{\textbf{65.6}} & 14.0/\textcolor{t2}{\textbf{53.3}} & \textcolor{t2}{\textbf{7.0}}/\textcolor{t2}{\textbf{39.8}} & \textcolor{t2}{\textbf{41.1}}/\textcolor{t2}{\textbf{67.8}} & 87.0 & 63.6  \\
				+SIA \cite{SIA} &  92.7/98.2 & 47.0/86.0 & \textcolor{t1}{\textbf{99.9}}/\textcolor{t1}{\textbf{99.9}} & 74.1/90.4 & 2.0/43.7 & 4.5/44.1 & 0.5/27.0 & 7.6/43.0 & 2.8/35.2 & 1.0/24.2 & 33.2/59.2 & 81.4 & 53.1  \\
				+DeCoWA \cite{decowa} & 88.1/96.9 & 56.4/\textcolor{t2}{\textbf{93.4}} & 97.8/\textcolor{t2}{\textbf{99.8}} & 78.0/92.8 & 4.7/47.8 & 8.5/50.2 & 0.8/\textcolor{t2}{\textbf{31.0}} & 18.7/59.9 & 9.3/49.6 & 4.4/36.2 & 36.7/65.8 & 78.4 & 52.8 \\
				+BSR \cite{BSR} & 91.3/97.8 & 54.4/88.7 & \textcolor{t2}{\textbf{99.5}}/\textcolor{t1}{\textbf{99.9}} & 74.0/91.2 & 2.8/44.6 & 6.3/47.9 & 0.5/27.1 & 11.6/52.0 & 5.6/42.6 & 2.6/31.3 & 34.9/62.3 & 82.6 & 51.5 \\
				+T-Aug$^{{\dag}}$ \cite{wei2023rethinking} &  92.6/\textcolor{t2}{\textbf{98.3}} & 56.2/91.1 & 98.6/\textcolor{t2}{\textbf{99.8}} & 81.1/93.5 & 5.7/\textcolor{t2}{\textbf{51.1}} & 10.8/51.0 & 0.5/29.5 & 24.6/62.4 & 11.5/50.4 & 6.5/37.0 & 38.8/66.4 & 87.3 & 62.6 \\
				+H-Aug$^{{\dag}}$ \cite{wei2023rethinking} &  \textcolor{t2}{\textbf{92.8}}/98.2 & 58.4/90.5 & 98.1/\textcolor{t2}{\textbf{99.8}} & 81.4/\textcolor{t2}{\textbf{94.7}} & \textcolor{t2}{\textbf{8.0}}/49.4 & \textcolor{t2}{\textbf{13.5}}/51.7 & 0.6/30.0 & \textcolor{t2}{\textbf{29.8}}/65.0 & \textcolor{t2}{\textbf{14.2}}/51.7 & 6.9/37.6 & 40.4/66.9 & \textcolor{t2}{\textbf{90.4}} & \textcolor{t2}{\textbf{73.3}} \\
				+S$^4$ST$^{{\dag}}$ (Ours)  &  \textcolor{t1}{\textbf{98.3}}/\textcolor{t1}{\textbf{99.8}} & \textcolor{t1}{\textbf{79.8}}/\textcolor{t1}{\textbf{96.1}} & \textcolor{t1}{\textbf{99.9}}/\textcolor{t1}{\textbf{99.9}} & \textcolor{t1}{\textbf{94.3}}/\textcolor{t1}{\textbf{98.5}} & \textcolor{t1}{\textbf{25.0}}/\textcolor{t1}{\textbf{63.8}} & \textcolor{t1}{\textbf{35.3}}/\textcolor{t1}{\textbf{65.8}} & \textcolor{t1}{\textbf{3.5}}/\textcolor{t1}{\textbf{34.4}} & \textcolor{t1}{\textbf{64.0}}/\textcolor{t1}{\textbf{83.9}} & \textcolor{t1}{\textbf{41.5}}/\textcolor{t1}{\textbf{70.9}} & \textcolor{t1}{\textbf{28.7}}/\textcolor{t1}{\textbf{60.0}} & \textcolor{t1}{\textbf{57.0}}/\textcolor{t1}{\textbf{77.3}} & \textcolor{t1}{\textbf{95.8}}& \textcolor{t1}{\textbf{83.7}} \\ \hline
				
				\multicolumn{14}{c}{Surrogate: ResNet-50 + ResNet-152 + DenseNet-121 + VGG16-bn (Cross-Model Ensemble)} \\ \hline
				TMI$_{\textit{Ens}}$ & 28.0/69.1 & 5.0/56.6 & 67.8/89.5 & 5.4/49.4 & 0.2/28.6 & 0.1/26.2 & 0.0/21.0 & 0.0/19.7 & 0.0/19.4 & 0.0/12.2 & 10.7/39.2 & 23.0 & 9.7 \\ 
				+ODI$^{{\dag}}$ \cite{ODI}  & 93.0/97.4 & 61.8/88.6 & 98.9/\textcolor{t2}{\textbf{99.8}} & 79.7/92.7 & 5.6/48.7 & 10.0/49.8 & 0.9/30.1 & 20.8/55.7 & 10.1/42.9 & 5.6/30.5 & 38.6/63.6 & 91.8 & 73.8 \\
				+SIA \cite{SIA}  & \textcolor{t2}{\textbf{98.4}}/99.5 & 68.5/92.5 & \textcolor{t2}{\textbf{99.8}}/\textcolor{t1}{\textbf{99.9}} & 90.2/97.1 & 6.9/50.4 & 13.0/52.8 & 1.0/30.8 & 23.5/57.4 & 9.9/43.1 & 5.4/30.2 & 41.7/65.4 & 95.9 & 83.9 \\
				+DeCoWA \cite{decowa} & 97.7/99.0 & \textcolor{t2}{\textbf{77.0}}/\textcolor{t2}{\textbf{96.2}} & 99.4/\textcolor{t1}{\textbf{99.9}} & \textcolor{t2}{\textbf{93.4}}/\textcolor{t2}{\textbf{98.0}} & \textcolor{t2}{\textbf{13.4}}/\textcolor{t2}{\textbf{57.7}} & \textcolor{t2}{\textbf{21.7}}/\textcolor{t2}{\textbf{59.8}} & \textcolor{t2}{\textbf{2.4}}/\textcolor{t2}{\textbf{35.4}} & \textcolor{t2}{\textbf{51.9}}/\textcolor{t2}{\textbf{79.7}} & \textcolor{t2}{\textbf{32.5}}/\textcolor{t2}{\textbf{65.6}} & \textcolor{t2}{\textbf{17.7}}/\textcolor{t2}{\textbf{49.0}} & \textcolor{t2}{\textbf{50.7}}/\textcolor{t2}{\textbf{74.0}} & 95.2 & 82.2 \\
				+BSR \cite{BSR} &  98.1/\textcolor{t2}{\textbf{99.6}} & 73.0/94.2 & \textcolor{t2}{\textbf{99.8}}/\textcolor{t1}{\textbf{99.9}} & 89.8/96.8 & 7.3/52.0 & 15.4/55.9 & 1.0/31.3 & 35.0/67.1 & 14.0/51.3 & 8.4/37.1 & 44.2/68.5 & 95.4 & 78.7 \\
				+T-Aug$^{{\dag}}$ \cite{wei2023rethinking} & 97.6/99.4 & 67.7/93.9 & 99.7/\textcolor{t1}{\textbf{99.9}} & 90.3/97.4 & 10.5/52.5 & 16.5/57.4 & 1.4/33.3 & 48.8/75.4 & 24.3/56.9 & 14.7/44.8 & 47.2/71.1 & 96.3 & 83.1 \\
				+H-Aug$^{{\dag}}$ \cite{wei2023rethinking} & 97.1/99.1 & 62.9/91.2 & 99.4/\textcolor{t1}{\textbf{99.9}} & 88.1/96.0 & 8.6/52.2 & 14.8/54.6 & 1.1/31.5 & 39.4/69.0 & 17.1/52.6 & 9.1/39.6 & 43.8/68.6 & \textcolor{t2}{\textbf{96.5}} & \textcolor{t2}{\textbf{85.3}} \\
				+S$^4$ST$^{{\dag}}$ (Ours) &  \textcolor{t1}{\textbf{99.3}}/\textcolor{t1}{\textbf{99.9}} & \textcolor{t1}{\textbf{81.6}}/\textcolor{t1}{\textbf{96.3}} & \textcolor{t1}{\textbf{99.9}}/\textcolor{t1}{\textbf{99.9}} & \textcolor{t1}{\textbf{96.6}}/\textcolor{t1}{\textbf{99.0}} & \textcolor{t1}{\textbf{27.2}}/\textcolor{t1}{\textbf{64.4}} & \textcolor{t1}{\textbf{39.3}}/\textcolor{t1}{\textbf{68.6}} & \textcolor{t1}{\textbf{4.5}}/\textcolor{t1}{\textbf{36.3}} & \textcolor{t1}{\textbf{70.8}}/\textcolor{t1}{\textbf{87.1}} & \textcolor{t1}{\textbf{47.0}}/\textcolor{t1}{\textbf{72.5}} & \textcolor{t1}{\textbf{30.5}}/\textcolor{t1}{\textbf{56.9}} & \textcolor{t1}{\textbf{59.7}}/\textcolor{t1}{\textbf{78.1}} & \textcolor{t1}{\textbf{98.6}} & \textcolor{t1}{\textbf{92.5}} \\
				\Xhline{1pt}
		\end{tabular}}
		\vspace{-3mm}
	\end{table*}
	
	\begin{table*}[tbp]
		\caption{Comparison with SoTA under the 10-Targets (all-source) setting. $\#$ Data means the number of samples involved in method training. $^{{\ddag}}$ means the data is from the same dataset of victim models. $^{\lozenge}$ means the victim models are involved for parameter tuning.}
		\vspace{-4mm}
		\label{agnostic_comp}
		\centering
		\resizebox{.95\linewidth}{!}{
			\begin{tabular}{l|c|ccccc|ccccc|ccccc|cc}
				\Xhline{1pt}
				\multirow{2}{*}{Attack} & \multirow{2}{*}{$\#$ Data}	& \multicolumn{5}{c|}{CNNs} & \multicolumn{5}{c|}{ViTs} &\multicolumn{5}{c|}{Robust mechanisms} & \multirow{2}{*}{Avg.} & \multirow{2}{*}{\textcolor{black}{$\ell_{2}$}} \\  \cline{3-17}
				& & CNX & INv3 & INv4 & IRv2 & Xcep & ViT & Swin  & PiT & TNT & DeiT  & Augmix & SIN & SIN-IN & AT$_{\ell_{2}=.5}$ & AT$_{\ell_{\infty}=.5}$  &  &  \\ \hline
				DFA$_\text{CVPR20}$ \cite{domainfeatureuap}& 50K$^{{\ddag\lozenge}}$ &  25.3 & 1.6 & 15.0 & 16.3 & 13.8 & 0.7 & 3.1 & 1.6 & 1.7 & 3.9 & 19.9 & 7.8 & 56.4 & 0.3 & 0.3 & 11.2 & 21.19 \\
				TTP$_\text{ICCV21}$ \cite{naseer2021generating}& 50K$^{{\ddag\lozenge}}$ & 54.7 & 38.4 & 56.6 & 57.9 & 60.6 & 30.0 & 35.6 & 24.4 & 28.6 & 11.4 & 68.7 & 27.8 & 84.1 & 2.9 & 5.4 & 39.1 & 19.87 \\		
				M3D$_\text{CVPR23}$ \cite{zhao2023minimizing}& 50K$^{{\ddag\lozenge}}$  &	84.0 & 65.9 & 83.5 & 85.9 & 84.4 & \textcolor{t2}{\textbf{60.8}} & 67.7 & 62.0 & 62.2 & 41.0 & 88.0 & 62.9 & 94.6 & 5.2 & 9.2 & 63.8 & 19.20 \\	
				C-GSP$_\text{ECCV22}$ \cite{yang2022boosting}& 1.2M$^{\ddag\lozenge}$ & 57.2 & 20.5 & 47.1 & 48.5 & 46.7 & 22.6 & 26.5 & 19.7 & 22.1 & 9.8 & 50.7 & 17.4 & 72.1 & 1.4 & 1.8 & 30.9 & 15.41 \\
				IAA$_\text{ICLR22}$ \cite{zhu2021rethinking}&  5K$^{{\ddag\lozenge}}$ &  63.2 & 30.7 & 55.2 & 58.7 & 54.6 & 15.9 & 33.2 & 24.6 & 27.5 & 17.0 & 81.6 & 42.7 & 98.7 & 1.4 & 2.4 & 40.5  & 22.82  \\
				DRA-DI-TMI$_\text{TIP22}$ \cite{zhu2022toward}&  1.2M$^{{\ddag\lozenge}}$ & 58.3 & 69.2 & 58.9 & 67.9 & 61.7 & \textcolor{t1}{\textbf{69.1}} & 55.9 & 63.1 & 55.6 & 46.6 & 86.5 & \textcolor{t1}{\textbf{90.2}} & 94.3 & \textcolor{t1}{\textbf{73.4}} & \textcolor{t1}{\textbf{71.8}} & 68.2 &  23.53 \\
				LRS-DI-TMI$_\text{NeurIPS21}$ \cite{springer2021little}& 1.2M$^{\ddag\lozenge}$ &  72.3 & 55.1 & 61.3 & 72.9 & 60.5 & 45.6 & 47.2 & 50.9 & 41.5 & 35.1 & 85.0 & 41.5 & 93.3 & \textcolor{t2}{\textbf{24.5}} & \textcolor{t2}{\textbf{31.7}} & 54.6 & 23.50  \\
				SASD-WS-DI-TMI$_\text{CVPR24}$ \cite{Wu_2024_CVPR}&  1.2M$^{{\ddag\lozenge}}$ & 88.6 & \textcolor{t2}{\textbf{85.2}} & \textcolor{t2}{\textbf{92.2}} & \textcolor{t2}{\textbf{93.3}} & \textcolor{t2}{\textbf{91.2}} & 58.2 & 70.4 & 68.8 & 61.6 & 47.2 & 95.3 & \textcolor{t2}{\textbf{88.3}} & \textcolor{t1}{\textbf{99.8}} & 12.4 & 20.8 & \textcolor{t2}{\textbf{71.6}} &  23.30 \\
				SU-DI-TMI$_\text{CVPR23}$ \cite{SelfU}& N/A$^{\lozenge}$ & 54.7 & 17.1 & 41.6 & 43.0 & 33.2 & 6.5 & 19.2 & 18.8 & 11.7 & 10.2 & 43.8 & 11.2 & 89.1 & 0.8 & 1.0 & 26.8 & 23.52  \\
				CFM-RDI-TMI$_\text{CVPR23}$ \cite{CFM}&  N/A$^{\lozenge}$ & \textcolor{t2}{\textbf{88.1}} & 81.7 & 89.0 & 91.0 & 88.2 & 56.3 & \textcolor{t2}{\textbf{72.3}} & \textcolor{t2}{\textbf{72.7}} & \textcolor{t2}{\textbf{65.3}} & \textcolor{t2}{\textbf{49.4}} & \textcolor{t1}{\textbf{96.0}} & 79.6 & \textcolor{t2}{\textbf{99.7}} & 17.0 & 26.5 & 71.5 & 23.55  \\
				ODI-TMI$_\text{CVPR22}$ \cite{ODI}  & N/A$^{\lozenge}$ & 74.3 & 62.6 & 73.6 & 75.1 & 70.8 & 37.6 & 53.3 & 55.5 & 42.8 & 32.2 & 78.1 & 45.2 & 96.8 & 6.1 & 9.6 & 54.2 & 23.51 \\
				H-Aug-TMI$_\text{ArXiv23}$ \cite{wei2023rethinking}& N/A$^{\lozenge}$ & 86.3 & 69.0 & 84.2 & 82.8 & 82.5 & 42.9 & \textcolor{t2}{\textbf{72.3}} & 65.2 & 57.2 & 40.9 & 80.8 & 36.4 & 92.7 & 5.5 & 8.7 & 60.5 & 23.41 \\
				S$^4$ST-TMI$_\text{Ours}$ &  N/A$^{\quad\!\!\!}$ & \textcolor{t1}{\textbf{96.0}} & \textcolor{t1}{\textbf{86.2}} & \textcolor{t1}{\textbf{94.5}} & \textcolor{t1}{\textbf{95.0}} & \textcolor{t1}{\textbf{94.7}} & \textcolor{t1}{\textbf{69.1}} & \textcolor{t1}{\textbf{88.0}} & \textcolor{t1}{\textbf{84.9}} & \textcolor{t1}{\textbf{80.1}} & \textcolor{t1}{\textbf{66.5}} & \textcolor{t2}{\textbf{95.8}} & 68.1 & 99.5 & 18.1 & 28.3 & \textcolor{t1}{\textbf{77.7}}& 23.56   \\			
				\Xhline{1pt}
		\end{tabular}}
		\vspace{-2mm}
	\end{table*}
	
	\begin{figure*}[tbp]
		\centering
		\includegraphics[width=.95\linewidth]{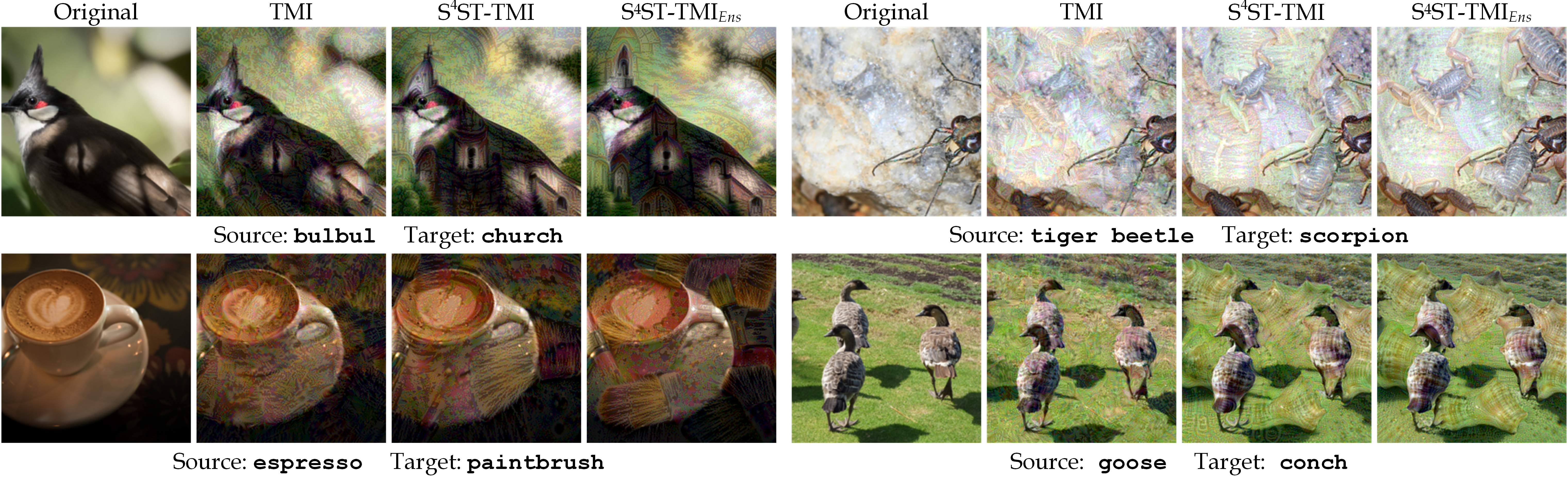}
		\vspace{-4mm}
		\caption{Targeted AEs, bolstered by S$^{4}$ST, unveil the potential mechanism for enhancing targeted transferability through image transformations---by strengthening the manipulation of objects and textures within individual images to generate semantics that align with the intended target label.}
		\label{vis}
		\vspace{-2mm}
	\end{figure*}
	
	\textbf{Quantitative Results.} S$^{4}$ST achieves a SoTA average tSuc of 77.7\%, outperforming the best data-free (CFM \cite{CFM}) and data-reliant (SASD-WS \cite{Wu_2024_CVPR}) methods by 6.2\% and 6.1\%, respectively, despite them already utilizing input transformations. Crucially, S$^{4}$ST operates strictly black-box. In contrast, SASD-WS post-trains surrogates using 1.2M ImageNet-1K samples, while CFM requires evaluated data and victim access for parameter tuning. Ultimately, our method offers unprecedented flexibility for any \textit{(sample, label)} pair without relying on victims or external data, underscoring the severe risks of strict black-box TTAs.

	\textbf{Qualitative Results.} Visualizations in Fig. \ref{vis} (and supplementary) demonstrate S$^4$ST effectively infuses target semantics into AEs, an effect amplified by ensemble surrogates. For example, a bird's crista morphs into church loft windows, and a goose flock's background adopts conch textures. Consistent with findings on unbounded \cite{logit} and data-agnostic \cite{naseer2021generating,domainfeatureuap} perturbations, this underscores the deliberate manipulation of local objects and textures. While robust transformations facilitate this semantic infusion, they inevitably expose an intrinsic tension between transferability and imperceptibility. This trade-off could inform specialized defenses by distinguishing targeted AEs from untargeted or clean images.
	
	\begin{table*}[tbp]
		\centering
		\caption{Evaluation setting and results of attacking real-world APIs.}
		\vspace{-4mm}
		\label{apis_labels}
		
		\text{(a) Target labels and semantically similar alternatives} \\ \vspace{1mm}
		\resizebox{0.9\linewidth}{!}{
			\begin{tabular}{c|ccccc}
				\Xhline{1pt}
				APIs & \texttt{Street sign} (T1) & \texttt{Goose} (T2) & \texttt{Parachute} (T3) & \texttt{Hippopotamus} (T4) & \texttt{Cannon} (T5) \\ \hline
				Google Cloud Vision & \makecell{\texttt{sign}\\\texttt{street name sign}\\\texttt{electronic signage}} &  \makecell{\texttt{canada goose}\\\texttt{ducks}} & \makecell{ \texttt{paragliding}\\\texttt{parachuting}}& \texttt{snout} & \texttt{self-propelled artillery}   \\ \hline
				Azure &  \makecell{\texttt{sign}\\\texttt{signage}} & \makecell{\texttt{geese and swans}\\\texttt{canada goose}} & \makecell{\texttt{paragliding}\\\texttt{parachuting}} & \texttt{hippo}  & \texttt{self-propelled artillery} \\  \hline
				Baidu & \cn{指路牌}, \cn{指示牌}, \cn{路标} & \makecell{\cn{鹅}, \cn{加拿大鹅},\\\cn{肉鹅}, \cn{野鹅}, \cn{天鹅}}  & \cn{降落伞}, \cn{动力伞}, \cn{滑翔伞} & \cn{河马} & \cn{火炮}   \\
				\Xhline{1pt}
		\end{tabular}}
		\\
		\vspace{2mm}

		\text{(b) Quantitative results} \\ \vspace{1mm}
		\resizebox{0.9\linewidth}{!}{
			\begin{tabular}{l|c|cccccc|cccccc|cccccc|c}
				\Xhline{1pt}
				\multirow{2}{*}{Attack} & \multirow{2}{*}{$\#$ Data} & \multicolumn{6}{c}{Google Cloud Vision} & \multicolumn{6}{c}{Azure} &\multicolumn{6}{c|}{Baidu} & \multirow{2}{*}{Avg. } \\  \cline{3-8}  \cline{9-14}  \cline{15-20} 
				&  &  T1 & T2 & T3 & T4 & T5 & Avg. &  T1 & T2 & T3 & T4 & T5  & Avg.  &  T1 & T2 & T3 & T4 & T5  & Avg.  &  \\ \hline
				
				SASD-WS-DI-TMI \cite{Wu_2024_CVPR} & 1.2M & \textcolor{t2}{\textbf{2}} & \textcolor{t2}{\textbf{68}} & \textcolor{t2}{\textbf{65}} & \textcolor{t2}{\textbf{63}} & \textcolor{t2}{\textbf{57}} & \textcolor{t2}{\textbf{51.0}} & \textcolor{t2}{\textbf{39}} & 65 & \textcolor{t1}{\textbf{62}} & \textcolor{t1}{\textbf{94}} & \textcolor{t2}{\textbf{65}} & \textcolor{t2}{\textbf{65.0}} & \textcolor{t1}{\textbf{79}} & \textcolor{t1}{\textbf{88}} & \textcolor{t2}{\textbf{83}} & 96 & \textcolor{t1}{\textbf{98}} & \textcolor{t1}{\textbf{89.0}}  & \textcolor{t2}{\textbf{68.3}}  \\			
				CFM-RDI-TMI \cite{CFM} & N/A & \textcolor{t2}{\textbf{2}} & 55 & 38 & 55 & 46 & 39.0 & 26 & \textcolor{t2}{\textbf{69}} & 50 & 84 & 64 & 58.6 & 70 & \textcolor{t2}{\textbf{86}} &  79 & \textcolor{t2}{\textbf{97}} & 93 & 85.0  &  60.9 \\			
				
				S$^4$ST-TMI & N/A & \textcolor{t1}{\textbf{7}} & \textcolor{t1}{\textbf{69}} & \textcolor{t1}{\textbf{71}} & \textcolor{t1}{\textbf{66}} & \textcolor{t1}{\textbf{62}} & \textcolor{t1}{\textbf{55.0}} & \textcolor{t1}{\textbf{52}} & \textcolor{t1}{\textbf{70}} & \textcolor{t2}{\textbf{56}} & \textcolor{t2}{\textbf{88}} & \textcolor{t1}{\textbf{68}} & \textcolor{t1}{\textbf{66.8}} & \textcolor{t2}{\textbf{71}} & 80 & \textcolor{t1}{\textbf{89}} & \textcolor{t1}{\textbf{99}}  & \textcolor{t2}{\textbf{97}}  & \textcolor{t2}{\textbf{87.0}} & \textcolor{t1}{\textbf{69.6}}   \\			\hline
				S$^4$ST-TMI$_{\textit{Ens}}$ & N/A & 54 & 96 & 95 & 88 & 95 & 86.0 & 89 & 91 & 89 & 98 & 98 & 93.0 & 86 & 96 & 99 & 99  & 100 & 96.0 & 91.7 \\					
				\Xhline{1pt}
			\end{tabular}
		}
		\vspace{-2mm}
	\end{table*}
	
	\subsection{In-domain Task Transferability}
	
	\textcolor{black}{We evaluate S$^4$ST's transferability across diverse natural image scenarios to validate its dataset- and task-agnostic nature. By directly transferring AEs crafted on an ImageNet-trained ResNet-50 to commercial APIs, VLMs, and dense prediction models without task-specific adaptation, we demonstrate S$^4$ST captures universal semantic features that transcend varying architectures, training distributions, and learning paradigms.}
	
	\subsubsection{Transferring Against Real-World API}
	
	\textcolor{black}{Following \cite{zhao2023minimizing}, we attack Google Cloud Vision\footnote{\url{https://cloud.google.com/vision/docs/labels}}, Azure\footnote{\url{https://portal.vision.cognitive.azure.com/demo/generic-image-tagging}}, and Baidu\footnote{\url{https://cloud.baidu.com/product/imagerecognition/general}} APIs using 5 target labels across 100 images (500 AEs per API). An attack succeeds if the API's returned labels include the exact target or pre-defined semantic equivalents (Table \ref{apis_labels}). As a negative control, we evaluate the success rate using clean images under the same protocol for Baidu; the baseline success rate is $0\%$, confirming that the observed tSuc is explicitly driven by our adversarial perturbations.} 
	
\begin{figure*}[tbp]
		\includegraphics[width=1\linewidth]{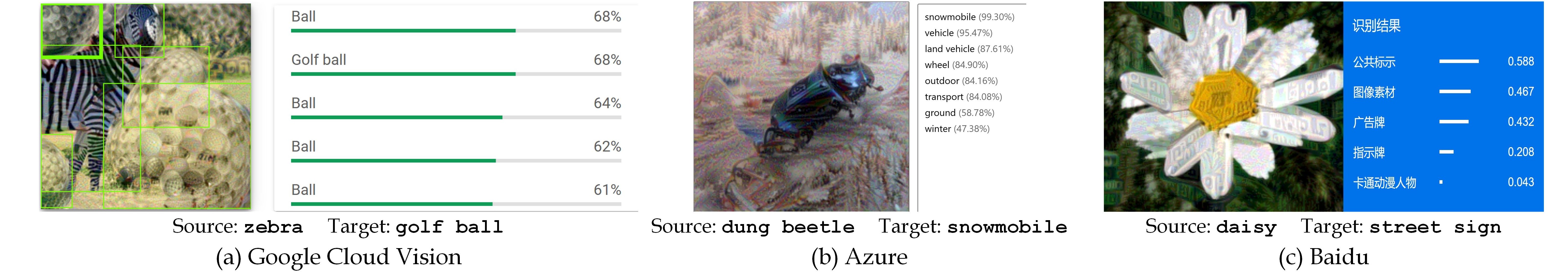}
		\vspace{-8mm}
		\caption{Real-world APIs-returned label list for targeted AEs generated by S$^{4}$ST-TMI.}
		\label{apiins}
		\vspace{-3mm}
	\end{figure*}

	Using a single ResNet-50, S$^{4}$ST achieves 69.6\% average tSuc, outperforming the data-reliant SoTA SASD-WS \cite{Wu_2024_CVPR} by 1.3\%. Employing surrogate ensembles further elevates tSuc to 91.7\%. Fig. \ref{apiins} provides several examples obtained from these APIs, demonstrating that targeted AEs can induce high-confidence erroneous predictions toward the target class. These results demonstrate that our S$^{4}$ST-empowered attacks hold significant potential for evaluating model vulnerabilities under unknown training distributions and data processing pipelines.
	
	\begin{table*}[htbp]
		\caption{Results of attacking VLMs.}
		\vspace{-4mm}
		\label{vlm}
		\centering
		\resizebox{1.\linewidth}{!}{
			\begin{tabular}{l|c|ccccc|ccccc|ccccc|ccccc|c}
				\Xhline{1pt}
				\multirow{2}{*}{Attack} & \multirow{2}{*}{$\#$ Data} & \multicolumn{5}{c|}{GLM-4V Flash \cite{hong2024cogvlm2}} & \multicolumn{5}{c|}{Gemini 1.5 Flash \cite{team2024gemini}} &\multicolumn{5}{c|}{Janus Pro 7B \cite{chen2025janus}} & \multicolumn{5}{c|}{\textcolor{black}{Qwen3-VL 8B} \cite{qwen3technicalreport}} &  \multirow{2}{*}{Avg.} \\  \cline{3-22}
				& & T1 & T2 & T3 & T4 & T5 & T1 & T2 & T3 & T4 & T5 & T1 & T2 & T3 & T4 & T5 & T1 & T2 & T3 & T4 & T5 &  \\ \hline
				
				SASD-WS-DI-TMI \cite{Wu_2024_CVPR} & 1.2M & \textcolor{t1}{\textbf{51}} & \textcolor{t1}{\textbf{74}} & \textcolor{t1}{\textbf{71}} & \textcolor{t2}{\textbf{80}} & \textcolor{t2}{\textbf{72}} & \textcolor{t1}{\textbf{33}} & \textcolor{t2}{\textbf{34}} & \textcolor{t1}{\textbf{36}} & \textcolor{t1}{\textbf{57}} & \textcolor{t1}{\textbf{50}} & \textcolor{t1}{\textbf{47}} & \textcolor{t2}{\textbf{56}} & \textcolor{t1}{\textbf{63}} & \textcolor{t1}{\textbf{74}} & \textcolor{t1}{\textbf{78}} & \textcolor{t1}{\textbf{43}} & \textcolor{t1}{\textbf{78}} & \textcolor{t1}{\textbf{69}} & \textcolor{t1}{\textbf{86}} & \textcolor{t1}{\textbf{75}} & \textcolor{t1}{\textbf{61.4}} \\			
				CFM-RDI-TMI \cite{CFM} & N/A & 28 & 66 & \textcolor{t2}{\textbf{64}} & 67 & 59 & 14 & 29 & 17 & 37 & 38 & 20 & 47 & 39 & 57 & 59 & \textcolor{t2}{\textbf{21}} & 65 & 49 & 79 & 65 & 46.0 \\			
				
				S$^4$ST-TMI  & N/A & \textcolor{t2}{\textbf{41}} & \textcolor{t2}{\textbf{72}} & \textcolor{t1}{\textbf{71}} & \textcolor{t1}{\textbf{81}} & \textcolor{t1}{\textbf{76}} & \textcolor{t2}{\textbf{29}} & \textcolor{t1}{\textbf{41}} & \textcolor{t2}{\textbf{21}} & \textcolor{t2}{\textbf{54}} & \textcolor{t2}{\textbf{44}} & \textcolor{t2}{\textbf{42}} & \textcolor{t1}{\textbf{62}} & \textcolor{t2}{\textbf{53}} & \textcolor{t2}{\textbf{72}} & \textcolor{t2}{\textbf{70}} & \textcolor{t1}{\textbf{43}} & \textcolor{t2}{\textbf{73}} & \textcolor{t2}{\textbf{62}} & \textcolor{t2}{\textbf{85}} & \textcolor{t2}{\textbf{72}} & \textcolor{t2}{\textbf{58.2}} \\			\hline
				S$^4$ST-TMI$_{\textit{Ens}}$ & N/A & 81 & 93 & 87 & 93 & 95 & 58 & 68 & 55 & 83 & 85 & 81 & 90 & 83 & 89 & 95 & 76 & 91 & 88 & 93 & 92 & 83.8 \\					
				\Xhline{1pt}
			\end{tabular}
		}
		\vspace{-2mm}
	\end{table*}
	
	\begin{figure*}[tbp]
		\includegraphics[width=1\linewidth]{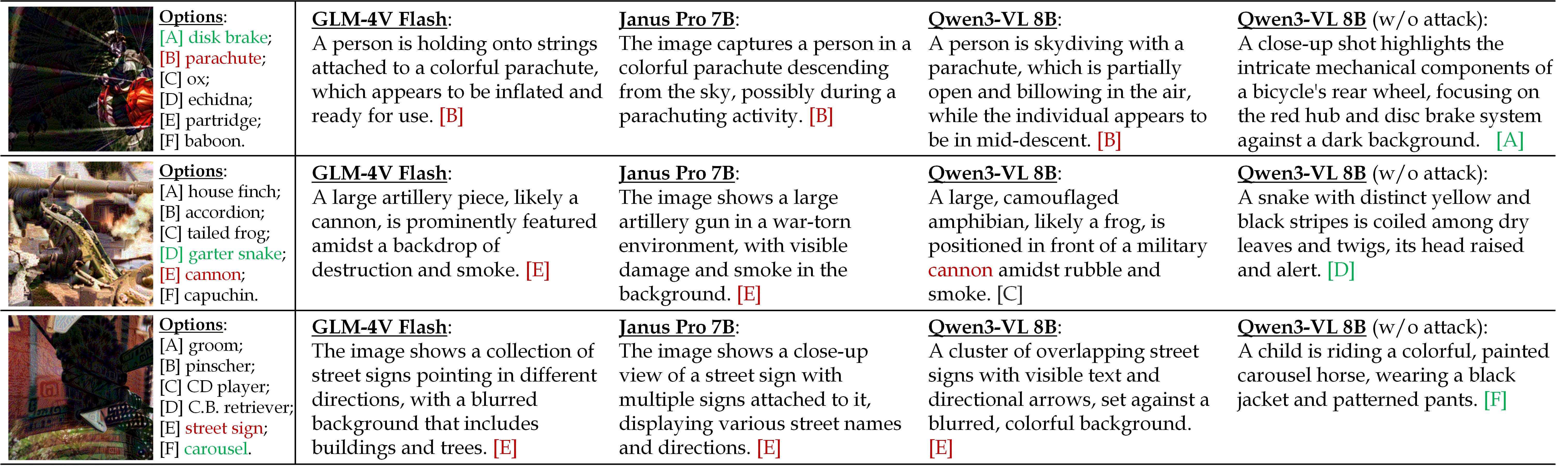}
		\vspace{-8mm}
		\caption{Responses obtained from VLMs for targeted AEs generated by S$^{4}$ST-TMI. Source labels are in \textcolor{s1}{green} and target labels are in \textcolor{s2}{red}.}
		\label{vlmins}
		\vspace{-3mm}
	\end{figure*}

	\subsubsection{Evaluation on VLMs}	
	
	\textcolor{black}{To further validate the scalability of our method to VLMs, we maintain the evaluation setting used for commercial APIs and design a structured prompt to assess attack effectiveness in an open-vocabulary scenario. Specifically, for each AE, we provide the model with the following standardized instruction:
		\begin{quote}\textit{“You are an image analysis expert. Follow these steps strictly: 1. \textbf{Description}: Concisely describe the main object(s) in the picture in 1-2 sentences. 2. \textbf{Selection}: Choose ONLY ONE best match option from the list below marked with [ ]. Do NOT explain your choice. Options: \{opt\}. Output format: Separate description and selection with '---'; Use '[ ]' to mark your final choice.”}
		\end{quote}
		The options $\{opt\}$ consist of six shuffled candidates: the source label, the target label, and four random ImageNet-1K labels. A successful attack forces the VLM to select the target label. As a negative control, we use the same prompt on clean images for Qwen3-VL 8B \cite{qwen3technicalreport}, where the baseline tSuc of selecting the target label is $0\%$ (accuracy of selecting the source label is $99\%$), confirming perturbations drive the deception.}
	
	Tested across GLM-4V Flash \cite{hong2024cogvlm2}, Gemini 1.5 Flash \cite{team2024gemini}, Janus Pro 7B \cite{chen2025janus}, and Qwen-VL 8B \cite{qwen3technicalreport} (Table \ref{vlm}), S$^{4}$ST achieves a competitive 58.2\% average tSuc. While marginally lower (3.2\%) than SASD-WS, S$^{4}$ST rigorously maintains data-free constraints, demonstrating potent textual-visual manipulation without external data dependencies (Fig. \ref{vlmins}).
	
	\subsubsection{\textcolor{black}{Evaluation on Dense Prediction Tasks}}
	
	\begin{table}[tbp]
		\vspace{-2mm}
		\caption{\textcolor{black}{Results of transferring ImageNet surrogate (ResNet-50, target label: \texttt{Model-T}) agasint COCO2017 victims (corresponding label: \texttt{car}).}}
		\vspace{-4mm}
		\label{det_seg}
		\centering
		\resizebox{1\linewidth}{!}{
			\begin{tabular}{l|cc|cc|cc}
				\Xhline{1pt}
				Victims$\rightarrow$ & \multicolumn{2}{c|}{\makecell[c]{M-RCNN\\(ResNet-101)}} & \multicolumn{2}{c|}{\makecell[c]{Cascade M-RCNN\\(CNX-S)}} & \multicolumn{2}{c}{\makecell[c]{Mask2Former\\(Swin-S)}} \\ \hline
				Attack$\downarrow$ & tSuc$_{\text{Det}}$ & tSuc$_{\text{Seg}}$ & tSuc$_{\text{Det}}$ & tSuc$_{\text{Seg}}$ & tSuc$_{\text{Det}}$ & tSuc$_{\text{Seg}}$  \\ \hline
				w/o attack & 10.2 & 3.0  & 9.1 & 2.6 & 8.8  & 2.8  \\ 
				SASD-WS &  \textcolor{t2}{\textbf{19.6 (9.4$\uparrow$)}} & \textcolor{t2}{\textbf{5.7 (2.7$\uparrow$)}}  & \textcolor{t2}{\textbf{44.0 (34.9$\uparrow$)}} & 24.1 (21.5$\uparrow$) & \textcolor{t2}{\textbf{48.0 (39.2$\uparrow$)}} & \textcolor{t2}{\textbf{26.1 (23.3$\uparrow$)}} \\ 
				CFM & 19.1 (8.9$\uparrow$) & 5.6 (2.6$\uparrow$) & 43.2 (34.1$\uparrow$) & \textcolor{t2}{\textbf{25.0 (22.4$\uparrow$)}} & 41.1 (32.3$\uparrow$) & 19.1 (16.3$\uparrow$) \\ 
				S$^4$ST & \textcolor{t1}{\textbf{37.5 (27.3$\uparrow$)}} & \textcolor{t1}{\textbf{10.7 (7.7$\uparrow$)}}  & \textcolor{t1}{\textbf{59.5 (50.4$\uparrow$)}} & \textcolor{t1}{\textbf{26.9 (24.3$\uparrow$)}} & \textcolor{t1}{\textbf{65.3 (56.5$\uparrow$)}} & \textcolor{t1}{\textbf{28.6 (25.8$\uparrow$)}} \\ 
				\Xhline{1pt}
			\end{tabular}
		}
		\vspace{-3mm}
	\end{table}
	
	\textcolor{black}{Instance segmentation demands precise spatial localization and pixel-level semantic discrimination. We transfer 1,000 AEs targeting \texttt{Model-T} (corresponding to \texttt{car} category of COCO \cite{lin2014microsoft}) to diverse architectures and detection paradigms trained on COCO: Mask R-CNN \cite{he2017mask} (ResNet-101), Cascade Mask R-CNN \cite{cai2019cascade} (CNX-S), and Mask2Former \cite{cheng2022masked} (Swin-S) (Table \ref{det_seg}). Metrics include $\text{tSuc}_{\text{Det}}$ (\% of images detecting $\geq$1 \texttt{car}) and $\text{tSuc}_{\text{Seg}}$ (mean \texttt{car} mask area ratio).}
	
\begin{figure}[tbp]
		\vspace{-2mm}
		\includegraphics[width=1\linewidth]{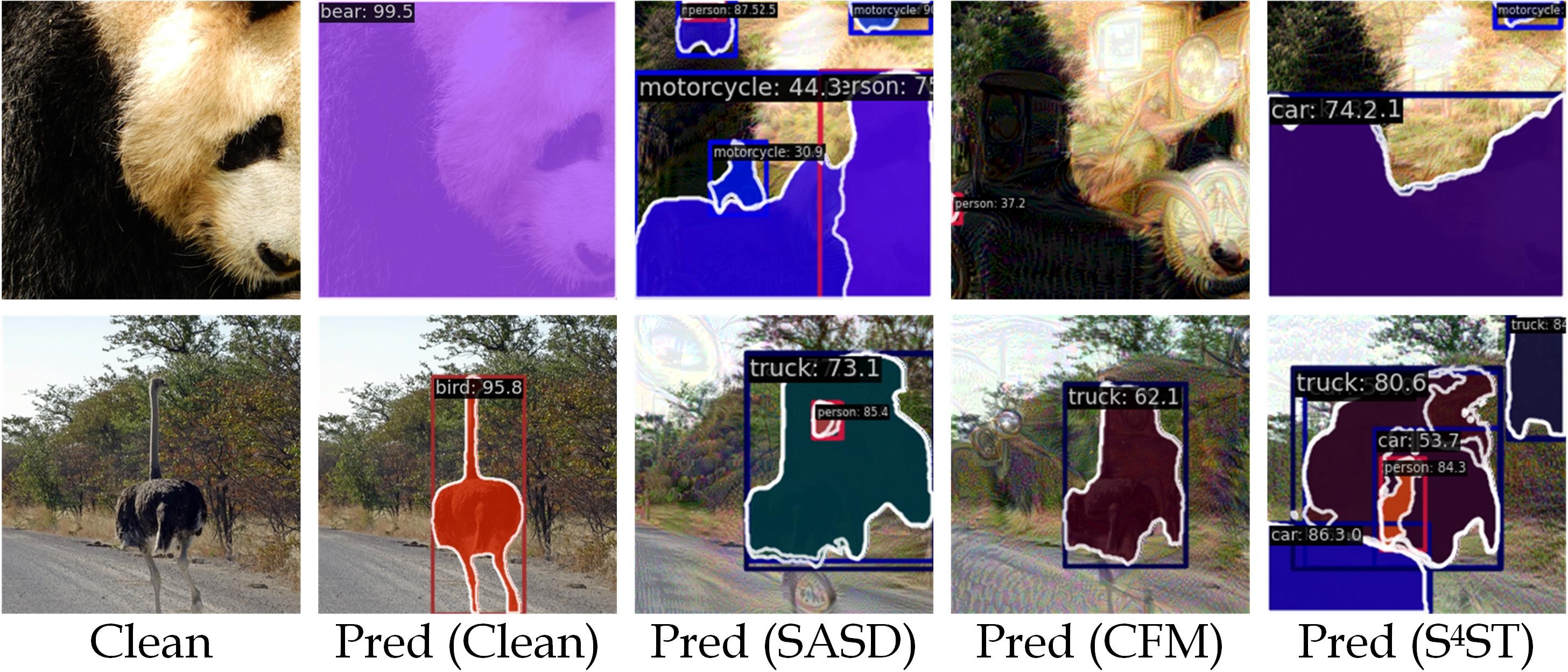}
		\vspace{-8mm}
		\caption{\textcolor{black}{Inference results by Mask2Former (Swin-S).}}
		\label{det_vis}
		\vspace{-3mm}
	\end{figure}

	\textcolor{black}{S$^4$ST exhibits remarkable cross-task transferability. On Mask2Former, it achieves 65.3\% $\text{tSuc}_{\text{Det}}$ and 28.6\% $\text{tSuc}_{\text{Seg}}$, significantly outperforming SASD-WS and CFM. Such high mask area ratios signify successful pixel-level contour manipulation (Fig. \ref{det_vis}). Consistent gains across diverse backbones and detection heads confirm S$^4$ST effectively exploits fundamental representation convergence within natural images, rather than merely overfitting to global semantics.}

	\subsection{Cross-domain Adaptability (Boundary Cases)\label{beyondnatural}}
	
	\textcolor{black}{To test our framework's boundaries, we investigate cross-domain adaptability on multi-label thoracic disease classification and face verification. Unlike natural images, these domains feature fixed fields-of-view and standardized object scales. Evaluating them reveals whether our scaling-centric transformations and analytic measures generalize to structurally distinct data distributions. Detailed analyses are in the supplementary, with top-performing methods summarized in Fig.~\ref{cxr_fv}.}
	
	\subsubsection{\textcolor{black}{Multi-Label Thoracic Disease Classification}}
	
	Evaluations regarding multi-label thoracic disease classification are conducted on the NIH CXR14 Dataset \cite{wang2017chestx}. It contains over 100,000 chest X-ray (CXR) images, annotated with labels for 14 different thoracic conditions in a multi-label classification setting. We evaluate the \texttt{Cardiomegaly} label using 500 random test images (33 positive, 467 negative).
	
	The surrogate model is an ImageNet-pretrained DenseNet-121 (DN121-IN). The victim models are \{DN121-CXR, RN50-IN, ViT-S-CXR, ViT-B-CXR\}, where -CXR denotes masked image modeling pretraining using CXR images \cite{xiao2023delving}. This selection reflects variations in both training data distributions and model architecture. We generate AEs on negative images via Logit loss \cite{logit} and BIM \cite{kurakin2016adversarial} ($\epsilon=8/255, \alpha=1/255, T=100$), and then evaluate them alongside clean positive samples, and the AUC metric is computed for the black-box model. A lower AUC value indicates stronger transferability of the attack method. 
	
\begin{figure*}[tbp]
		\centering
		\includegraphics[width=\linewidth]{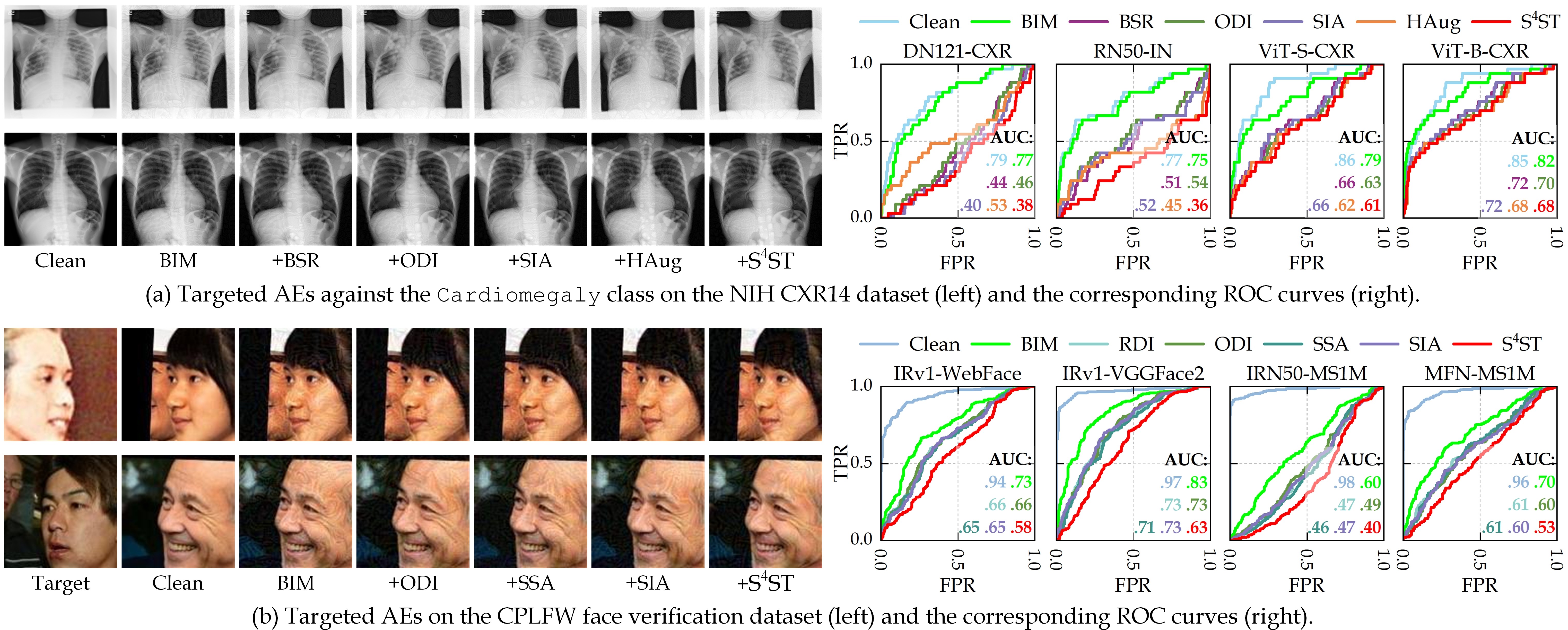}
		\vspace{-8mm}
		\caption{\textcolor{black}{Visualization of targeted AEs and corresponding attack results on (a) NIH CXR14 thoracic disease classification and (b) CPLFW face verification.}}
		\label{cxr_fv}
		\vspace{-5mm}
	\end{figure*}

	For this task, the self-alignment analysis reveals that scaling transformations remain competitively effective. Meanwhile, the correlation analysis based on self-transferability yields conclusions consistent with those observed in natural image classification tasks. Therefore, we directly apply the aforementioned S$^{4}$ST parameters without further modification. As shown in Fig.~\ref{cxr_fv}(a), S$^{4}$ST significantly degrades the performance of classifiers with different pre-training data and network architectures, achieving superior performance.
	
	\subsubsection{\textcolor{black}{Face Verification}\label{faceveri}}
	
	We evaluate the face verification task using the CPLFW \cite{CPLFWTech} dataset, which contains 12,000 aligned face images with diverse poses and is widely adopted in related works. We select 600 image pairs (300 positive, 300 negative) for evaluation.
	
	The surrogate model is an Improved ResNet-101 trained with CurricularFace loss \cite{huang2020curricularface} on the MS1M \cite{deng2019arcface} dataset (CurricularFace-IRN101-MS1M). The victim models consist of \{FaceNet-IRv1-WebFace \cite{schroff2015facenet}, FaceNet-IRv1-VGGFace2, ArcFace-IRN50-MS1M \cite{deng2019arcface}, and MobileFaceNet (MFN)-MS1M \cite{deng2019arcface,chen2018mobilefacenets}\}. We attack one image per negative pair to maximize feature cosine similarity using BIM \cite{kurakin2016adversarial} ($\epsilon=16/255, \alpha=2/255, T=100$). The attacked negative pairs are then evaluated together with all positive clean pairs to obtain the ROC curves and corresponding AUC values of the victim models. Applying S$^{4}$ST-\textit{Base} ($p_{\textit{r}}=1.0, r=1.4, p_{\textit{aug}}=0, m=1$), our method demonstrates a clear performance gap, severely degrading victim AUCs (Fig.~\ref{cxr_fv}b).

	\textcolor{black}{\textbf{Adaptation Strategy}. In this case, the underlying task transitions from label-based classification to metric learning dominated by feature space distances, involving significantly different image distributions and data augmentation strategies \cite{deng2019arcface}. Consequently, it becomes necessary to re-evaluate the overall transformation framework using our proposed self-alignment and self-transferability measures before conducting the BO search. Specifically, we observe a narrower tolerance for transformation intensity, along with strong correlations between geometric and color transformations. This likely stems from the limited scale diversity in fixed region of face images and the common practice of using minimal data augmentations. Our analysis reveals that directly applying the BO strategy for natural images (averaging self-transferability over 12 basic transformations) yields a weak PCC of 0.37 between the proxy and actual black-box transferability. However, by leveraging our self-transferability observations to simplify the framework, specifically by removing the redundant S$^4$ST-\textit{Block} and S$^4$ST-\textit{Aug} to focus solely on S$^4$ST-\textit{Base}, the PCC rises to 0.78. Further, by substitute the proxy by the self-transferability against scaling (findings provided in Fig. \ref{lineartest}), the PCC reaches 0.97. This ensures that S$^4$ST remains a principled and effective framework even under severe domain shifts. We also provide a streamlined adaptation strategy for such boundary cases. Please see supplementary.}
	
	\section{Discussion: Why Scaling Excels?\label{analysis_why_scaling}}\label{discussion}
	
	\textcolor{black}{In this section, we analyze why scaling transformations are distinctively effective in enhancing targeted transferability.}
	
	\subsection{\textcolor{black}{Why Transformations Benefit Transferability}} 
	
	As Section \ref{model_alignment_intro} establishes, the core function of transformations in gradient-based attacks is to restore the surrogate model's feature distribution toward that of clean samples. But \textit{why do transformed samples maintain feature alignment?}
	
	\textbf{The key is data variety and training-time augmentation}. Neither CNNs nor ViTs possess inherent scale invariance; their transformation tolerance stems from the multi-scale nature of visual data and inductive biases learned via extensive augmentations. First, since natural images inherently contain objects at various scales, orientations, and perspectives, models must develop some degree of transformation tolerance to achieve generalizability. Second, training-time augmentations, exemplified by a scaling variation,  RandomResizedCrop\footnote{\url{https://github.com/pytorch/vision/blob/main/torchvision/transforms/transforms.py\#L855}} (RRC), serve as a critical component across a wide spectrum of visual training paradigms, ranging from supervised to self-supervised learning (\eg MAE \cite{he2022masked}, DINO \cite{caron2021emerging}), from vision-only to vision-language tasks (\eg CLIP \cite{radford2021learning}), and from classification to segmentation and detection, making it the de facto \textit{universal key} to generalization capability in visual representation learning.
	
	The shared data variety and dependency on a universal training component, designed to capture essential multi-scale data properties, rather than architectural priors, explains why scaling uniquely acts as a model-agnostic \textit{universal key} for transferable AEs, a vulnerability S$^{4}$ST explicitly exploits.
	
	\begin{table}[tbp]
		\caption{\textcolor{black}{Accuracy of RN50 variants.}}
		\vspace{-4mm}
		\label{rn50_variants_acc}
		\centering
		\resizebox{1\linewidth}{!}{
			\begin{tabular}{l|ccc}
				\Xhline{1pt}
				Model & RN50-\textit{stdRRC} & RN50-\textit{halfRRC}  & RN50-\textit{woRRC}  \\ \hline
				ImageNet val. set & 76.1 & 75.2 & 69.0 \\
				ImageNet Compatible & 93.2 & 90.6  & 79.3 \\
				\Xhline{1pt}
			\end{tabular}
		}
		\vspace{-5mm}
	\end{table}
	
	\subsection{\textcolor{black}{Experimental Validation}}
	
	\textcolor{black}{Controlled experiments are conducted with three RN50 variants to validate our hypothesis:}
	\begin{enumerate}
		\item \textcolor{black}{RN50-\textit{stdRRC}: official weights (v1) released by \textit{torchvision}, trained with standard RRC augmentation (\texttt{scale=(0.08,1.)}).}
		\item \textcolor{black}{RN50-\textit{halfRRC}: the scaling intensity used is half that of the standard RRC (\texttt{scale=(0.41,1.)}).}
		\item \textcolor{black}{RN50-\textit{woRRC}: trained without RRC but the standard resize and center crop pipeline.}
	\end{enumerate}
	\textcolor{black}{Their corresponding accuracies are reported in Table~\ref{rn50_variants_acc}. As expected, the model trained without RRC augmentation exhibits a significant drop in accuracy.}
	
\begin{figure*}[tbp]
		\centering
		\includegraphics[width=\linewidth]{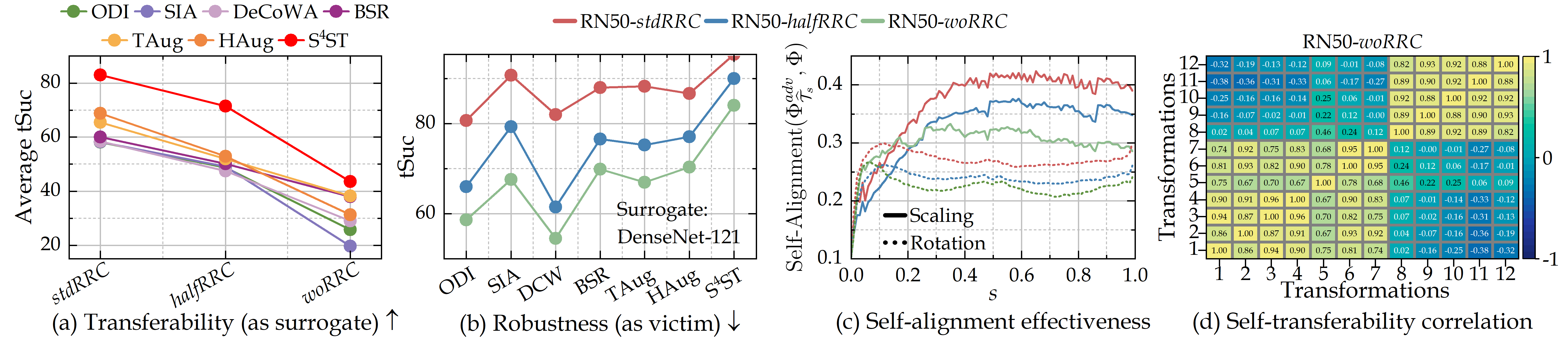}
		\vspace{-8mm}
		\caption{
			\textcolor{black}{While stronger RRC augmentation enhances attack transferability and accuracy (a) at the cost of model robustness (b), models trained entirely without it still exhibit transformation tolerance (c) and the same correlation effects (d). This demonstrates that while training-time augmentation is a powerful modulator of adversarial properties, the fundamental basis is imparted by the intrinsic diversity of the data itself.
		}}
		\label{rn50_variants_fig}
		\vspace{-5mm}
	\end{figure*}

	\textcolor{black}{We evaluate the transferability of these models when used as surrogates and their robustness when attacked as victims. Fig.~\ref{rn50_variants_fig}(a) shows that the effectiveness of transformation-based attacks diminishes as the intensity of RRC used in training surrogate decreases. Meanwhile, Fig.~\ref{rn50_variants_fig}(b) exhibits that model robustness improves with weaker RRC intensity. These significant variance validates the crucial impact of RRC augmentation during training.}
	
	Notably, even without RRC (RN50-\textit{woRRC}), scaling-related transformations like S$^{4}$ST and H-Aug still yield the best transferability as attackers. Furthermore, when RN50-\textit{woRRC} acts as a victim, it remains significantly more vulnerable to these scaling-based attacks than to other transformations. Additionally, Fig.~\ref{rn50_variants_fig}(c) and (d) demonstrate that even without applying RRC during training, transformations can still help capture model alignment during attack, and strong internal correlations persist among geometric and color transformations. This indicates that the inherent diversity in the data, such as multi-scale, multi-orientation, and multi-perspective, plays a fundamental role in shaping the transformation-related adversarial properties.
	
	\subsection{\textcolor{black}{Implications and Discussion}}
	
	\textcolor{black}{\textbf{Why scaling excels}. Our findings suggest that \textit{both the intrinsic diversity of the data itself and the use of training-time RRC augmentation collectively contribute to the efficacy of transformation-based TTAs, or more intrinsically, the transferability of AEs}. And, the contribution of scale diversity appears to be significant than that of others.} Scaling's distinctive effectiveness originates from its direct correspondence to this data nature and coupled universal training practices established over decades of vision model development. This explains why prior attacks relying on arbitrary or partial aspects of data variety achieved limited success compared to S$^{4}$ST.
	
	\textcolor{black}{\textbf{The Dilemma of Generalization-Robustness Tradeoff}.  
		It also presents a dilemma analogous to that observed in adversarial training \cite{madry2018towards} and pre-processing defense \cite{guo2018jpeg}, the generalization-robustness tradeoff. Specifically, models trained without RRC would likely suffer from compromised generalization capability while exhibiting stronger resistance to transfer attacks. Conversely, models trained with RRC achieve higher accuracy at the cost of increased vulnerability. This dual nature establishes data augmentation as a double-edged sword in visual representation learning. It connects adversarial vulnerability with broader robustness considerations and suggests that future secure model development must explicitly address this tradeoff.}
	
	\section{Conclusion}\label{conclusion}
	
	We proposed two blind estimators, self-alignment and self-transferability, to analyze basic transformations' effectiveness and synergies under strict black-box constraints. These insights drove the design of S$^{4}$ST, a scaling-centered transformation that significantly boosts data-free TTA transferability without external data or model training. S$^{4}$ST excels in efficiency and black-box adherence, outperforming data-reliant SoTA TTAs (0 \textit{vs.} 1.2M training samples) across CNNs, ViTs, robust models, commercial APIs, and VLMs.
	
	Furthermore, we revealed that inherent visual data diversity fundamentally underpins transformation-enhanced transferability, an effect amplified by universally adopted training-time scale augmentations. This exposes a critical generalization-robustness tradeoff in data augmentation practices. Ultimately, this work highlights the untapped potential of data-free TTAs and the severe threat black-box targeted attacks pose to real-world systems.
	
	\appendices
	
	\section{Gradient Vanishing of Data-free TTAs}
	
	While effective in untargeted attacks, the cross-entropy loss faces challenges in TTAs due to gradient vanishing issues \cite{PoTrip,logit}. Let $\bm{\phi}$ denote the representation produced by $f$, such as features preceding the final linear layer. The logit values are computed as $l_i=w_i^\top\bm{\phi}+b_i$, with $w_i$ and $b_i$ being the class-specific weight and bias, respectively. The normalized probability with respect to each category is given by the SoftMax function: $p_i=\frac{e^{l_i}}{\sum e^{l_j}}$. The gradient
	\begin{equation}
		\label{vanishing}
		\centering
		\frac{\partial \mathcal{L}_{\text{CE}}}{\partial \bm{x}} = \frac{\partial \mathcal{L}_{\text{CE}}}{\partial \bm{\phi}}\frac{\partial \bm{\phi}}{\partial \bm{x}} =  [\left(p_t-1\right) w_t^\top + \sum_{i\neq t} p_i w_i^\top] \frac{\partial \bm{\phi}}{\partial \bm{x}}
	\end{equation}
	tends to vanish as the model becomes confident ($p_{i\neq t}\rightarrow0$). To address this, Li \etal introduced a Poincaré distance-based triplet \cite{PoTrip} loss to facilitate effective optimization. The Logit loss \cite{logit} proposed by Zhao \textit{et al.} directly maximizes the target logit and provides a direct solution to the gradient vanishing dilemma. Subsequent studies have explored ways to soften $\mathcal{L}_{\text{CE}}$ to stabilize the optimization process, such as through margin-based logit calibration \cite{marginangleT} or an adversarial optimization scheme \cite{AOSBLL}.
	
	\section{Input Transformations for Transferability}
	
	Image transformation is an effective and necessary strategy in training deep vision models, effectively mitigating the need for extremely large quantities of data and annotations \cite{shorten2019survey,mikolajczyk2018data,taylor2018improving}. For example, horizontal flips and random resized cropping are commonly used in natural image understanding tasks, significantly boosting generalization performance. Furthermore, transformation strategies are typically data-dependent and can be conveniently implemented in a plug-and-play manner across different model architectures. The same benefits extend to the domain of adversarial attacks. There are two primary streams of transformation strategies for enhancing adversarial transferability: \textit{1)} Dynamically searching for optimal combinations from a vast pool of transformations during the attack iteration \cite{yuan2021automa,yuan2022adaptive,Zhu_2024_CVPR}; and \textit{2)} Consistently applying a predefined transformation in each iteration. In this paper, we focus exclusively on the latter to avoid the need to access black-box models for optimization. In addition to TTA-oriented transformation methods \cite{wei2023rethinking,ODI}, we also examine several prominent works in the field of untargeted attacks. This is to validate whether the prevalent consensuses in transformation design are applicable in the context of TTA. An overview of the methods involved is presented in Fig. 4 (in the main file), introduced below.
	
	Xie \etal first introduced the DI method \cite{inputdiversity2019xie}, which involves random resizing and padding, and has shown a significant boost in transferability. The Resized DI (RDI) approach \cite{RDI} rescales the DI-transformed image back to the original size, creating a reduced-scale version. The Scale-Invariant (SI) technique \cite{nestrov2019lin} computes the average gradient with respect to images scaled linearly by powers of two, repeated five times. Admix \cite{admix} blends the original image with those from other classes. The Spectrum Simulation Attack (SSA) \cite{SSA} introduces noise within the frequency domain. The Structure Invariant Attack (SIA) \cite{SIA} and Block Shuffle \& Rotation (BSR) \cite{BSR} employ random transformations and shuffling \& rotation operations on image blocks, respectively. The Deformation-Constrained Warping (DeCoWa) method \cite{decowa} applies elastic deformation coupled with an internal min-max optimization step.

	\section{Quantification of Existing Consensuses}
	The consensuses under investigation and their corresponding measures are as follows:
	\begin{itemize}
		\item \textbf{Diversity}. Initially, researchers emphasized the need for transformations that preserve loss \cite{admix,inputdiversity2019xie,nestrov2019lin,evading2019dong}, ensuring that the surrogate model's output remains relatively constant post-transformation. The intensity of transformations was thus limited to a narrow scope to maintain this characteristic. However, more recent advancements have shifted towards utilizing gradient averaging over multiple transformed samples (20$\times$) to achieve stable gradient updates \cite{BSR,SIA,decowa,SSA}. This shift implies that increased diversity might lead to improved transferability. We quantify \textbf{\textit{diversity}} by calculating the average cross-entropy loss values of the surrogate model across 50,000 images from the ImageNet-1K validation set.
		\item \textbf{Attention Deviation}. Dong \etal first pointed out that discrepancies in discriminative regions between models and defensive strategies can hinder adversarial transferability \cite{evading2019dong}. They advocated for the use of translation transformations to extend the impact of adversarial perturbations to wider attention regions. Following this, other studies have also utilized attention as a universal objective for crafting transferable attacks \cite{chen2020universal,wang2021feature}. More recently, \cite{BSR,wei2023rethinking,10251606,11018095} have set the new SoTAs for untargeted and targeted attacks by disrupting attention through image block shuffling and random attention-deviation transformations, respectively. We adopt the \textit{\textbf{attention deviation}} metric proposed by Wei \etal \cite{wei2023rethinking}, which computes the Intersection over Union (IoU) of Grad-CAM \cite{selvaraju2017grad} activations concerning the target label between original and transformed images. The results are averaged over 1,000 images with distinct target labels.
		\item \textbf{Gradient Magnitude}. A significant portion of research on data-free TTAs has been devoted to addressing the issue of gradient vanishing, particularly in studies focusing on loss functions \cite{logit,SelfU,AOSBLL,marginangleT}. The magnitude of gradients has often been used as an indicator to gauge the efficacy of different approaches. It is hypothesized that larger gradient magnitudes could lead to more efficient updates of perturbations. To test this hypothesis, we examine the gradient magnitudes produced by various transformations under an identical loss function. Specifically, the \textit{\textbf{gradient magnitude}} is measured by the mean $\ell_{2}$-norm of gradients over 900 $\mathcal{T}$-TMI iterations and is averaged across 1,000 images with distinct target labels.
	\end{itemize}

	\begin{figure*}[tbp]
		\centering	
		\includegraphics[width=1\linewidth]{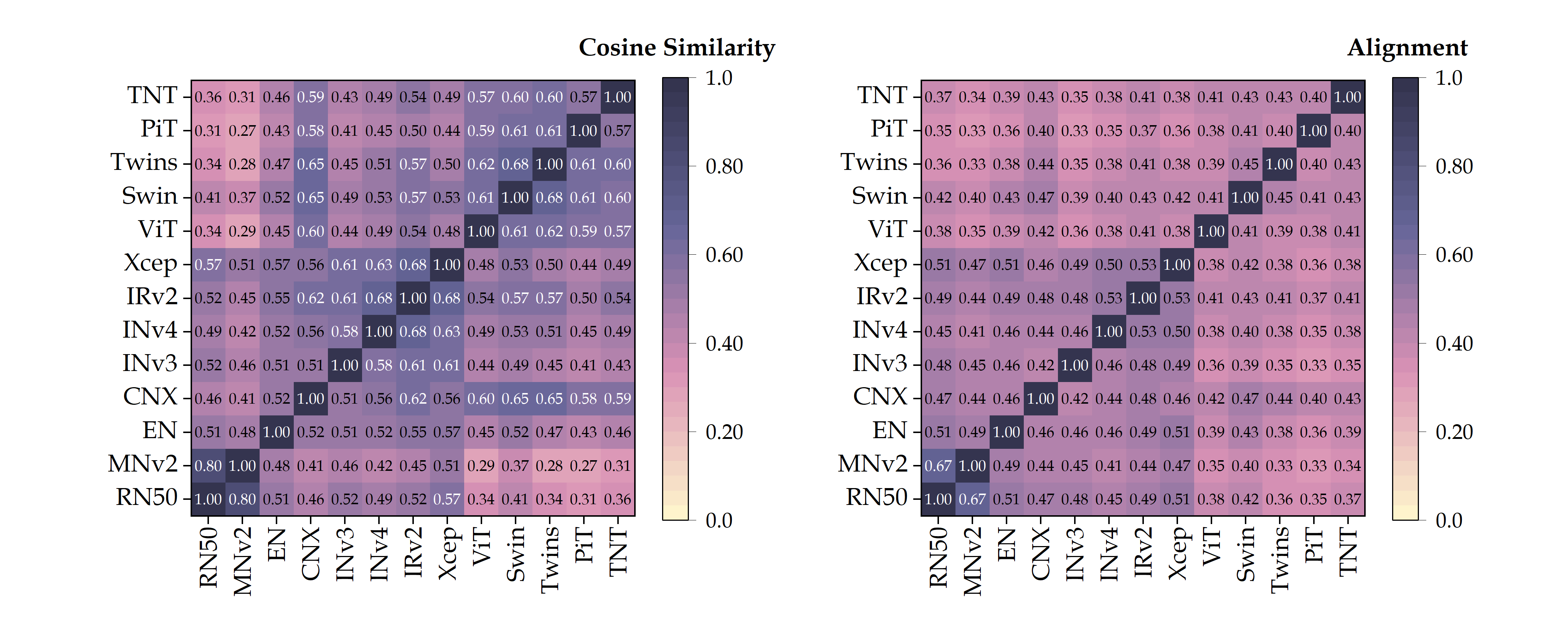}
		\vspace{-8mm}
		\caption{\textcolor{black}{Feature space alignment among ImageNet-trained architectures.}}
		\label{fig:in_models_alignment}
		\vspace{-5mm}
	\end{figure*}
	
	\section{Details of Evaluated Basic Transformations}\label{detailsoftrans}
	
	A range of basic transformations that are routinely utilized in training neural networks are investigated. We quantify the intensity of all transformations using a unified factor, denoted as $s \in [0,1]$. We establish the intensity factor set $\mathcal{S}$ as an arithmetic sequence comprising 100 equally spaced samples within the interval $[0,1]$. For transformations with multiple directions, we compile all variations at the same intensity to compute the average outcomes. For example, regarding rotation, we consider images rotated by an identical angle in both the clockwise and counterclockwise directions. Two special cases are translate and shear, for which we combine the variations in both horizontal and vertical directions and both the positive and negative angles. To encapsulate their multidirectional nature, we uniformly sample 100 grid points $(s_1, s_2)$ within the $[0,1] \times [0,1]$ space, yielding four variants at the same intensity level. We ensure a sufficiently broad range of variation for each transformation to secure more precise outcomes.
	
	These transformations are executed using \textit{torchvision}\footnote{\url{https://github.com/pytorch/vision/tree/main/torchvision/transforms}}, conforming to its parameter definitions. For an input $\bm{x}$ with dimensions $H \times W$, we succinctly describe the evaluation settings as follows.
	
	\begin{itemize}
		\item \textbf{Rotation} returns two images respectively rotated by $\texttt{angle}=s \times 180^{\circ}$ (clockwise) and $\texttt{angle}=-s \times 180^{\circ}$ (counterclockwise) degrees.
		\item \textbf{Scaling} returns two images respectively scaled to the \texttt{size} of $(1+1.5s)H\times (1+1.5s)W$ (enlargement) and $\frac{1}{(1+1.5s)}H\times \frac{1}{(1+1.5s)}W$ (reduction).
		\item \textbf{Shear} returns four images respectively sheared by $(s_1\times 90^{\circ}, s_2\times 90^{\circ})$, $(-s_1\times 90^{\circ}, s_2\times 90^{\circ})$, $(s_1\times 90^{\circ}, -s_2\times 90^{\circ})$, and $(-s_1\times 90^{\circ}, -s_2\times 90^{\circ})$ degrees parallel to the x-axis and y-axis.
		\item \textbf{Perspective} returns one image applied perspective transformation with $\texttt{distortion\_scale}=s$.
		\item \textbf{Flip} returns two images respectively flipped vertically and horizontally.
		\item \textbf{Crop} returns a square patch of $\bm{x}$ with an area of $(1-0.95s)\times H \times W$.
		\item \textbf{Translate} returns four images respectively translated by $(s_1\times H, s_2\times W)$, $(-s_1\times H, s_2\times W)$, $(s_1\times H, -s_2\times W)$, and $(-s_1\times H, -s_2\times W)$
		pixels along the x-axis and y-axis.
		\item \textbf{Solarize} returns one image with pixels values $x_{i,j}\geq(1-s)$ been inverted to $1-x_{i,j}$.
		\item \textbf{Hue} returns two images with the \texttt{hue} factor jittered to $-0.5s$ and $0.5s$, respectively.
		\item \textbf{Brightness}/\textbf{Contrast}/\textbf{Saturation} returns two images with the \texttt{brightness}/\texttt{contrast}/\texttt{saturation} factor jittered to $(1+4s)$ and $\frac{1}{1+4s}$, respectively.
	\end{itemize}

	\section{\textcolor{black}{Supporting Evidence for Self-Alignment}}
	
	\subsection{\textcolor{black}{Theoretical Intuition for Self-Alignment}}
	
	\textcolor{black}{Based on the inherent alignment between the surrogate $f$ and black-box $g$, we suggest that $\bm{\phi}(\bm{x}) \approx \bm{\psi}(\bm{x})$. For an AE $\bm{x}^{adv}$ crafted using a weakly-transferable algorithm (\eg TMI), the impact on the black-box model $g$ is limited, implying $\bm{\psi}(\bm{x}^{adv}) \approx \bm{\psi}(\bm{x}) \approx \bm{\phi}(\bm{x})$. Conversely, $\bm{x}^{adv}$ is specifically optimized to be markedly different from $\bm{\phi}(\bm{x})$ in the surrogate's space. By measuring how well a transformation restore this "distance" within the surrogate's own space (Self-alignment), we can effectively estimate its ability to maintain the adversarial direction when transferred to the black-box $g$ (Black-box Alignment). This mechanism allows self-alignment to perform blind effectiveness assessment of basic transformations.}
	
	\textcolor{black}{The core assumption lies in that well-trained models within the same semantic domain (\eg natural images) exhibit representation convergence. This implies that despite differences in network architectures or training objectives, their high-level feature spaces tend to align, resulting in a positive cosine similarity between their representations: $\text{CosSim}(\bm{\phi}(\bm{x}), \bm{\psi}(\bm{x})) \gg 0$. To substantiate this, we provide a broader spectrum of evaluations across diverse models and learning paradigms.}
	
	\begin{figure}[tbp]
		\centering	
		\includegraphics[width=1\linewidth]{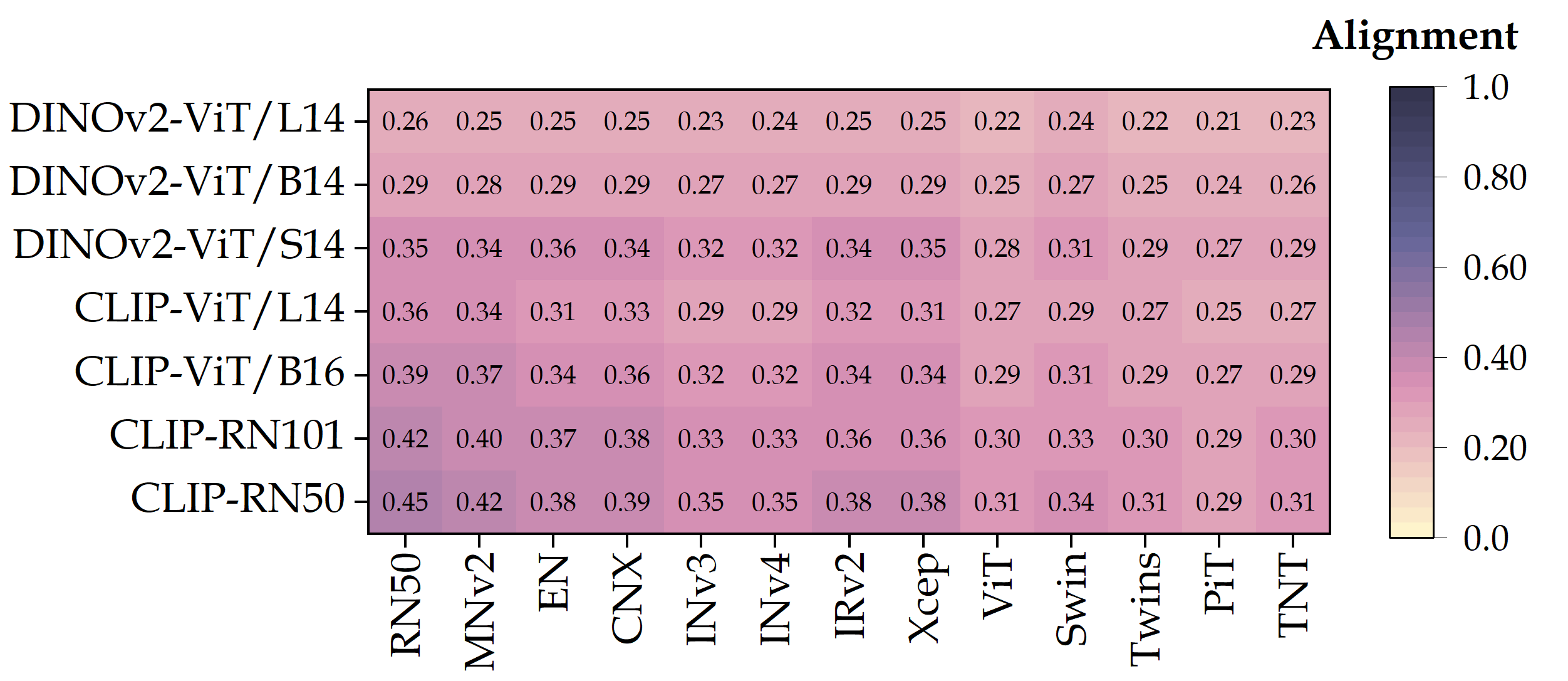}
		\vspace{-8mm}
		\caption{\textcolor{black}{Feature space alignment among distinct training paradigms and distributions.}}
		\label{fig:clip_dino_alignment}
		\vspace{-5mm}
	\end{figure}
	
	\subsection{\textcolor{black}{Evaluation Across Diverse Architectures}}
	
	\textcolor{black}{As illustrated in Fig. \ref{fig:in_models_alignment}, we evaluate both the Cosine Similarity and our proposed Alignment metric across various ImageNet-trained standard architectures. The results consistently show that the feature representations of these models are significantly aligned. This empirical evidence supports our hypothesis that the self-alignment of a surrogate model $f$ can serve as a reliable "proxy" for the actual alignment towards unknown black-box model $g$, provided they are trained on a similar domain.}
	
	\subsection{\textcolor{black}{Robustness to Different Training Distributions and Paradigms}}
	
	\textcolor{black}{A critical concern is whether this alignment hinges on the models being trained on identical datasets or under the same supervision. We investigate this by involving models with fundamentally different training regimes: CLIP \cite{radford2021learning} (trained on 400M image-text pairs via contrastive learning) and DINOv2 \cite{oquab2023dinov2} (trained via large-scale self-supervised learning without labels). Both models differ significantly from the ImageNet supervised training in terms of data distribution, scale, and objective. }
	
	\textcolor{black}{As shown in Fig. \ref{fig:clip_dino_alignment}, even under such substantial model discrepancies, the inter-model alignment remains at a considerable level. Since CLIP and DINOv2 lack classification heads compatible with the ImageNet label set, we directly utilize their high-level feature representations for alignment assessment. The results demonstrate that domain consistency, rather than identical training data, is the primary driver of representation alignment. This validates the broad applicability of self-alignment: even when the surrogate and victim models are trained with different paradigms and distributions, they still share a fundamental semantic understanding of the domain, allowing self-alignment to effectively assess transformations blindly.}
	
	\section{Details for  Self-Transferability Correlation}
	
	When performing self-transferability analysis for the exploration of inter-transformation correlations, we utilize ResNet-50 as the surrogate model and generate AEs using 12 existing attack methods. These attacks used for testing, along with their parameter configurations, remain consistent with those in Section 2.3 of the main manuscript. This includes the default parameters of TMI \cite{momentum2018dong,evading2019dong}, Admix \cite{admix}, SI \cite{nestrov2019lin}, RDI \cite{RDI}, BSR \cite{BSR}, ODI \cite{ODI}, T/H-Aug \cite{wei2023rethinking}, and the extended parameter combinations listed in Table \ref{extened_hyp}.

	We then calculate the average self-transferability of these AEs against the surrogate ResNet-50 coupled with the 12 basic transformations under investigation:
	\begin{equation}
		\textit{ST}_{\widehat{\mathcal{T}}} = \mathbb{E}_{\bm{x}\in\mathcal{X}, s\in\mathcal{S}}\left(f\circ \widehat{\mathcal{T}}_s(\bm{x}_{\mathcal{T}}^{adv},t) - f\circ \widehat{\mathcal{T}}_s(\bm{x},t)\right).
		\label{selftrans}
	\end{equation}
	Specifically, we uniformly sample 100 points within the transformation intensity intervals defined in Section \ref{detailsoftrans} and compute the average self-transferability across these sampled intensities.
	
	This process yields a $40\times12$ matrix where each row represents the self-transferability of an attack method against various basic transformations, and each column reflects the \textit{robustness} of a basic transformation against complex transformation-based attacks. The self-transferability correlation is analyzed by examining how different basic transformations are affected by various complex transformation-based attacks, thereby revealing their redundancy or correlation. In practice, the self-transferability correlation results is obtained by computing column-wise Pearson correlations across this matrix.
	
	\begin{table}[tbp]
		\caption{Extended hyperparameter configurations for self-transferability correlation analysis. Please refer to the corresponding references for detailed definitions. Default configurations are highlighted in \textcolor{blue}{\textbf{blue}}.}
		\vspace{-2mm}
		\label{extened_hyp}
		\centering
		\resizebox{\linewidth}{!}{
			\begin{tabular}{ll}
				\Xhline{1pt}
				Method	& Configurations \\
				\hline
				SSA \cite{SSA} & \texttt{tuning factor} $\in [0.1, 0.2, 0.3, 0.4, \textcolor{blue}{\textbf{0.5}}, 0.6, 0.7]$  \\
				DI \cite{inputdiversity2019xie} & \texttt{relative scale}  $\in [1.02, 1.04, 1.06, 1.08, \textcolor{blue}{\textbf{1.1}}, 1.3, 1.5, 1.7, 1.9, 2.1]$ \\
				SIA \cite{SIA} & \texttt{\#blocks} $\in [1, 2,  \textcolor{blue}{\textbf{3}}, 4, 5]$  \\
				DeCoWA \cite{decowa} & \texttt{noise scale} $\in [0.2, 0.4, 0.6, 0.8, 1.0, 1.2, 1.4, 1.6, 1.8, \textcolor{blue}{\textbf{2.0}}]$ \\
				\Xhline{1pt}
		\end{tabular}}
		\vspace{-2mm}
	\end{table}
	
	\begin{figure}[tbp]
		\centering	
		\includegraphics[width=1\linewidth]{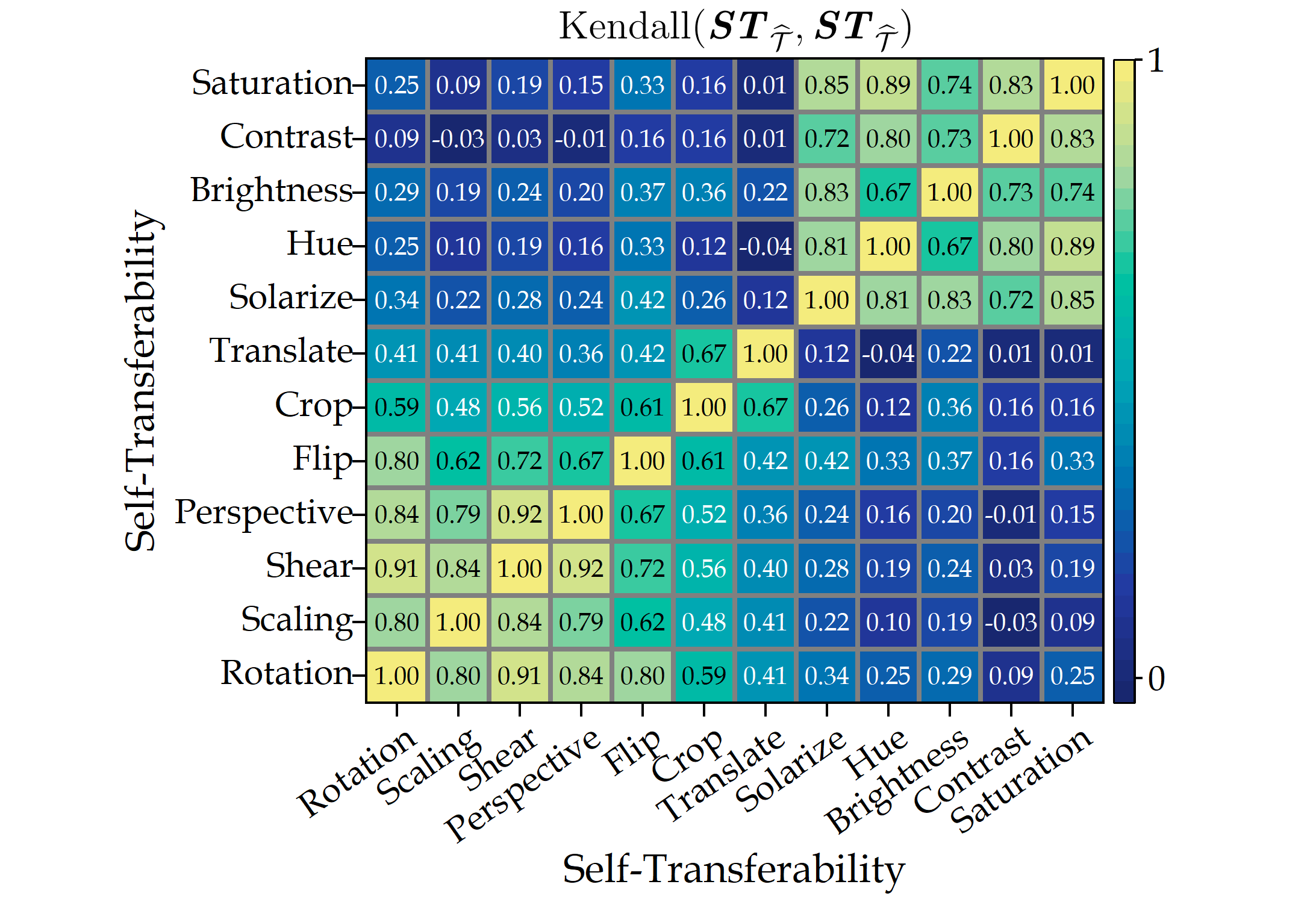}
		\vspace{-8mm}
		\caption{Robustness check of transformation redundancy via Kendall’s $\tau$ correlation matrix.}
		\label{fig:kendall}
		\vspace{-5mm}
	\end{figure}
	
	\textcolor{black}{\textbf{Robustness Check with Alternative Correlation Metrics.} To further validate the redundancy patterns observed in our self-transferability analysis, we provide a robustness check using Kendall’s $\tau$ rank correlation. While the Pearson correlation coefficient utilized in the main text is ideal for capturing the linear strength of associations between self-transferability, Kendall’s $\tau$ offers a complementary perspective by assessing the ordinal consensus (rank consistency) without assuming a linear relationship. As shown in Fig. \ref{fig:kendall}, the Kendall-based matrix exhibits clustering patterns that are highly consistent with the Pearson-based results. Specifically, the internal redundancy within geometric and color transformation categories remains evident even under this rank-based metric. This sanity check confirms that the observed transformation redundancies are an inherent characteristic of the optimization landscape rather than an artifact of a specific statistical measure, further solidifying the principled design of the S$^4$ST framework.}

	\begin{figure*}[tbp]
		\centering
		\includegraphics[width=0.45\linewidth]{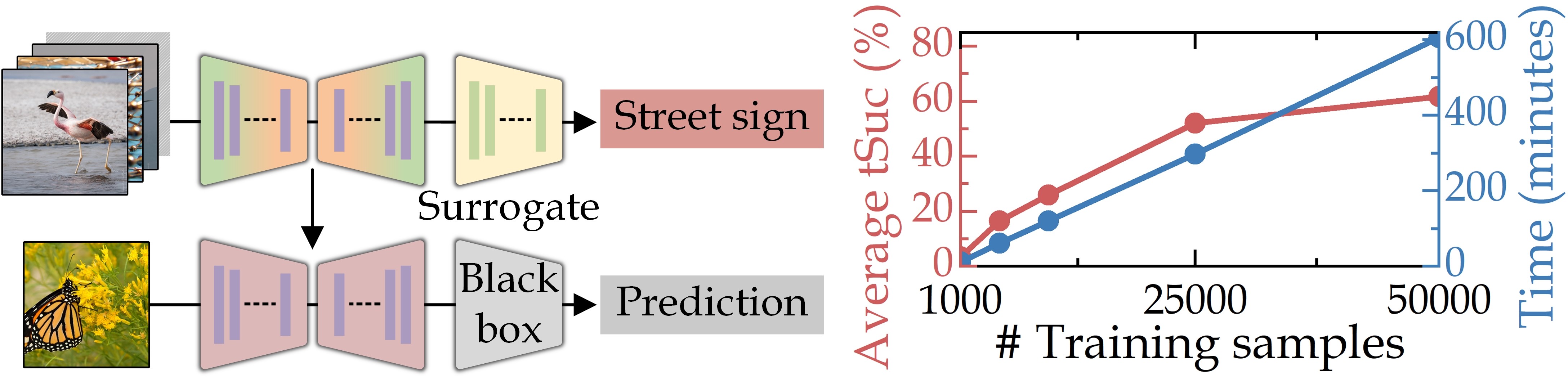} \quad
		\includegraphics[width=0.45\linewidth]{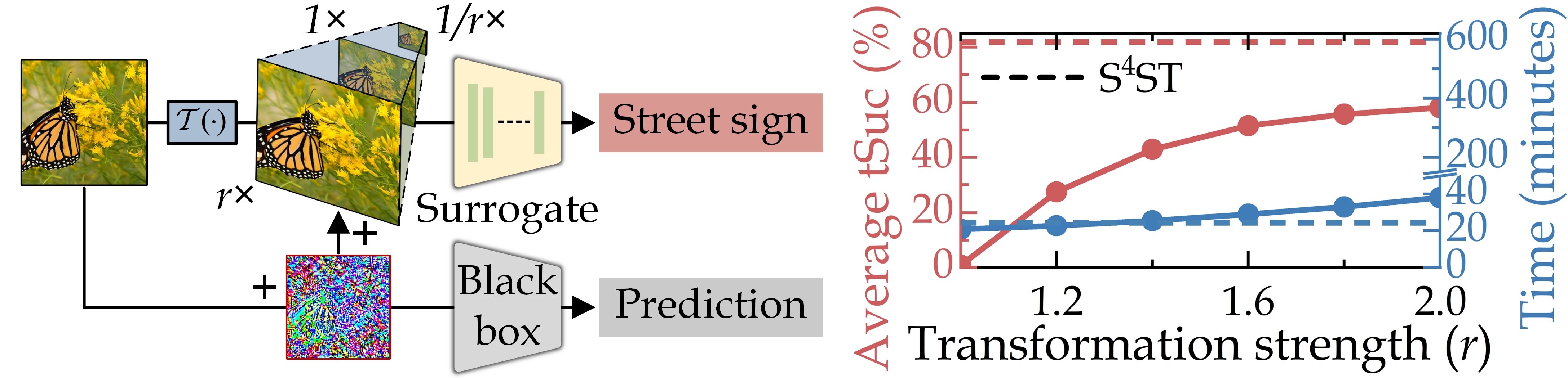}
		\caption{Comparison between M3D (left, data-reliant) and scaling-based data-free (right, gradient calculation) on a target label of \texttt{street sign}. Evaluating M3D across 1,000 target labels would require approximately 10,000 RTX 4090 hours—this is also a key reason we describe our method as fast and simple.}
		\label{m3d}
	\end{figure*}
	
	\section{Why 10-Targets (all source) Setting? Resource Comparison with Generative Attack}

	While generative TTAs generate adversarial examples quickly without gradient computation during inference, they require extensive data, computational resources, and time to train a separate generator for each target label. If adversarial examples for other labels are needed, retraining is necessary. This is why they cannot be applied to the full ImageNet-Compatible benchmark ($\sim$1000 target labels) and are instead evaluated on only 10 selected labels.
	
	We provide additional experiments comparing our method with M3D \cite{zhao2023minimizing}, an improved version of the classic generative TTA method TTP \cite{naseer2021generating}. As shown in Fig. \ref{m3d}, under default settings, M3D requires 10 hours of training on 50K ImageNet samples to achieve ~60\% tSuc for a single label (tested on an RTX 4090 GPU). In contrast, a data-free method based on simple scaling achieves comparable performance in ~40 minutes, where the increased time is mainly due to additional computation caused by enlarged input dimensions. Our S$^{4}$ST maintains the original input size, achieves over 80\% tSuc, and incurs minimal overhead.

	\section{Additional Visual Examples}
	
	\begin{figure*}[tbp]
		\centering
		\label{vis_supp}
		\includegraphics[width=\linewidth]{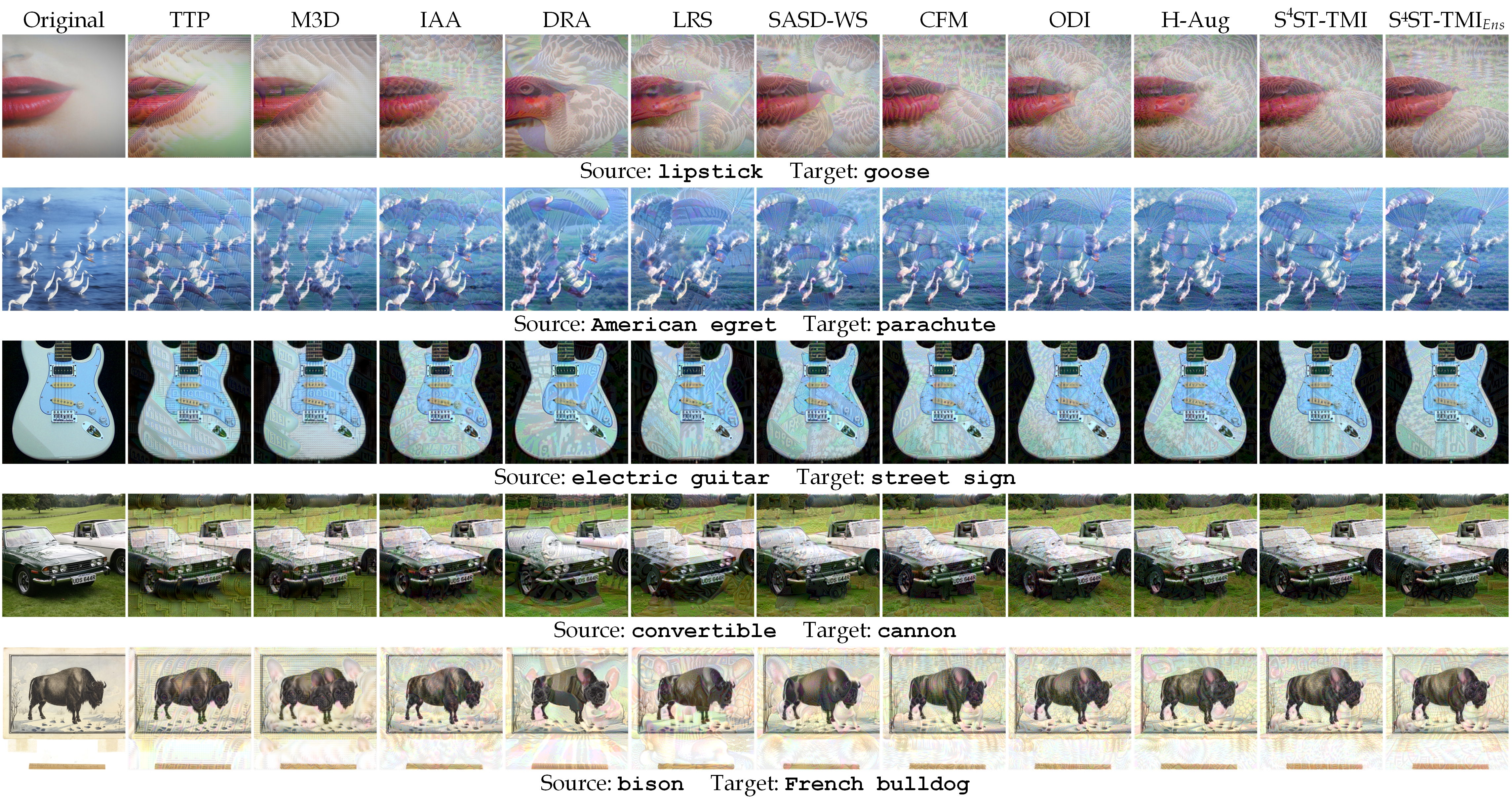}
		\vspace{-8mm}
		\caption{Visual comparisons among investigated TTAs.}
		\vspace{-3mm}
	\end{figure*}
	
	Fig. \ref{vis_supp} provides a comparison of AEs generated by various methods under the 10-Targets (all source) setting. Consistent with the visual inspection in the main manuscript, all methods demonstrate the behavior of incorporating original visual details to depict visual semantic patterns of the target classes. For optimal observation of these patterns, we recommend viewing the figure at high magnification on a color display.

	\section{Results on CXR Classification and Face Verification}
	
	This section provides detailed evaluation results for multi-label thoracic disease classification and face verification tasks. 
	
	\subsection{Self-Alignment Analysis}

	The premise of using feature-space model alignment rests on the assumption that the surrogate model $f$ and black-box model $g$ exhibit inherent alignment on the clean sample distribution, while showing misalignment on AEs overfitted to the surrogate model. That is, we expect:
	\begin{equation}
		\centering
		\begin{aligned}
			\operatorname{Alignment}(\Phi,\Psi) \sim \operatorname{Alignment}(&\Psi,\Psi^{adv}) > \\ \operatorname{Alignment}(\Psi^{adv},\Phi^{adv}&) \sim \operatorname{Alignment}(\Phi^{adv},\Phi)
		\end{aligned}
	\end{equation}
	The test sets for these two tasks contain 500 and 600 images, respectively, of which 467 and 300 images were attacked to generate adversarial examples. With the number of \textit{k}-nearest neighbors set to 20, the results are presented in Table \ref{modelalignment_nihcxr14_fv}. The outcomes generally align with expectations across both tasks. It is worth noting that on the CPLFW dataset, $\operatorname{Alignment}(\Psi^{adv},\Phi^{adv})$  shows a relatively high level, suggesting that overfitting may be less severe in this scenario.  This is validated in Table \ref{fv_results}, where the baseline attack method BIM demonstrates strong attack effectiveness: the average AUC across four victim models drops from 0.96 to 0.72 (0.24 $\downarrow$), while in the CXR classification task, the result decreases from 0.82 to 0.78 (0.04 $\downarrow$).

	Fig. \ref{alignment} presents the results of self-alignment per-transformation effectiveness analysis for both tasks. For the CXR classification task, we observe phenomena similar to those in natural images: scaling transformations remain superior across a wider intensity range. However, in the face verification task, geometric transformations generally exhibit a narrower effective range. This may be attributed to the fact that face images are typically aligned using keypoint detection and resized to a fixed scale during training, and usually do not employ RandomResizedCrop augmentation, significantly differing from natural image classification. Meanwhile, perspective transformation shows promising potential in pursuing self-alignment, which likely stems from the multi-view (pose) variations inherent in face images.
	
	\begin{table}[tbp]
		\caption{Mutual alignment between surrogate model and black-box models.}
		\vspace{-2mm}
		\label{modelalignment_nihcxr14_fv}
		\centering
		\begin{minipage}[b]{.45\linewidth}
			\centering
			\text{(a) NIH CXR14 classification}\\
			\resizebox{\linewidth}{!}{
				\begin{tabular}{lc}
					\Xhline{1pt}
					Case	& Value \\
					\hline
					$\operatorname{Alignment}(\Phi,\Psi)$ & 0.34 \\
					$\operatorname{Alignment}(\Psi,\Psi^{adv})$ & 0.19 \\
					$\operatorname{Alignment}(\Psi^{adv},\Phi^{adv})$ & 0.07 \\
					$\operatorname{Alignment}(\Phi^{adv},\Phi)$ & 0.06 \\
					\Xhline{1pt}
			\end{tabular}}
		\end{minipage} 
		\quad
		\begin{minipage}[b]{.45\linewidth}
			\centering
			\text{(b) CPLFW face verification}\\
			\resizebox{\linewidth}{!}{
				\begin{tabular}{lc}
					\Xhline{1pt}
					Case	& Value \\
					\hline
					$\operatorname{Alignment}(\Phi,\Psi)$ & 0.24 \\
					$\operatorname{Alignment}(\Psi,\Psi^{adv})$ & 0.35 \\
					$\operatorname{Alignment}(\Psi^{adv},\Phi^{adv})$ & 0.18 \\
					$\operatorname{Alignment}(\Phi^{adv},\Phi)$ & 0.09 \\
					\Xhline{1pt}
			\end{tabular}}
		\end{minipage} 
		
	\end{table}

	\begin{figure*}[tbp]
		\centering	
		\begin{minipage}[b]{.45\linewidth}
			\centering
			\subfloat[][NIH CXR14 classification]{\label{cxr_alignment}\includegraphics[width=1\linewidth]{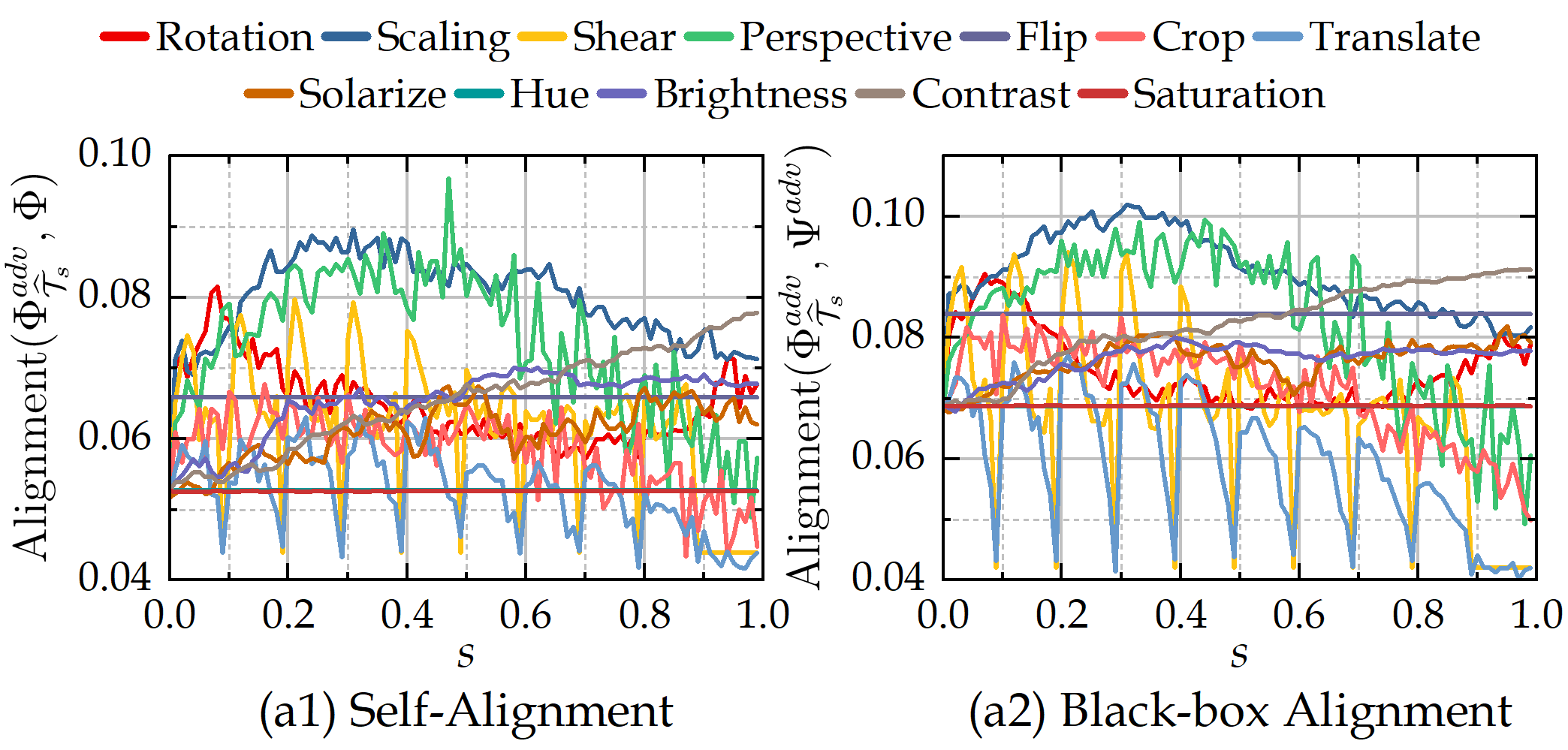}\vspace{-7mm}}
		\end{minipage}\quad\quad\quad\quad
		\begin{minipage}[b]{.45\linewidth}
			\centering
			\subfloat[][CPLFW face verification]{\label{fv_alignment}\includegraphics[width=1\linewidth]{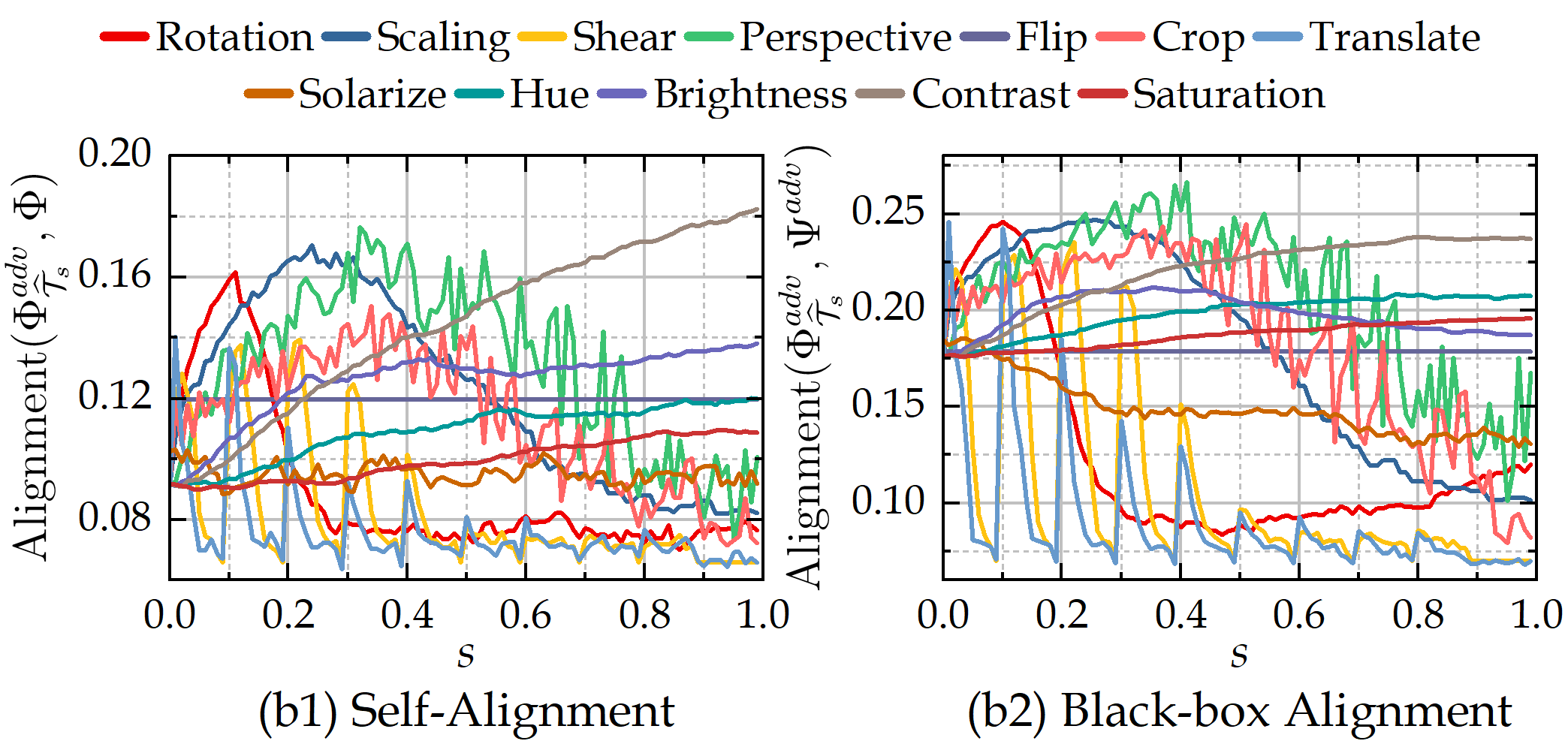}\vspace{-7mm}}
		\end{minipage}
		\caption{Results of \textbf{\textit{self-alignment}} per-transformation effectiveness analysis.}
		\label{alignment}
		\vspace{-5mm}
	\end{figure*}
	
	\begin{figure*}[tbp]
		\centering	
		\begin{minipage}[b]{.45\linewidth}
			\centering
			\subfloat[][NIH CXR14 classification]{\label{cxr_matrix}\includegraphics[width=1\linewidth]{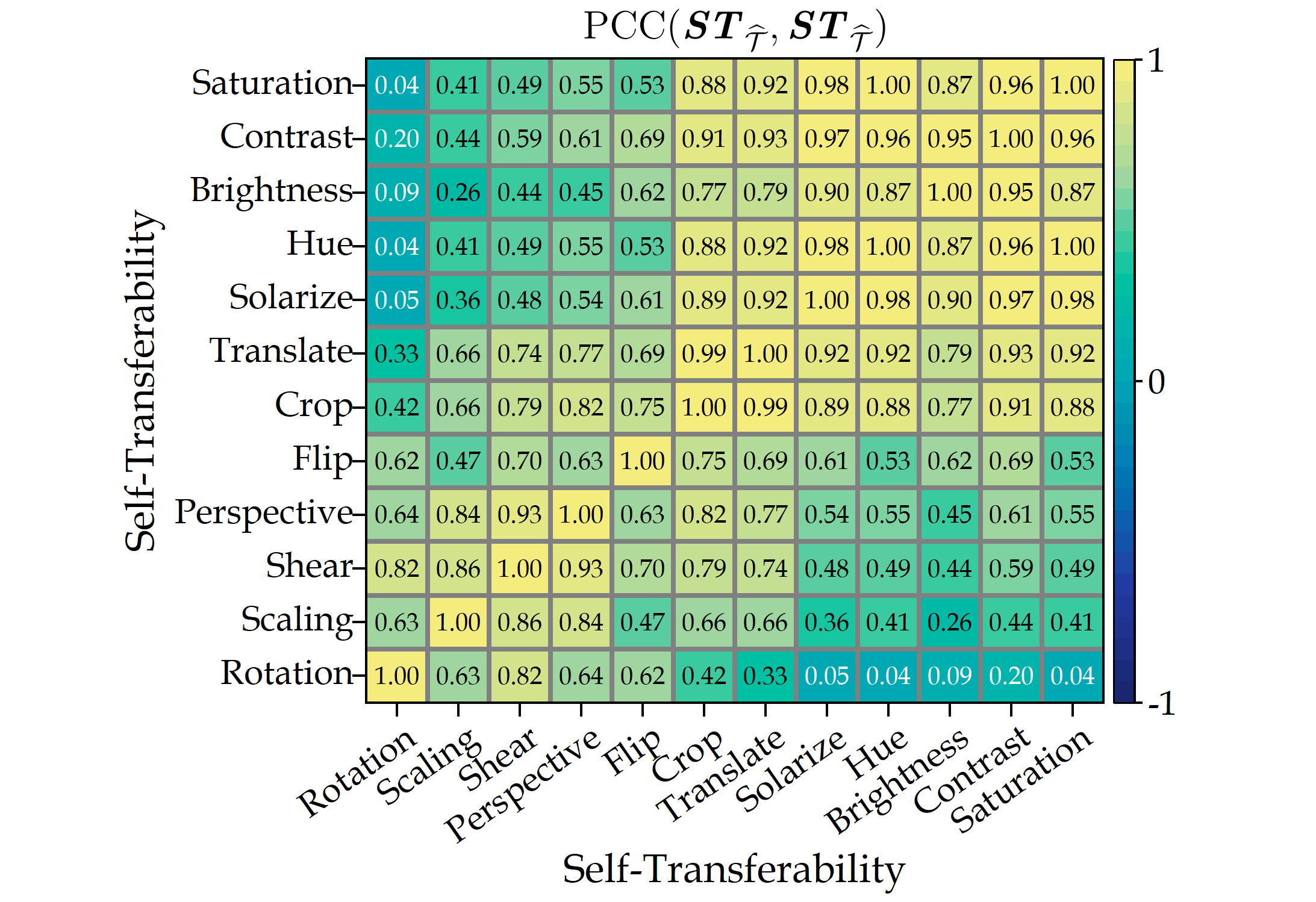}\vspace{-7mm}}
		\end{minipage}\quad\quad\quad\quad
		\begin{minipage}[b]{.45\linewidth}
			\centering
			\subfloat[][CPLFW face verification]{\label{fv_matrix}\includegraphics[width=1\linewidth]{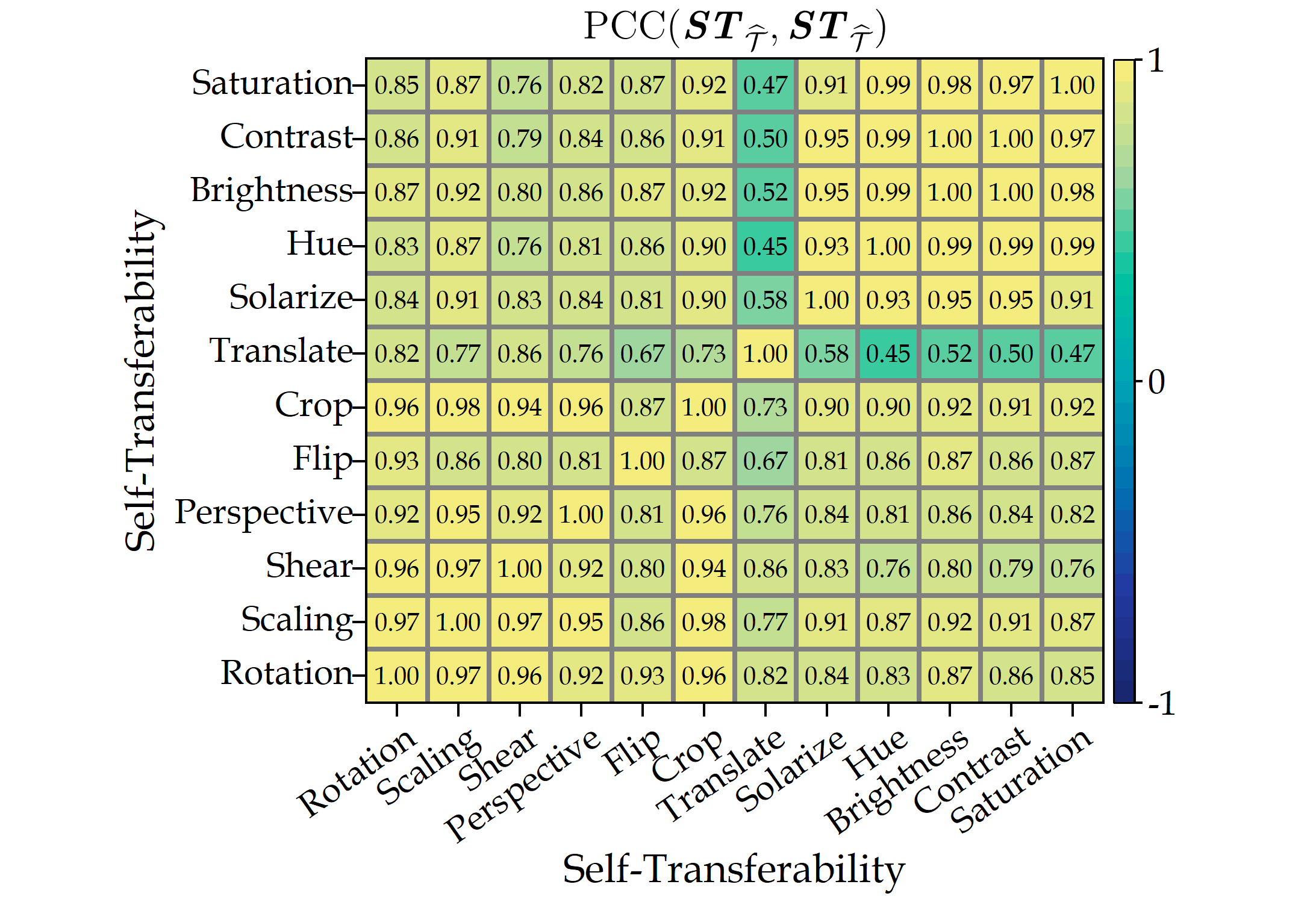}\vspace{-7mm}}
		\end{minipage}
		\caption{Results of \textbf{\textit{self-transferability}} inter-transformation correlation analysis.}
		\label{matrix}
		\vspace{-3mm}
	\end{figure*}
	
	\begin{figure*}[tbp]
		\centering
		\includegraphics[width=\linewidth]{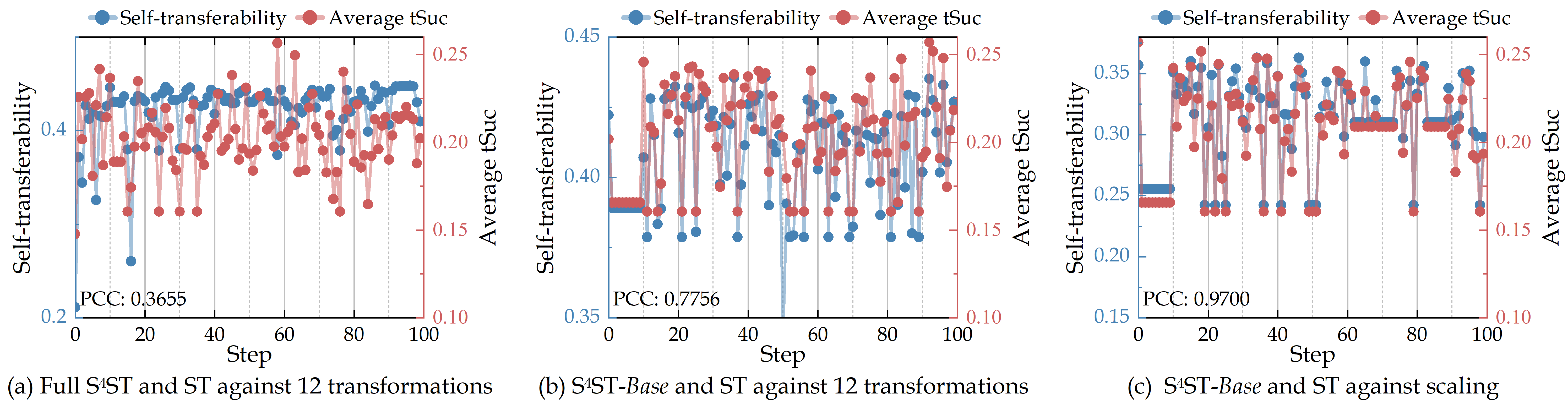}
		\vspace{-6mm}
		\caption{\textcolor{black}{The optimization proxy (self-transferability) \textit{vs.} actual black-box transferability during BO process of S$^4$ST adaption for face verification task.}
			\label{r2_fv_bo}}
		\vspace{-3mm}
	\end{figure*}
	
	\subsection{Self-Transferability Analysis}
	
	Fig. \ref{matrix} presents the results of the self-transferability inter-transformation correlation analysis. The CXR classification task exhibits patterns similar to natural images, showing intra-category redundancy and inter-category correlations. In contrast, the face verification task displays significant correlations across all transformations. This indicates that neither the characteristics of the training data nor the data augmentation methods employed during training have induced a clear bias in the model's tolerance toward specific transformations.
	
	\subsection{\textcolor{black}{Tuning Strategy for Face Verification}}
	
	\textcolor{black}{Guided by the distinct observations from our self-alignment and self-transferability analyses in the face verification task, we provide a further investigation and an adaptation strategy for S$^4$ST in such boundary cases. Unlike natural images with rich data diversity, face recognition models are typically trained on fixed facial regions cropped via landmark detection and often employ minimal data augmentation. Consequently, directly adopting the tuning strategy designed for natural images, which combines S$^4$ST-\textit{Base}, S$^4$ST-\textit{Block}, and S$^4$ST-\textit{Aug} using the average self-transferability of 12 basic transformations as the objective, fails to achieve effective optimization, as illustrated in Fig. \ref{r2_fv_bo}(a). }
	
	\textcolor{black}{To address this, we first simplify the framework by removing S$^4$ST-\textit{Block} and S$^4$ST-\textit{Aug} based on the observed weak tolerance to various transformations in our self-alignment analysis. By focusing solely on the parameters of S$^4$ST-\textit{Base} (\ie $r$ and $p_{\textit{r}}$) for optimization, the process is significantly improved, as shown in Fig. \ref{r2_fv_bo}(b), where the PCC between the optimization proxy and the actual black-box transferability reaches 0.78. Furthermore, leveraging the observations from Fig. 10 in the main text, we utilize the self-transferability specifically for scaling as the optimization proxy, which yields an ideal optimization trajectory. It is worth noting that this empirical prior was not utilized during the tuning for natural images to maintain a strict black-box setting. Similarly, according to the results in Fig. \ref{alignment}(b), using the self-transferability against perspective transformation as a proxy also produces desirable results. This aligns with the intuition that facial image distributions are inherently sensitive to perspective variations, while simultaneously reaffirming the cross-domain efficacy of our proposed scaling transformation.}
	
	\textcolor{black}{We also provide a streamlined adaptation strategy: specifically, we fix $p_{\text{r}}=1$ and determine the transformation intensity $r$ for S$^4$ST-\textit{Base} according to the tolerance observed in the self-alignment analysis. Simultaneously, the self-transferability analysis is utilized to decide whether to apply $S^4$ST-\textit{Block} and $S^4$ST-\textit{Aug}. In the face verification task, this simplified strategy yields the parameters $p_{\textit{r}}=1$ and $r=1.4$. Notably, this configuration performs fairly well and is highly consistent with the optimal parameters obtained through the full BO process ($p_{\textit{r}}=0.9, r=1.5$). This demonstrates that our proposed blind estimation measures can serve as efficient and reliable alternatives to exhaustive hyperparameter tuning in diverse and challenging domains.}
	
	\subsection{Detailed Results}
	
	Based on the above analysis, we finalized the S$^{4}$ST configurations for these two tasks. For the CXR task, we directly apply the version optimized for natural images. For the face verification task, we employ S$^{4}$ST-\textit{Base} with $p_{\textit{r}}=1$ and $r=1.4$, without incorporating complementary transformations or block-wise scaling.
	
	As shown in Tables \ref{cxr_results} and \ref{fv_results}, S$^{4}$ST achieves the best performance among all investigated competing methods. These results validate the generalization capability of the proposed self-alignment, self-transferability, and S$^{4}$ST design methodology across various image domains and tasks.

	\begin{table}[tbp]
		\centering
		\vspace{-1mm}
		\caption{Detailed results (AUC value) for NIH CXR14 classification. The surrogate is DN121-IN.}
		\label{cxr_results}
		\vspace{-2mm}
		\resizebox{\linewidth}{!}{
			\begin{tabular}{l|cccc|c}
				\Xhline{1pt}	
				Clean & 0.79 & 0.77 & 0.85  & 0.86 & 0.82  \\  \hline
				Attack & DN121-CXR & RN50-IN & ViTS-CXR & ViTB-CXR & Avg. \\ \hline \hline
				BIM & 0.77 (0.02$\downarrow$) & 0.75 (0.02$\downarrow$) & 0.82 (0.03$\downarrow$) & 0.79 (0.07$\downarrow$) & 0.78 (0.04$\downarrow$)  \\
				+Admix & 0.77 (0.02$\downarrow$)  & 0.77  & 0.82 (0.03$\downarrow$)  & 0.78 (0.08$\downarrow$)  & 0.79 (0.03$\downarrow$)  \\
				+SI & 0.58 (0.21$\downarrow$)  & 0.69 (0.08$\downarrow$)  & 0.77 (0.08$\downarrow$)  & 0.74 (0.12$\downarrow$)  & 0.70 (0.12$\downarrow$)  \\
				+SSA & 0.60 (0.19$\downarrow$)  & 0.59 (0.18$\downarrow$)  & 0.74 (0.11$\downarrow$)  & 0.71 (0.15$\downarrow$)  & 0.66 (0.16$\downarrow$)   \\
				+DI & 0.69 (0.10$\downarrow$)  & 0.72 (0.05$\downarrow$)  & 0.80 (0.05$\downarrow$)  & 0.76 (0.10$\downarrow$)  & 0.74 (0.08$\downarrow$)   \\
				+RDI & 0.61 (0.18$\downarrow$)  & 0.72 (0.05$\downarrow$)  & 0.78 (0.07$\downarrow$)  & 0.74 (0.12$\downarrow$)  & 0.71 (0.11$\downarrow$)   \\
				+ODI & 0.46 (0.33$\downarrow$)  & 0.54 (0.23$\downarrow$)  & \textcolor{t2}{\textbf{0.70 (0.15$\downarrow$)}}  & 0.63 (0.23$\downarrow$)  & 0.58 (0.24$\downarrow$)  \\
				+SIA & \textcolor{t2}{\textbf{0.40 (0.39$\downarrow$)}}  & 0.52 (0.25$\downarrow$)  & 0.72 (0.13$\downarrow$)  & 0.66 (0.20$\downarrow$)  & 0.57 (0.25$\downarrow$)   \\
				+DeCoWA & 0.49 (0.30$\downarrow$)  & 0.56 (0.21$\downarrow$)  & 0.72 (0.13$\downarrow$)  & 0.67 (0.19$\downarrow$)  & 0.61 (0.21$\downarrow$)   \\
				+BSR & 0.44 (0.35$\downarrow$)  & 0.51 (0.26$\downarrow$)  & 0.72 (0.13$\downarrow$)  & 0.66 (0.20$\downarrow$)  & 0.58 (0.24$\downarrow$)   \\
				+T-Aug & 0.49 (0.30$\downarrow$)  & 0.53 (0.24$\downarrow$)  & \textcolor{t2}{\textbf{0.70 (0.15$\downarrow$)}}  & 0.64 (0.22$\downarrow$)  & 0.59 (0.23$\downarrow$)   \\
				+H-Aug & 0.53 (0.26$\downarrow$)  & \textcolor{t2}{\textbf{0.45 (0.32$\downarrow$)}}  & \textcolor{t1}{\textbf{0.68 (0.17$\downarrow$)}}  & \textcolor{t2}{\textbf{0.62 (0.24$\downarrow$)}}  & \textcolor{t2}{\textbf{0.53 (0.29$\downarrow$)}}  \\
				+S$^{4}$ST & \textcolor{t1}{\textbf{0.38 (0.41$\downarrow$)}}  & \textcolor{t1}{\textbf{0.36 (0.41$\downarrow$)}}  & \textcolor{t1}{\textbf{0.68 (0.17$\downarrow$)}}  & \textcolor{t1}{\textbf{0.61 (0.25$\downarrow$)}}  & \textcolor{t1}{\textbf{0.51 (0.31$\downarrow$)}}   \\ 
				\Xhline{1pt}
		\end{tabular}}
		\vspace{-3mm}
	\end{table}
	
	\begin{table}[tbp]
		\centering
		\vspace{-1mm}
		\caption{Detailed results (AUC value) for CPLFW face verification. The surrogate is CurricularFace-IRN101-MS1M \cite{huang2020curricularface}.}
		\label{fv_results}
		\vspace{-2mm}
		\resizebox{\linewidth}{!}{
			\begin{tabular}{l|cccc|c}
				\Xhline{1pt}	
				Clean & 0.94 & 0.97 & 0.98  & 0.96 &  0.96 \\  \hline
				& \multicolumn{2}{c|}{FaceNet \cite{schroff2015facenet}} & \multicolumn{2}{c|}{ArcFace \cite{deng2019arcface}} &  \\ \hline
				Attack & IRv1-WebFace & IRv1-VGGFace2 & IRN50-MS1M & MFN-MS1M  & Avg. \\ \hline \hline
				BIM & 0.73 (0.21$\downarrow$) & 0.83 (0.14$\downarrow$) & 0.60 (0.38$\downarrow$) & 0.70 (0.26$\downarrow$) & 0.72 (0.24$\downarrow$)  \\
				+Admix & 0.71 (0.23$\downarrow$) & 0.81 (0.16$\downarrow$) & 0.55 (0.43$\downarrow$)  & 0.67 (0.29$\downarrow$) & 0.69 (0.27$\downarrow$) \\
				+SI  & 0.67 (0.27$\downarrow$) & 0.73 (0.24$\downarrow$) & 0.48 (0.50$\downarrow$) & 0.61 (0.35$\downarrow$) & 0.62 (0.34$\downarrow$) \\
				+SSA & \textcolor{t2}{\textbf{0.65 (0.29$\downarrow$)}} & \textcolor{t2}{\textbf{0.71 (0.26$\downarrow$)}} & \textcolor{t2}{\textbf{0.46 (0.52$\downarrow$)}} & 0.61 (0.35$\downarrow$) & \textcolor{t2}{\textbf{0.61 (0.35$\downarrow$)}} \\
				+RDI & 0.66 (0.28$\downarrow$) & 0.73 (0.24$\downarrow$) & 0.47 (0.51$\downarrow$) & 0.61 (0.35$\downarrow$) & 0.62 (0.34$\downarrow$)  \\
				+ODI & 0.66 (0.28$\downarrow$) & 0.73 (0.24$\downarrow$) & 0.49 (0.49$\downarrow$) & \textcolor{t2}{\textbf{0.60 (0.36$\downarrow$)}} & 0.62 (0.34$\downarrow$) \\
				+SIA & \textcolor{t2}{\textbf{0.65 (0.29$\downarrow$)}} & 0.73 (0.24$\downarrow$) & 0.47 (0.51$\downarrow$) & \textcolor{t2}{\textbf{0.60  (0.36$\downarrow$)}} & \textcolor{t2}{\textbf{0.61 (0.35$\downarrow$)}} \\
				+DeCoWA & 0.82 (0.12$\downarrow$) & 0.88 (0.09$\downarrow$) & 0.74 (0.24$\downarrow$) & 0.75 (0.21$\downarrow$) & 0.79 (0.17$\downarrow$) \\
				+BSR & 0.76 (0.18$\downarrow$) & 0.83 (0.14$\downarrow$) & 0.67 (0.31$\downarrow$) & 0.71 (0.25$\downarrow$) & 0.74 (0.22$\downarrow$) \\
				+T-Aug & 0.66 (0.28$\downarrow$) & 0.73 (0.24$\downarrow$) & 0.53 (0.45$\downarrow$) & 0.61 (0.35$\downarrow$) & 0.63 (0.33$\downarrow$) \\
				+S$^{4}$ST-\textit{Base} & \textcolor{t1}{\textbf{0.58 (0.36$\downarrow$)}} & \textcolor{t1}{\textbf{0.63 (0.34$\downarrow$)}} & \textcolor{t1}{\textbf{0.40 (0.58$\downarrow$)}} & \textcolor{t1}{\textbf{0.53 (0.43$\downarrow$)}} & \textcolor{t1}{\textbf{0.54 (0.42$\downarrow$)}} \\
				\Xhline{1pt}
		\end{tabular}}
		\vspace{-3mm}
	\end{table}
	
	\bibliographystyle{IEEEtran}    
	\bibliography{s4st_refs_new}         
	
	\begin{IEEEbiography}[{\includegraphics[width=1in,height=1.25in,clip,keepaspectratio]{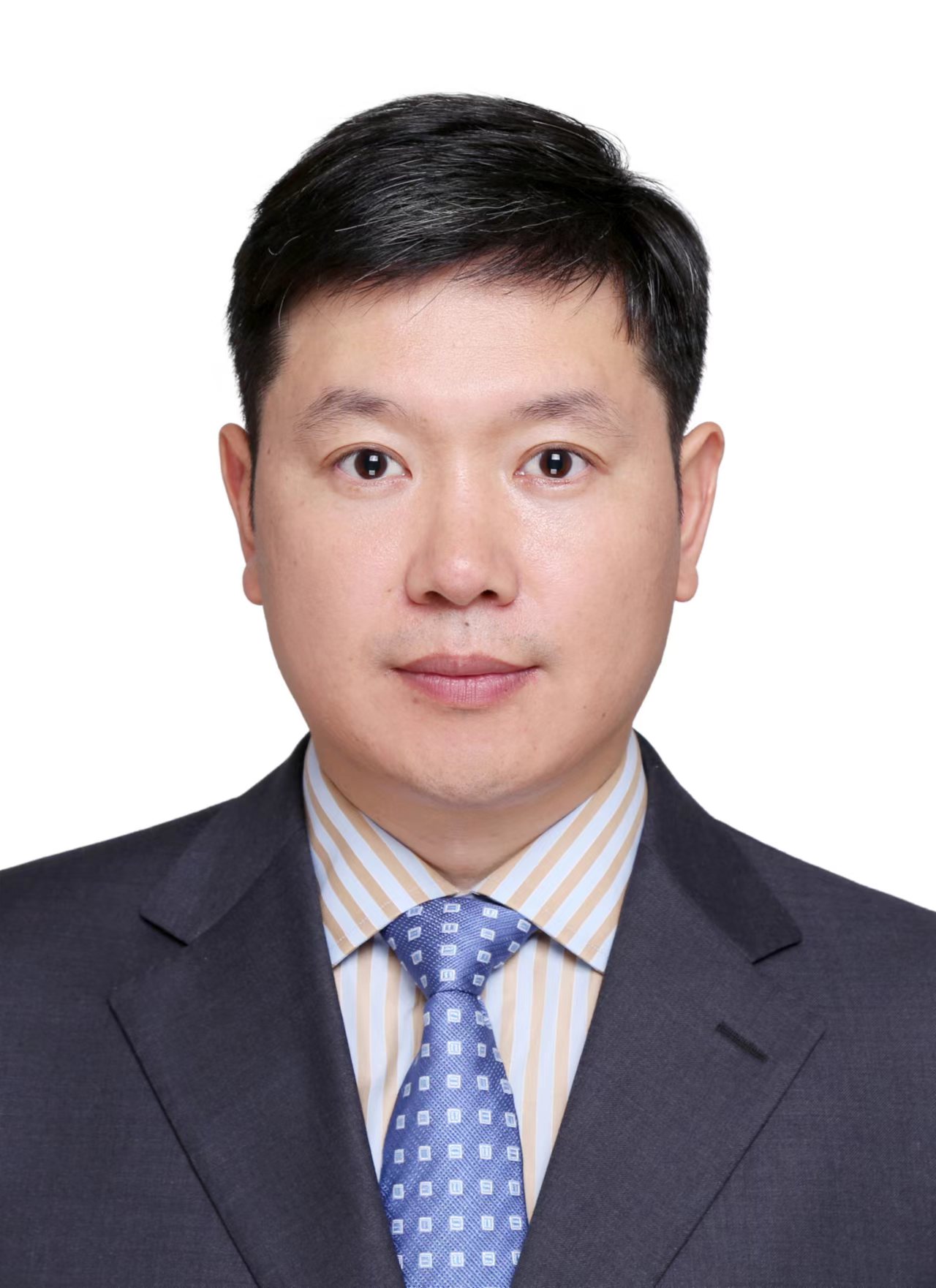}}]{Yongxiang Liu}
		received his Ph.D. degree in Information and Communication Engineering from the National University of Defense Technology (NUDT), Changsha, China, in 2004. Currently, He is a Full Professor in the College of Electronic Science and Technology, NUDT. His research interests mainly include remote sensing imagery analysis, radar signal processing, object recognition and Inverse SAR (ISAR) imaging, and machine learning. 
	\end{IEEEbiography}

	\begin{IEEEbiography}[{\includegraphics[width=1in,height=1.25in,clip,keepaspectratio]{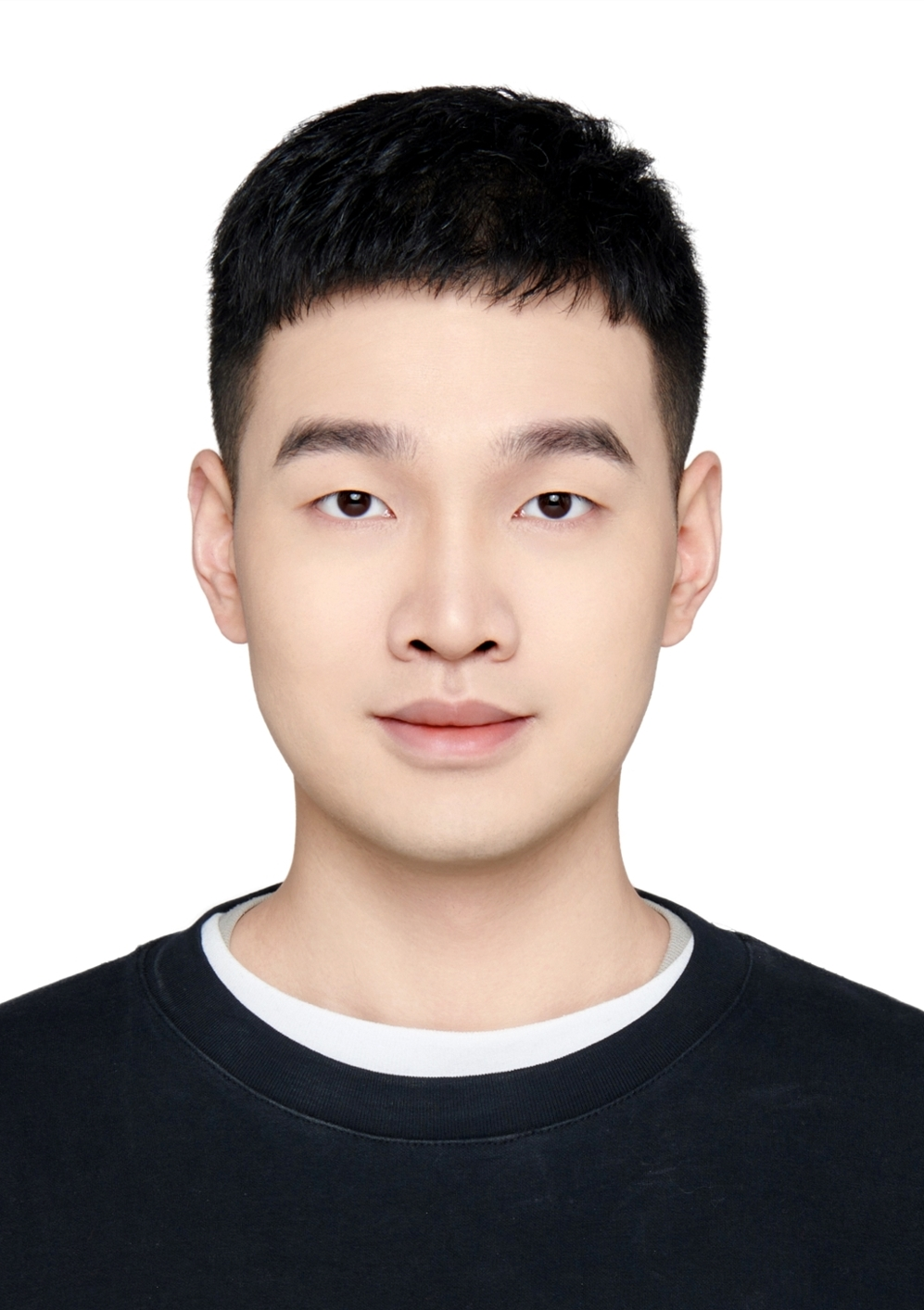}}]{Bowen Peng}
		received his B.S. and M.S. degrees from the National University of Defense Technology (NUDT), Changsha, China, respectively in 2020 and 2022. He is pursuing the Ph.D. degree with the Comprehensive Situation Awareness Group, College of Electronic Science and Technology, NUDT. His research interests include adversarial robustness of deep learning and trustworthy remote sensing object recognition.
	\end{IEEEbiography}
	
	\begin{IEEEbiography}[{\includegraphics[width=1in,height=1.25in,clip,keepaspectratio]{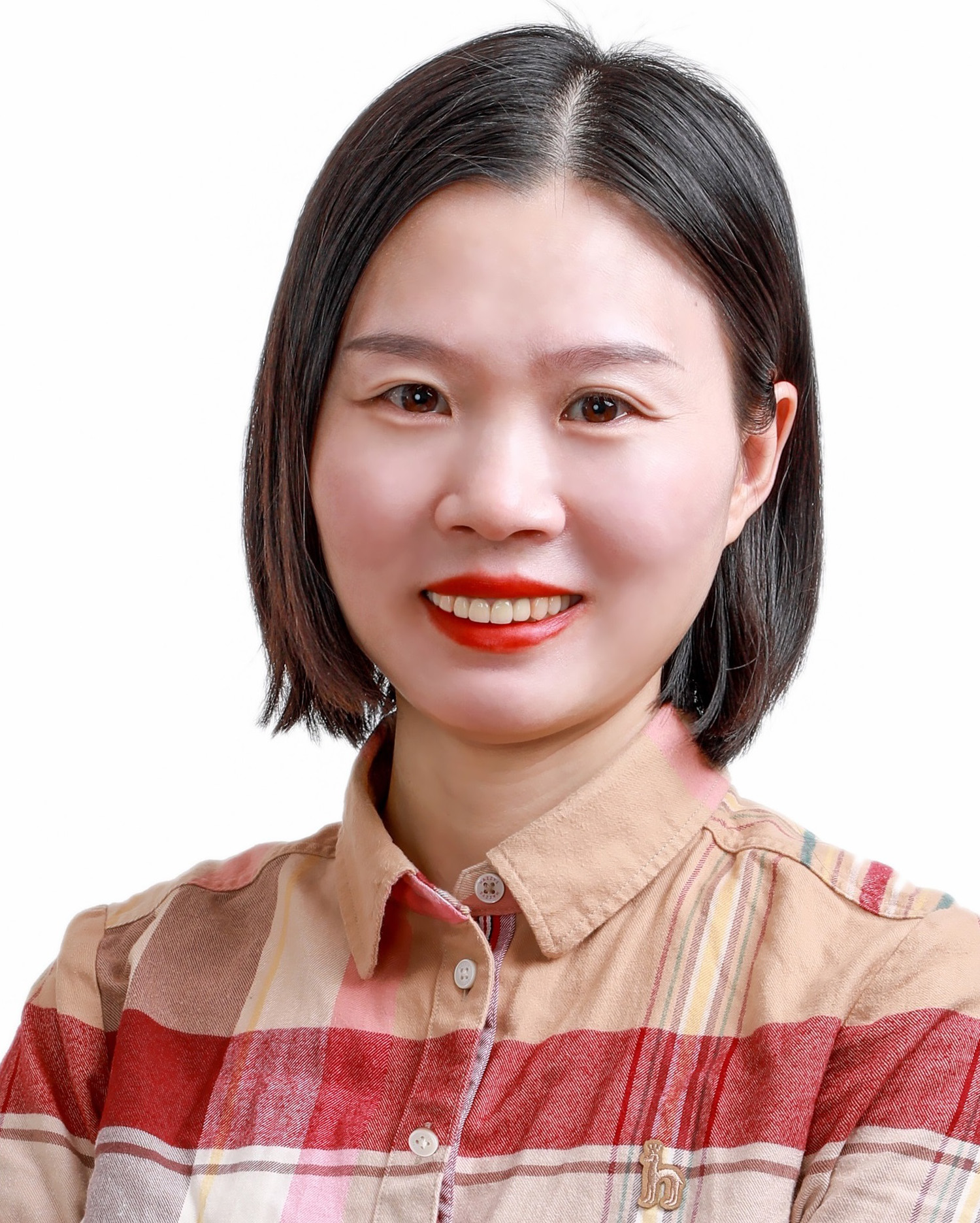}}]{Li Liu}
		(SM' 19) received her Ph.D. degree in Information and Communication Engineering from the National University of Defense Technology (NUDT), China, in 2012. She is now a full professor with the College of Electronic Science and Technology, NUDT. During her Ph.D. study, she spent more than two years as a visiting student with the University of Waterloo, Canada, from 2008 to 2010. From 2015 to 2016, she spent ten months visiting the Multimedia Laboratory at the Chinese University of Hong Kong. From 2016 to 2018, she worked as a senior researcher with the Machine Vision Group, University of Oulu, Finland. She served as the leading guest editor for special issues in IEEE Transactions on Pattern Analysis and Machine Intelligence (IEEE TPAMI) and International Journal of Computer Vision. Her current research interests include computer vision, pattern recognition, remote sensing image analysis, and machine learning. Her papers currently have more than 20,000 citations according to Google Scholar. She currently serves as associate editor for IEEE Transactions on Geoscience and Remote Sensing (IEEE TGRS), IEEE Transactions on Circuits and Systems for Video Technology (IEEE TCSVT), and Pattern Recognition.
	\end{IEEEbiography}
	
	\begin{IEEEbiography}[{\includegraphics[width=1in,height=1.25in,clip,keepaspectratio]{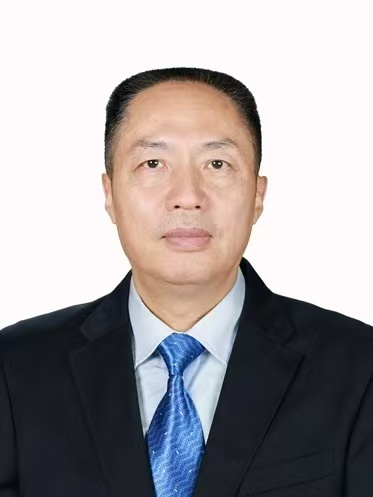}}]{Xiang Li} received his B.S. degree from Xidian University, Xi'an, China, in 1989, and the M.S. and Ph.D. degrees from the National University of Defense Technology (NUDT), Changsha, China, in 1995 and 1998, respectively. He is currently a Professor with the NUDT and was elected as an academician of the Chinese Academy of Sciences in 2022. His research interests include automatic target recognition, signal detection, nonlinear signal processing, and machine learning.
	\end{IEEEbiography}

	\vfill

\end{document}